\documentclass[preprint]{aastex}

\usepackage{isotope}
\usepackage{rotating}
\usepackage{capt-of}
\usepackage{multirow}
\usepackage{longtable}
\usepackage{lscape}
\usepackage{morefloats}
\usepackage{color}
\usepackage{graphicx}
\usepackage[section] {placeins}
\newcommand{\spr}{\mbox{$s$-process}}

\newcommand{\cdr}{\isotope[13]{C}}

\newcommand{\msun}{\ensuremath{\, M_\odot}}

\newcommand{\mppnp}{\textsf{mppnp}}

\newcommand{\beq}{\begin{equation}}
\newcommand{\beqa}{\begin{eqnarray}}
\newcommand{\eeq}{\end{equation}}
\newcommand{\eeqa}{\end{eqnarray}}
\newcommand{\bedis}{\begin{displaymath}}
\newcommand{\edis}{\end{displaymath}}

\graphicspath{{figures/}}

\slugcomment{DRAFT: \today}

\shorttitle{$s$-process in low-mass AGB stars}
\shortauthors{}

\begin{document}

\title{NuGrid stellar data set - III. Updated low-mass AGB models and s-process nucleosynthesis with metallicities Z=0.01, Z=0.02 and Z=0.03}
\author{U. Battino\altaffilmark{1,6},
  A. Tattersall\altaffilmark{1,6},
C. Lederer-Woods\altaffilmark{1,6},
F. Herwig\altaffilmark{4,5,6},
P. Denissenkov\altaffilmark{4,5,6},
R. Hirschi\altaffilmark{3,6,8},
R. Trappitsch\altaffilmark{6,7},
J. W. den Hartogh\altaffilmark{6,9},
M. Pignatari\altaffilmark{2,5,6,9}
}

\altaffiltext{1}{School of Physics and Astronomy, University of Edinburgh, UK}
\altaffiltext{2}{E.A. Milne Centre for Astrophysics, Dept of Physics and Mathematics, University of Hull, HU6 7RX, United Kingdom}
\altaffiltext{3}{Astrophysics group, Lennard-Jones Laboratories, Keele University, ST5 5BG, UK}
\altaffiltext{4}{Department of Physics \& Astronomy, University of Victoria, Victoria, BC, V8P5C2 Canada.}
\altaffiltext{5}{Joint Institute for Nuclear Astrophysics, USA.}
\altaffiltext{6}{The NuGrid collaboration, \url{http://www.nugridstars.org}}
\altaffiltext{7}{Nuclear and Chemical Sciences Division, Lawrence Livermore National Laboratory, Livermore, CA 94551}
\altaffiltext{8}{Kavli IPMU (WPI), The University of Tokyo, Kashiwa, Chiba 277-8583, Japan}
\altaffiltext{9}{Konkoly Observatory, Research Centre for Astronomy and Earth Sciences, Hungarian Academy of Sciences, Konkoly Thege M. út 15-17, 1121, Budapest, Hungary}

\begin{abstract}
  The production of the neutron-capture isotopes beyond iron that we observe today in the solar system is the result
  of the combined contribution of the $r$-process, the $s$-process and possibly the $i$-process.
  Low-mass AGB (2 $<$ M/M$_\odot$ $<$ 3) and massive (M $>$ 10 M$_\odot$) stars have been identified as the sites of the $s$-process.
  In this work we consider the evolution and nucleosynthesis of low-mass AGB stars.
  We provide an update of the NuGrid Set models,
  adopting the same general physics assumptions but using an updated convective-boundary mixing model
  accounting for the contribution from internal gravity waves.
  The combined data set includes the initial masses M$_{ZAMS}$/M$_\odot$  = 2, 3
  for Z = 0.03, 0.02, 0.01.
  These models are computed with the MESA stellar code and
  the evolution is followed up to the end of the AGB phase.
  The nucleosynthesis was calculated for all isotopes in post-processing with the NuGrid \mppnp\ code.
  The convective boundary mixing model leads to the formation of a $^{13}$C-pocket three times wider
  compared to the one obtained in the previous set of models,
  bringing the simulation results now in closer agreement with observations.
  We also discuss the potential impact of other processes inducing mixing, like rotation, adopting parametric models
  compatible with theory and observations.
  Complete yield data tables, derived data products and online analytic data access are provided.
\end{abstract}


\section{Introduction}
\label{intro}

Around half of the elements beyond Fe, are the result of $s$-process nucleosynthesis
\citep['slow' neutron-capture process][]{cameron:57,burbidge:57,gallino:98}
taking place in massive stars  ( M $>$ 10 M$_{\odot}$) and low mass AGB stars (2 $<$ M/M$_\odot$ $<$ 3).
In particular, low-mass AGB stars are the main site of the main $s$-process component \citep{gallino:98},
i.e. the nucleosynthesis process mainly responsible for around half of the neutron-capture element abundances between Zr and Bi in the Solar System.
The AGB phase starts when the star has exhausted both H and He in the center,
leaving an inert degenerate carbon-oxygen (CO-) core surrounded by a thin He-intershell and a H-burning shell where nuclear energy is released and the structure is maintained in equilibrium. These shells are surrounded by an extended H-rich convective envelope.
For the majority of the AGB lifetime, nuclear energy is released in the H-burning shell.
At the same time He and other H-burning ashes are accumulated on the top of He-intershell underneath,
until He-burning starts and thin-shell instability occurs \citep{kippenhahn:90}, 
triggering a violent thermonuclear-runaway known as thermal-pulse (TP) at typical temperature around 3$\times$10$^{8}$ K,
enough to activate the neutron release via $^{22}$Ne($\alpha$,n)$^{25}$Mg with high density (about 10$^{11}$ neutrons cm$^{-3}$) lasting a few years. In these conditions the neutron exposure (defined as the total neutron flux integrated over time) is low because of the very short timescales, preventing the neutron-capture flow to feed anything beyond the Sr-peak, but leaving a clear fingerprint in the isotopic ratios around branching points (unstable nuclei whose lifetimes are comparable to the neutron-capture timescale). An example is the production of $^{96}$Zr, which requires high neutron densities to trigger neutron captures on $^{95}$Zr that has a half-life of 64 days \citep{lugaro:14}. 
The TP will develop a pulse-driven convective zone (PDCZ), which mixes in the whole intershell the neutron-capture isotopes just synthetised and causes the expansion of the outer convective envelope \citep{herwig:05}.
The temperature in the outer layer of the expanding convective envelope will thus decrease and opacity increase,
which will make the convective motions more efficient.
This last fact has two main consequences:
1) C and heavy element-rich material from the intershell is mixed into the convective envelope and brought to the surface \citep[this event is known as third dredge-up, hereafter TDU][]{straniero:95,herwig:05};
2) p-capture reactions are triggered on the abundant $^{12}$C which will produce $\sim$10$^{-4}$ M$_{\odot}$  
$^{13}$C-rich material at the top of the intershell, known as $^{13}$C-pocket.
This represents the main neutron source for the $s$-process via the $^{13}$C($\alpha$,n)$^{16}$O reaction
(at typical intershell temperature T $\sim$ 0.09 GK) \citep{straniero:95, gallino:98}.
For these reasons, the $s$-process is very sensitive to how convective boundaries and hence chemical mixing across them are described.
Because of the about three times lower temperature compared to typical He-flash conditions,
the $s$-process in the $^{13}$C-pocket is characterized by low neutron densities (N$_{n}$ $\sim$10$^{7}$cm$^{-3}$),
but lasts for several thousand years, achieving high-neutron exposure and producing second (Ba-region) and third (Pb-region) elements \citep{herwig:05}.

Over the last 20 years, many efforts were dedicated to clarify the mixing mechanism at the boundary between the convective envelope and the He-intershell responsible for the formation of the  $^{13}$C-pocket.
\cite{herwig:97}, guided by multi-D simulations by \cite{freytag:96}, proposed an exponentially-decaying diffusion mixing  operating during the TDU.
Later on, \cite{langer:99} investigated the impact of rotational induced mixing, which was shown by \cite{herwig:03} to not produce a large
enough $^{13}$C-pocket.
\cite{denissenkov:03} proposed a model based on internal gravity waves (IGW) induced by the convective motion in the envelope.
Moreover, \cite{straniero:06} and \cite{cristallo:09} proposed an advection scheme as an alternative to the diffusion scheme.
Finally, \cite{nucci:14} suggested magnetic-buoyancy as a physical mechanism to transport H from the envelope into the He-C rich intershell.

Recently, \cite{ritter:18} (hereafter RI18)
computed a grid of stellar evolution and full-nucleosynthesis models over a wide range of both initial mass and metallicity,
from 1 M$_{\odot}$ to 25 M$_{\odot}$.
The same stellar evolution code, post-processing code and nuclear reaction network was adopted over the whole initial mass range,
ensuring a high degree of internal consistency.
The overshoot model by \cite{herwig:97} and \cite{herwig:00} was adopted to describe the convective-boundary-mixing (CBM) processes.
This formed \cdr-pockets producing a surface $s$-process enrichment between three and four times weaker than the highest abundances observed on C-stars \citep{busso:01,abia:02,zamora:09} and barium stars \citep{pereira:11,cseh:18}.
This motivated \cite{battino:16} to develop a new CBM prescription guided by the model proposed by \cite{denissenkov:03},
and tested it at the bottom of the convective envelope during TDU episodes.
The main result was a large increase of the pocket size up to around 10$^{-4}$ M$_{\odot}$ .

In this work, we provide an update of the NuGrid data set presented in RI18,
focusing on low-mass AGB models with initial metal content around solar value.
In particular,
we want to apply the \cite{battino:16} CBM model to increase the $s$-process production of NuGrid's Set II models \citep{ritter:18},
keeping the other initial settings and stellar evolution code the same (MESA , revision 3709. See \cite{paxton:10} for details)
and using the post-processing nucleosynthesis code \mppnp\ \citep{herwig:08,pignatari:16}.
This work is organized as follow. In Section \ref{sec:tool} we describe the stellar code and post-processing nucleosynthesis tools.
In Section \ref{sec:mesa} the stellar models are presented,
while in Section \ref{sec:postpro} we present our results, comparing them with a large set of observables. 
Our conclusions are given in Section \ref{sec:conclusions}.

\section{Computational methods}
\label{sec:tool}


The stellar models presented in this section are computed using the stellar code MESA \citep[revision 3709,][]{mesa}.
We used the solar distribution from \cite{grevesse:93}.
The modelling assumptions are the same as in RI18, except we also computed Z=0.03 models.
We adopted for Z=0.03 models the same modelling inputs as for Z=0.02 and Z=0.01 models,
including the same mass-loss formula  \citep{blocker:95} and efficiency parameter $\eta_{R}$ during the C-rich phase.
After the TDU event that makes the surface C/O ratio larger than 1.15,
we choosed the $\eta_{R}$ value only depending on the initial mass,
being $\eta_{R}$ = 0.04 and $\eta_{R}$ = 0.08 for the 2 and 3 M$_{\odot}$ models respectively.
For the simulations the MESA  nuclear network $agb.net$ is used,
including 18 isotopes from protons to $^{22}$Ne linked by nuclear-reactions as in RI18.
Here we also included $^{20}$Ne, $^{24}$Mg, $^{28}$Si and $^{56}$Fe in order to avoid mass-conservation issues at the beginning of the simulations, without linking them to the other isotopes with nuclear reactions.
The CBM modelling is included the same way as in \cite{battino:16}.

The post-processing code \mppnp\ was used, which is described in detail in \cite{pignatari:16}.
The stellar structure evolution data are computed and saved with MESA for all zones at all time
steps, and then used as input and processed with \mppnp\ .
This means that the stellar structure and the full nucleosynthesis are computed separately,
hence requiring less computing time and resources.

The network is the same as in RI18.
Exceptions relevant for this work are the neutron-capture cross sections of \isotope[90,91,92,93,94,95,96]{Zr}:
for which we adopted rates recommended by \cite{lugaro:14}, based on recent experimental measurements \citep[][]{tagliente:12}.

\section{Description of the stellar models}
\label{sec:mesa}

Table \ref{tab:model_def} lists the six stellar models calculated in this work, corresponding to three different initial metallicities
(Z=0.01, Z=0.02 and Z=0.03) and two initial masses (M=2,3 M$_{\odot}$ ).
All models' name start with a $'m'$ followed by a number indicating the initial mass in solar masses.
After this, initial metallicity is expressed by what follows $'z'$.
For example, considering m3z2m2 $'m3'$ means that this is a 3 M$_{\odot}$ model,
$'z2m2'$ is to be read as Z=2$\times$10$^{-2}$ , where $'m2'$ means $'minus$ $two'$ referring to the exponent.
Key global features like core masses and lifetimes are given for all the models,
which have all been computed with the same stellar code and input physics of RI18,
but with the CBM model by \cite{battino:16} during TDUs.
This is why we also included the values from RI18 in Table \ref{tab:model_def}
(with the exception of Z=0.03 models which were not considered in RI18) and we compared our results to it all along the present study.

\begin{figure}[htbp]
\begin{center}
\includegraphics[scale=0.4]{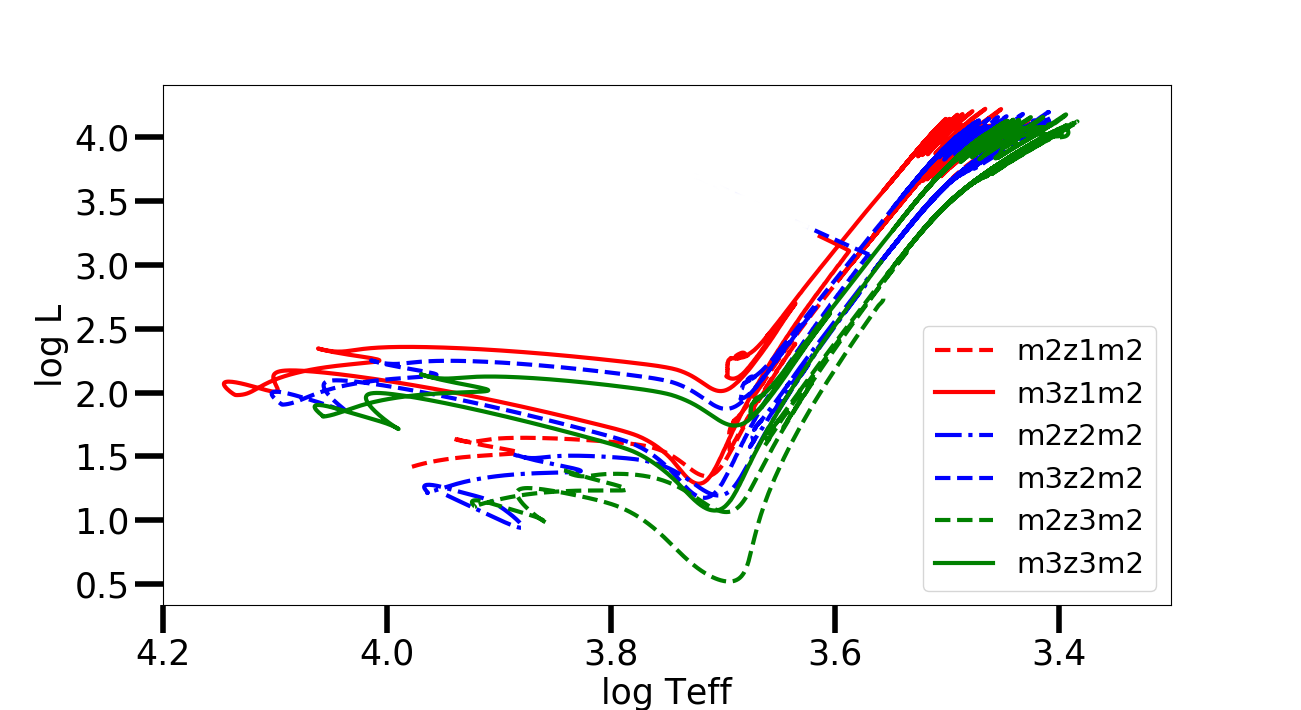}
\end{center}
\caption{HR diagram of tracks from all models listed in Table \ref{tab:model_def} from the pre main-sequence to the tip of the AGB phase.}
\label{hrds}
\end{figure}

\begin{figure}[htbp]
\centering
\resizebox{11.8cm}{!}{\rotatebox{0}{\includegraphics{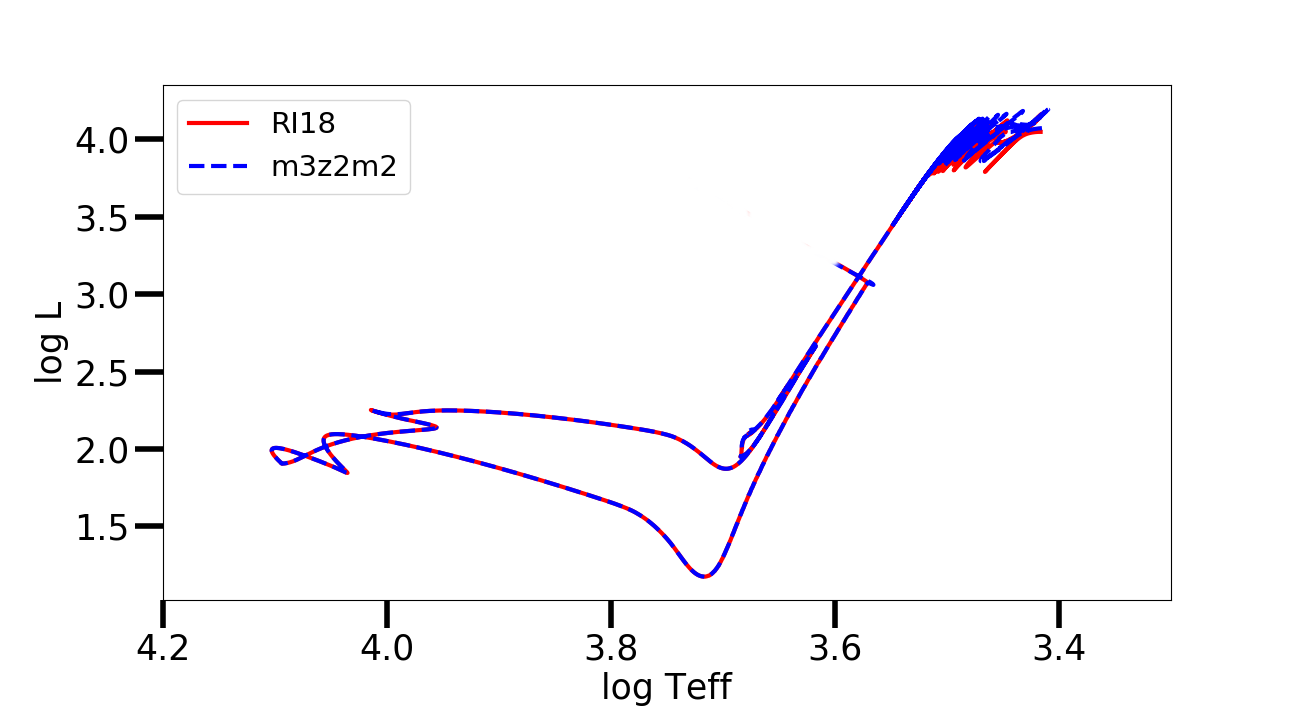}}}
\resizebox{11.8cm}{!}{\rotatebox{0}{\includegraphics{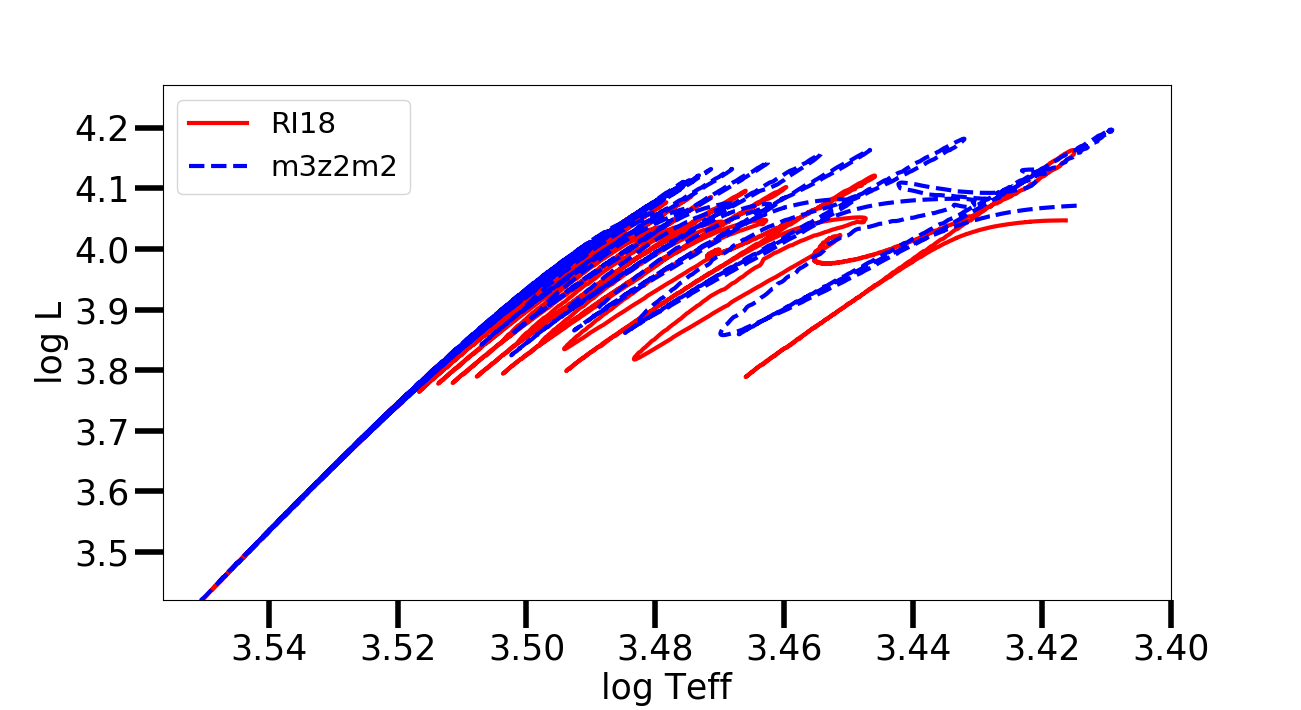}}}
\caption{Upper panel: Comparison between HR diagrams of our m3z2m2 model and the corresponding one (same initial mass and metallicity)
from RI18. Lower panel: Zoom on the AGB phase.}
\label{hrds:compSet1}
\end{figure}

Fig. \ref{hrds} shows the HR diagram tracks from all the models listed in Table \ref{tab:model_def}
from the pre main-sequence to the tip of the AGB phase.
Additionally, a comparison between HR diagrams of our m3z2m2 model and the corresponding one (same initial mass and metallicity)
from RI18 is given in Fig. \ref{hrds:compSet1}.
The two models are globally consistent along the evolution towards the AGB phase,
where it is evident that the TP events experienced by the m3z2m2 model are more luminous than RI18 by Log(L)$\sim$0.7 L$_{\odot}$.
Given the relation between core mass and luminosity during the AGB phase \citep{paczynski:70},
this is consistent with core masses listed in Table \ref{tab:model_def} being significantly larger than in RI18.

\subsection{The impact of different third-dredge-up efficiencies}
\label{sec:TDUs}

\begin{figure}[htbp]
\begin{center}
\includegraphics[scale=0.4]{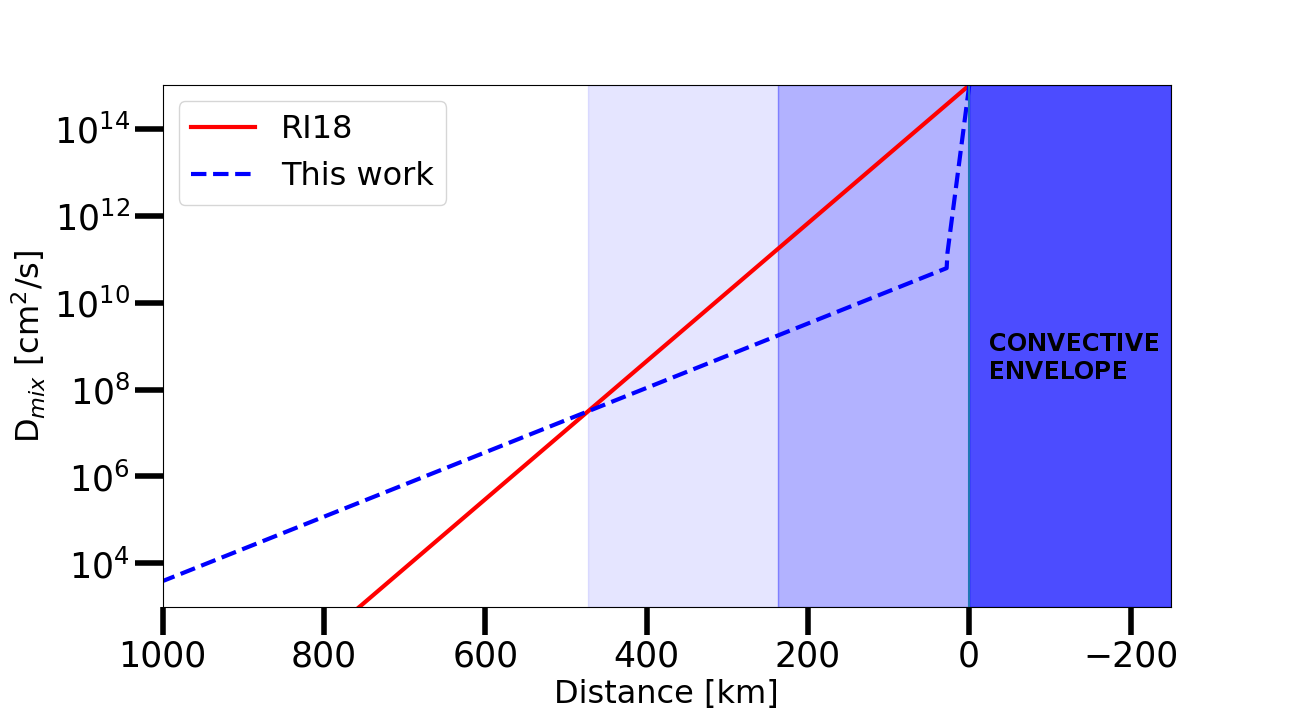}
\end{center}
\caption{CBM profiles from the m3z2m2 model and RI18 are shown.
  The dark-shaded area represents the convective envelope, with the Schwarzschild convective boundary being the left border.
RI18 mixing dominates over our prescription in the mid- and light-shaded areas, with an efficiency higher than 100 times in the mid-shaded one.}
\label{cbms}
\end{figure}

\begin{figure}[htbp]
\centering
\includegraphics[width=1.0\textwidth]{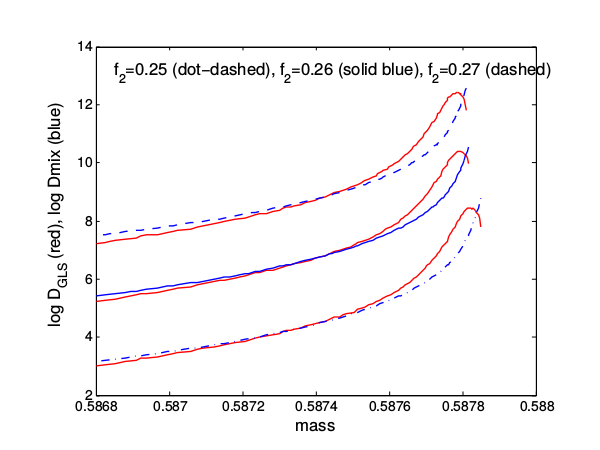}
\caption{Comparison between the diffusion coefficient profile calculated using the GLS prescription for the IGW mixing from
  \cite{denissenkov:03} (the red curves) and the one derived for the CBM with the parameterization used in this work (the blue curves).
  The dot–dashed, solid and dashed blue curves with their adjacent red curves show comparisons for the cases of
  f$_{2}$=0.25, f$_{2}$=0.26 and f$_{2}$=0.27. To make them more visible,
  the dashed and dot-dashed lines are shifted along the vertical axis by log(D) = 2 up and down relative to the solid line.
  The bump on the log(D$_{GLS}$) profile near the convective boundary is due to a rapid decrease of the thermal diffusivity K with depth
  accompanied by a fast increase of the buoyancy,
  and by the fact that D$_{GLS}$ is proportional to NK (Equation (15) in \cite{denissenkov:03}).
  The f$_{2}$=0.27 case has been selected as standard since it provides the best-fit of the IGW profile
  in the layers where the $^{13}$C-pocket forms (7$<$Log(D$_{mix}$)$<$8).}
\label{pavel:fit}
\end{figure}

The difference in core mass between RI18 and the present study is linked to the different CBM model here during TDUs.
TDU affects the core-mass growth along the thermally-pulsing AGB (TP-AGB) phase.
We recall here that the efficiency of the TDU is usually expressed with

\begin{equation}
\lambda = (\delta M_{DUP} / \delta M_{c} )
\end{equation}

defined as the fraction of the dredged-up mass ($\delta$M$_{DUP}$) over the core mass increment along an
inter-pulse period ($\delta$ M$_{c}$). Every time a dredge-up episode takes place with an efficiency $\lambda$,
the core mass decreases over the TDU duration by $\lambda$ $\delta$M$_{c}$ \citep[see][]{marigo:12,kalirai:14}.
As a consequence, the growth of the core-mass is smaller in models adopting CBM than in models not including it,
or adopting a less efficient CBM. This aspect can be clarified looking at Fig. \ref{cbms}.
The CBM profiles from the m3z2m2 model and RI18  are shown.
The dark-shaded area represents the convective envelope, with the Schwarzschild boundary placed at the left border.
Typically, the convective mixing coefficient in the envelope is around Log(D/(cm$^2$s)) $\sim$ 15.
In RI18 the mixing coefficient decays exponentially as a function of distance from the Schwartchild boundary,
using the exponential overshooting formalism of \cite{herwig:00}.
In order to consider the IGW contribution, which is not included in RI18,
we adopt the double-exponential CBM of \cite{battino:16}.
The CBM input parameters for all models in Table \ref{tab:model_def} are given in Table \ref{tab:cbm}.
All three input parameters were calibrated to fit the IGW-mixing profile by \cite{denissenkov:03}
in the layers where the $^{13}$C-pocket forms, as shown in Fig. \ref{pavel:fit}.
Hence, no fine-tuning was done to directly match observables, since this calibration is purely theory and simulations-based.
In Table \ref{tab:cbm}, we include the 3M$_{\odot}$ , Z=0.02 model from RI18 for comparison,
which only required the f1 parameter since it was calculated with the single exponential overshooting scheme of \cite{herwig:00}.
We also add two models, m3z2m2-hCBM and m3z3m2-hCBM, calculated with a $D_2$ parameter 4.3 times larger than the others,
consistently with the typical IGW mixing uncertainty described by \cite{denissenkov:03}.
As a consequence, it experiences more efficient TDUs and form larger $^{13}$C-pockets.

\begin{figure}[htbp]
\centering
\resizebox{14.8cm}{!}{\rotatebox{0}{\includegraphics{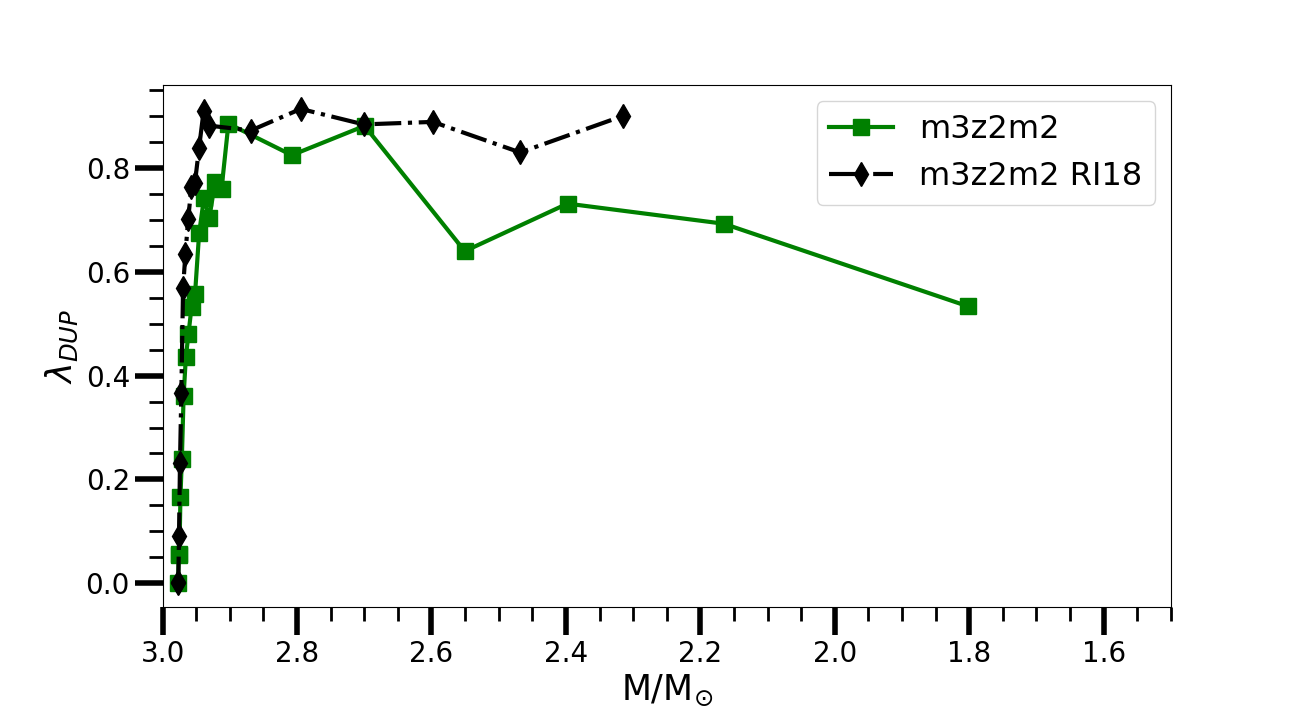}}}
\resizebox{14.8cm}{!}{\rotatebox{0}{\includegraphics{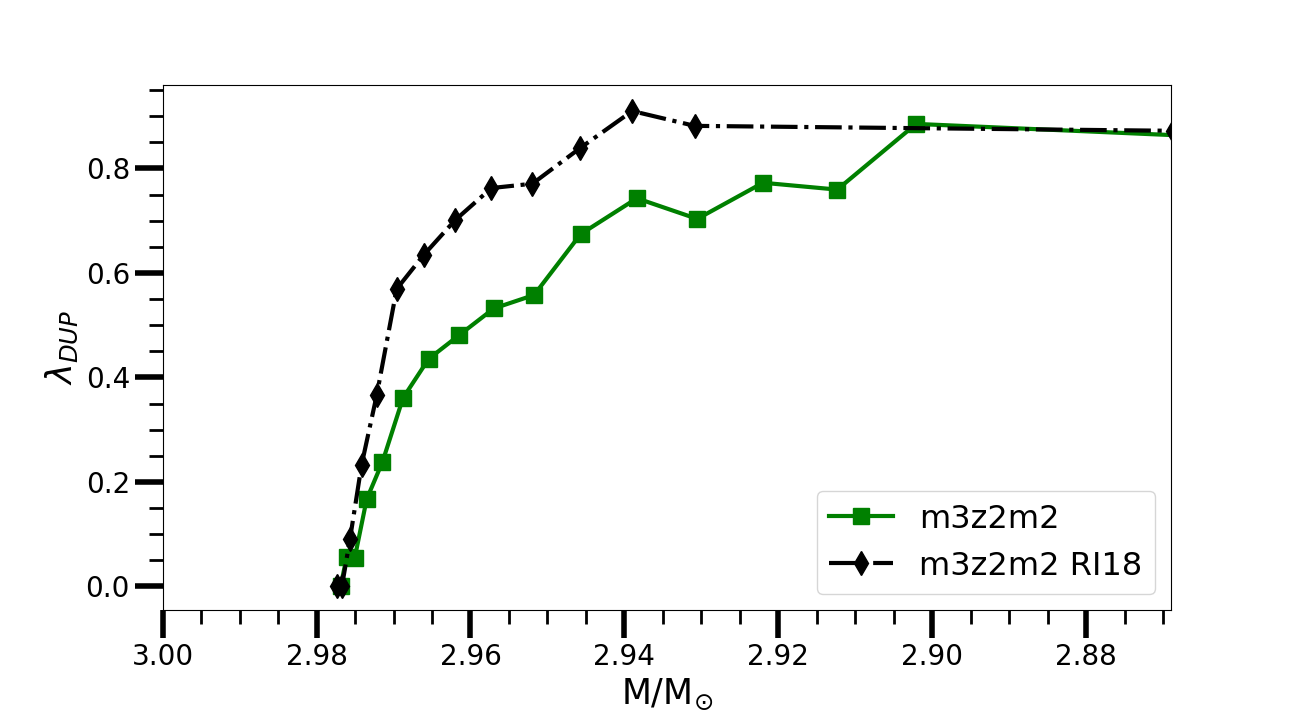}}}
\caption{TDUs efficiency ($\lambda$) temporal evolution.
  A comparison between m3z2m2 and RI18 is shown in the upper panel,
while a zoom into the early AGB-phase is shown in the lower panel.}
\label{lambdas:compSet1}
\end{figure}

\begin{figure}[htbp]
\centering
\resizebox{13.8cm}{!}{\rotatebox{0}{\includegraphics{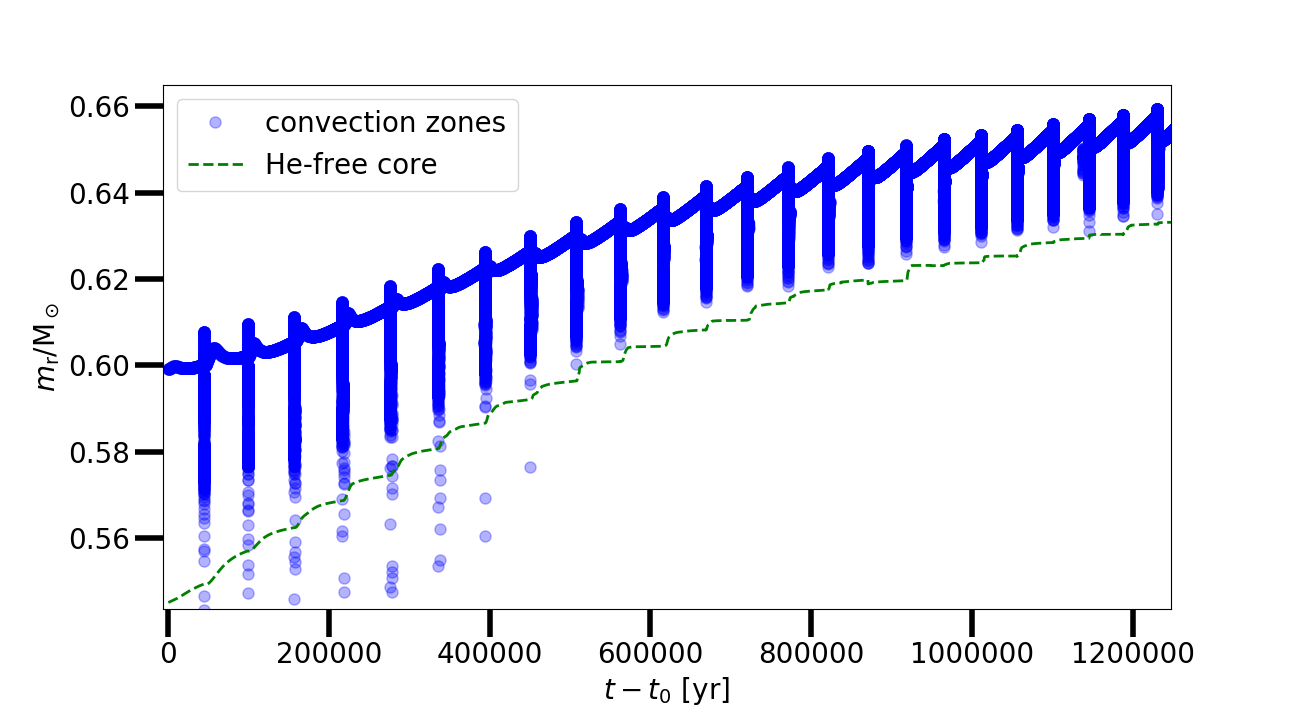}}}
\resizebox{13.8cm}{!}{\rotatebox{0}{\includegraphics{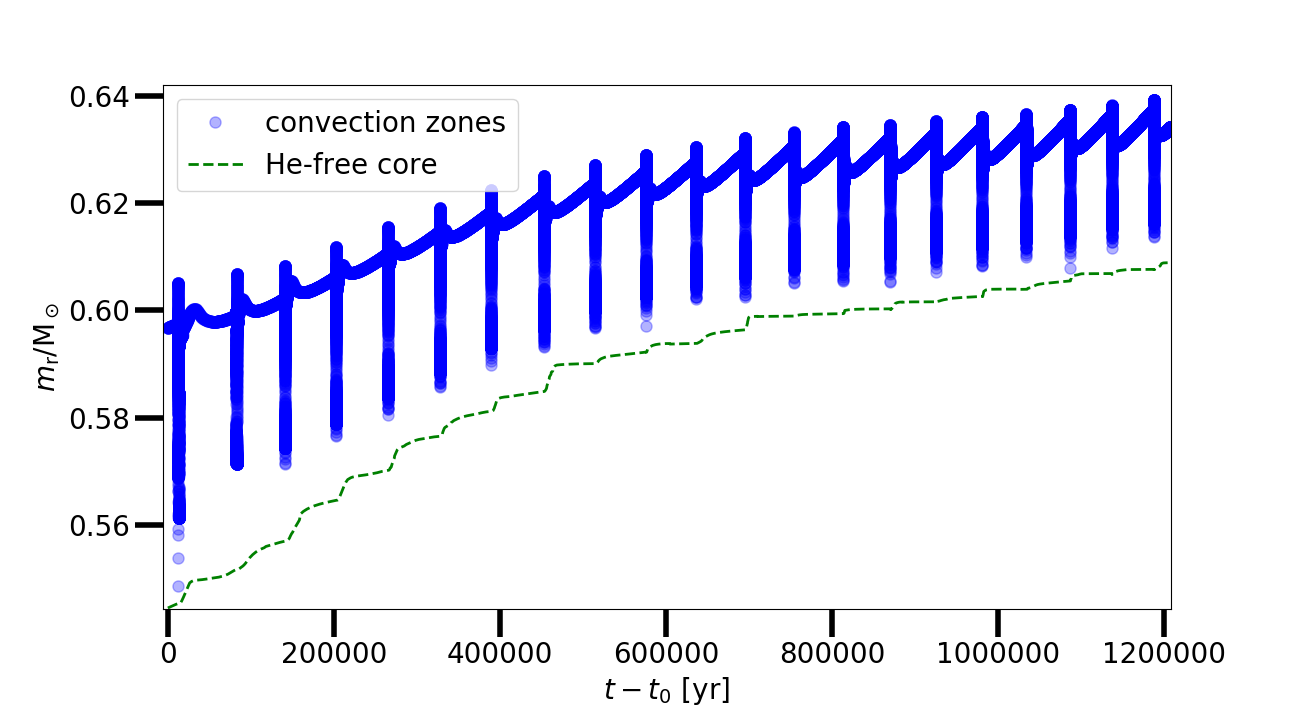}}}
\caption{Upper panel: Kippenhahn diagram of m3z2m2. The whole AGB phase is presented zoomed in the He-intershell.
  Lower panel: same as in the upper panels, but for RI18.}
\label{kippe:compSet1}
\end{figure}

\begin{figure}[htbp]
\centering
\resizebox{14.8cm}{!}{\rotatebox{0}{\includegraphics{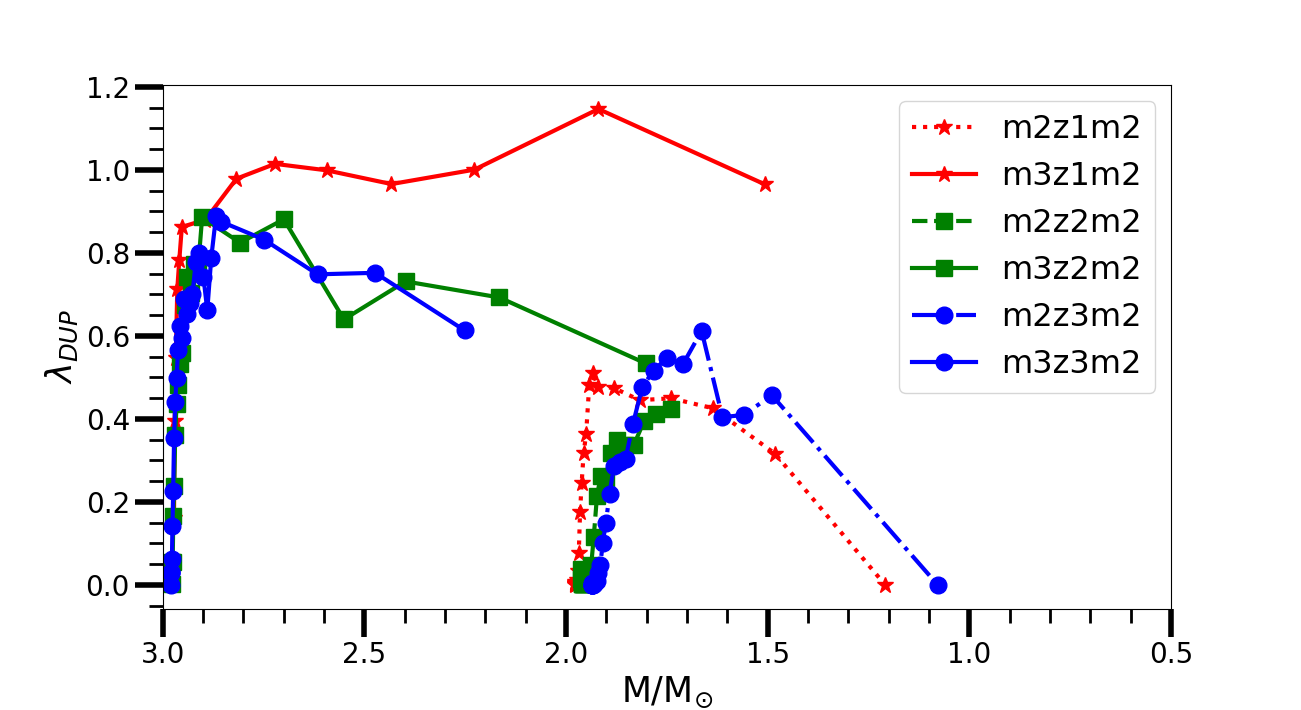}}}
\caption{TDU efficiency $\lambda$ temporal evolution as a function of total mass for all the models listed in Table \ref{tab:model_def}.}
\label{lambdas}
\end{figure}

In Fig. \ref{cbms} we compare the CBM profiles from the m3z2m2 model and RI18.
It is important to notice how the CBM profile in RI18 is more than two orders of magnitude higher in the medium-shaded area compared to the CBM adopted here, i.e. in the intershell zone immediately below the convective envelope.
In this area the mixing coefficient in RI18 is still high enough to impact the TDU $\lambda$ value, hence directly lowering the core-mass.
This picture is consistent with the $\lambda$ temporal evolution shown in Fig. \ref{lambdas:compSet1}.
A dedicated comparison between m3z2m2 and RI18 is shown in the upper panel,
while a zoom into the early AGB-phase is shown in the lower panel.
The plot shows how TDUs in the RI18 model has a systematically higher $\lambda$ (starting already in the early stage of the AGB)
because of the higher CBM efficiency in the stellar layers right below the convective envelope.
This indeed impacts as well the core-mass in an indirect way: every TDU causes a surface enrichment in primary carbon,
causing the surface C/O ratio to increase. As soon as the number of carbon atoms exceeds that of oxygen
(passing from the oxygen-rich phase to the carbon-rich phase, i.e. C/O $>$ 1)
a sudden rise in the atmospheric opacity occurs \cite{marigo:02}.
This results in an envelope expansion, lower effective temperatures and increased mass-loss from dust-driven
winds \citep{marigo:07,mattsson:10,nanni:18}.
Therefore, the AGB lifetime is shorter and consequently also the number of TPs and TDUs experienced by the star,
making the growth of the core mass smaller than otherwise predicted in models with a slower carbon surface enrichment
due to less efficient TDUs \citep{kalirai:14}.
Table \ref{tab:model_agb} shows the total number of TPs and number of TPs occurring during the oxygen-rich AGB phase.
Also the 3 M$_{\odot}$ , Z=0.02 model from RI18 is shown as a comparison with m3z2m2,
showing a significant reduction of the total number of TPs. This is visible already during the oxygen-rich phase,
during which RI18 model needs two TPs less to become carbon-rich.
The same conclusions can be reached looking at the Kippenhahn diagrams in Fig. \ref{kippe:compSet1},
where our m3z2m2 is again compared to RI18:
location  of convective boundaries and core mass as a function of time are presented.
In particular, the formation of the PDCZ is visible every time a TP occurs.
$\lambda$ temporal evolution for all the other models listed in Tables \ref{tab:model_def}
and \ref{tab:model_agb} is shown in Fig. \ref{lambdas}.

\subsection{ $^{13}$C-pocket formation and intershell abundances}
\label{sec:poc_int}

\begin{figure}[htbp]
\centering
\resizebox{13.8cm}{!}{\rotatebox{0}{\includegraphics{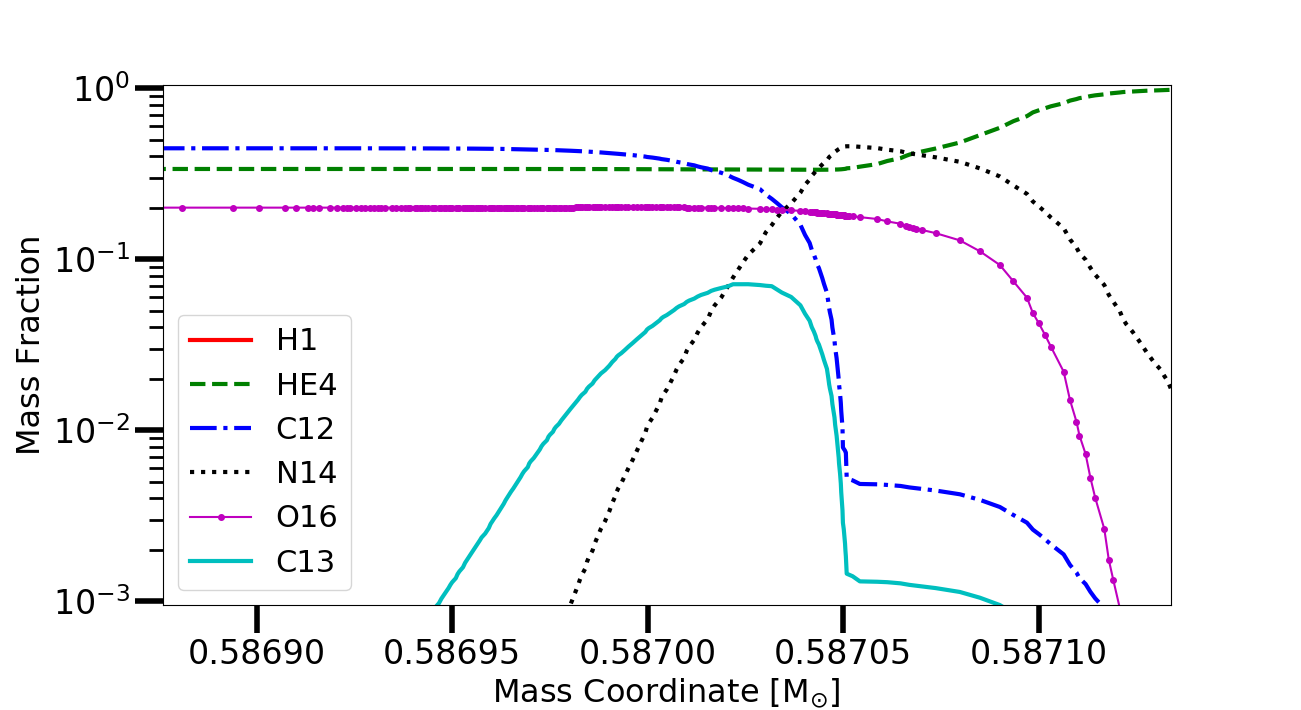}}}
\resizebox{13.8cm}{!}{\rotatebox{0}{\includegraphics{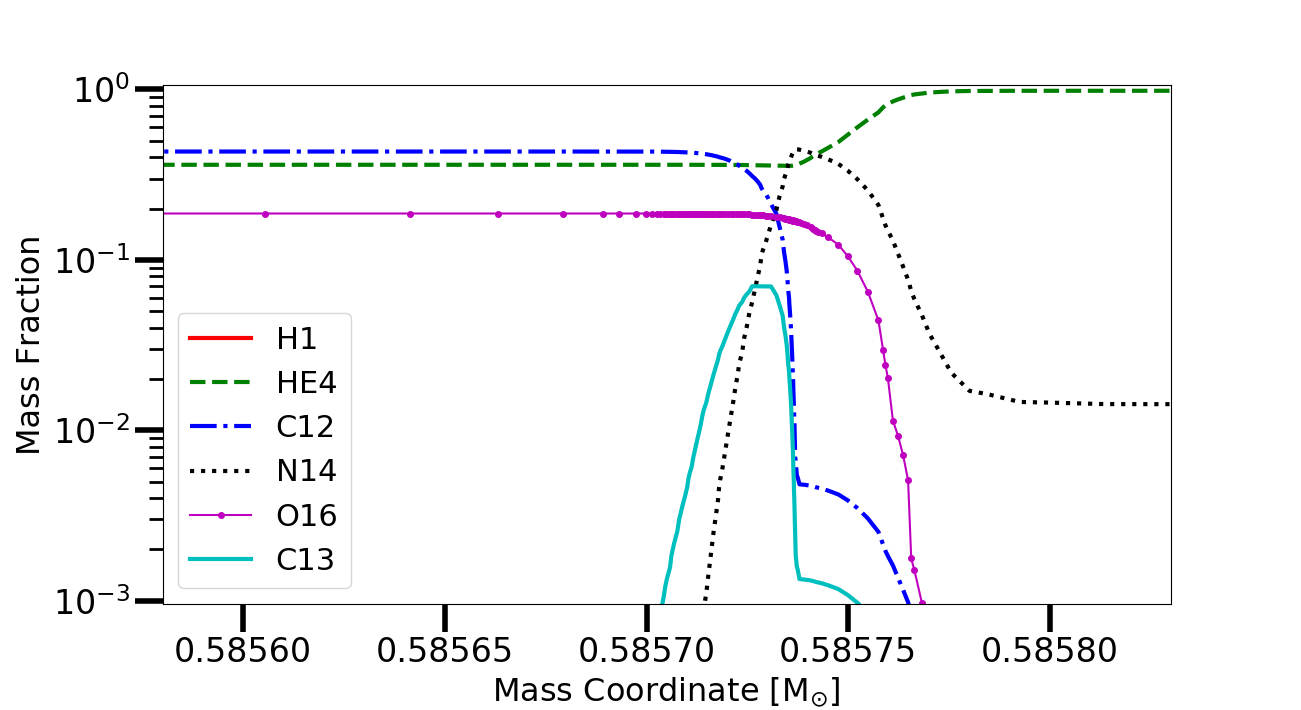}}}
\caption{Upper panel: $^{13}$C-pocket at the beginning of the carbon-rich phase from m2z1m2.
  Lower panel: $^{13}$C-pocket at the beginning of the carbon-rich phase from RI18 at the same mass coordinate as in the upper panel. The comparison shows a much larger pocket compared to RI18.}
\label{cpoc_compSet1}
\end{figure}

\begin{figure}[htbp]
\centering
\resizebox{14.8cm}{!}{\rotatebox{0}{\includegraphics{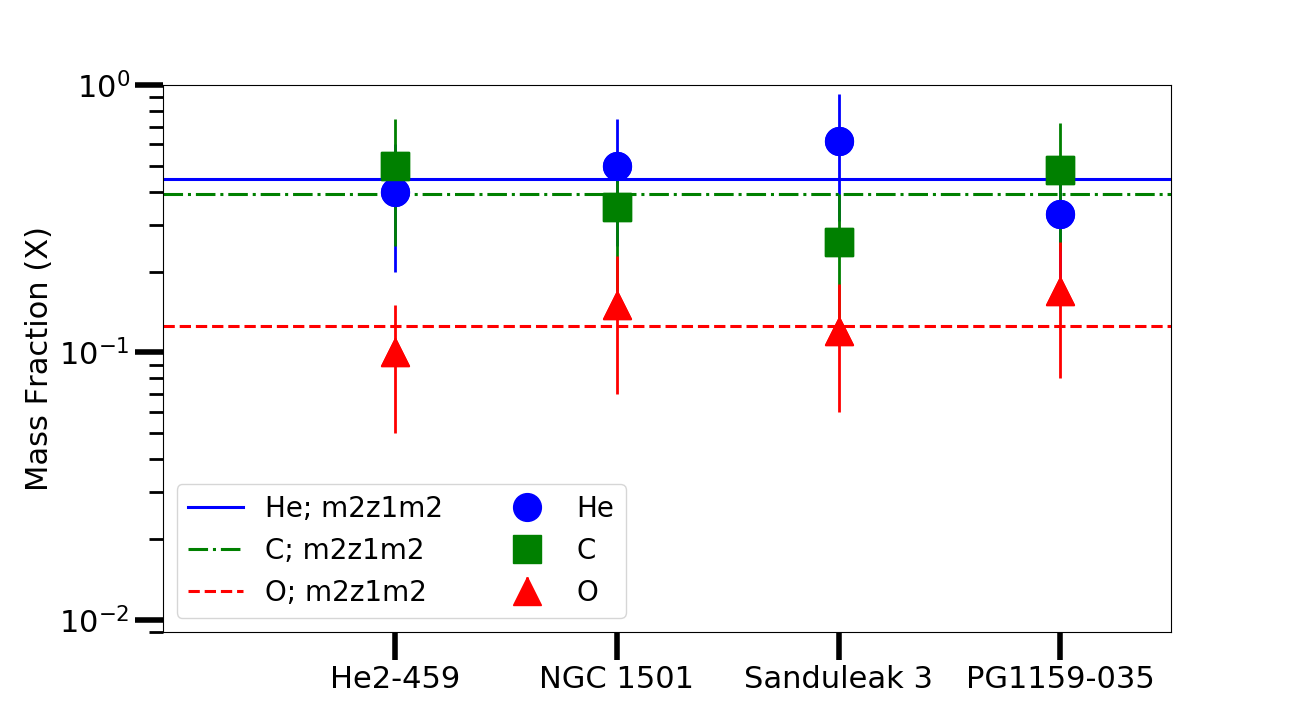}}}
\caption{Final intershell abundances of m2z1m2, which are representatives for all the other models,
are compared to surface abundance of four representatives H-deficient post-AGB stars.}
\label{postAGB}
\end{figure}

As mentioned in Section \ref{intro} and as described in \cite{battino:16},
the most direct impact of our CBM model is an increased $^{13}$C-pocket size
(defined as the mass-coordinate difference between the points where the mass fraction of $^{13}$C, X($^{13}$C), exceeds 0.001 and X($^{13}$C)$>$($^{14}$N)) compared to RI18, where the classic single-exponentially decaying diffusion mixing scheme is adopted.
Fig. \ref{cpoc_compSet1} compares two $^{13}$C-pockets, from our m2z1m2 model and the corresponding model in RI18,
around the same mass coordinate and at the beginning of the carbon-rich phase.
It shows how the pocket-size is larger by around a factor of three.
As we will see in the following sections, this will have profound consequences for $s$-process nucleosynthesis.
In addition to the pocket-size, another important feature, from the comparison between our models and RI18,
is the very similar abundance peak value of $^{13}$C inside the pocket.
This is directly linked to the almost identical $^{12}$C abundance in the intershell during the interpulse period,
which comes from the same CBM adopted at the intershell bottom during TPs.
Moreover, including CBM during TPs at the intershell bottom is very important to reproduce key observables like surface abundances of H-deficient post-AGB stars \citep{werner:06,battino:16},
as this is to date the only way to reproduce the observed enrichment in carbon and oxygen, at the expenses of helium abundance,
in the intershell at the end of the AGB phase.
This is shown in Fig. \ref{postAGB},
where final intershell abundances of m2z1m2 are compared to surface abundance of four representatives H-deficient post-AGB stars,
showing a very good agreement between our model with observations.
High abundances of $^{12}$C in the He-intershell region cause high $^{13}$C abundances
and a more efficient neutron flux in the $^{13}$C-pocket.
At the same time, lower abundances of $^{4}$He lead to higher temperatures during the TP,
leading to a stronger activation of the $^{22}$Ne neutron source \citep{lugaro:03b,lugaro:18}.

\subsection{Approximating rotationally induced mixing: models with additional constant mixing coefficient}
\label{sec:rot}

\cite{busso:01} presented a compilation of $s$-process observational data,
including the ratio of the $s$-process production around the barium peak ($hs$) over the nucleosynthesis around the strontium peak ($ls$).
In particular, -0.6 $<$ [$hs$/$ls$] $<$ 0.0 characterizes stars of solar metallicity, adopting the square-bracket notation defined as:

\begin{equation}
  [X/Y] = \log((X_{*}/Y_{*})/(X_{\odot}/Y_{\odot}))
\label{square_formalism}
\end{equation}

with X$_{*}$/Y$_{*}$ and X$_{\odot}$/Y$_{\odot}$ being the ratios of two quantities measured in a given star and in the Sun respectively.
It also seems that models applying CBM at the bottom of the helium-intershell during TPs
can reproduce only the largest observed $hs$/$ls$ ratios,
suggesting a neutron exposure in the  $^{13}$C-pocket at the maximum of the observed range \citep{lugaro:03b,herwig:05}.
The first study where the IGW-driven CBM was tested and calibrated was done by \cite{battino:16}:
the stellar models presented were all non-rotating and [$hs$/$ls$]$~$0.0 was obtained.
On the other hand, \cite{herwig:03}, \cite{siess:04} and \cite{piersanti:13}, have shown that by considering
rotation in AGB models the final [$hs$/$ls$] ratio tends to be reduced compared to non-rotating models.
The reason for this is that during the AGB phase the slowly rotating envelope and the fast-rotating compact core are in contact.
Hence, shear mixing sets in during the interpulse period polluting the  $^{13}$C-pocket with the neutron poison  $^{14}$N
from the  $^{14}$N-pocket just above (also visible in Fig. \ref{cpoc_compSet1}), reducing the neutrons available for the $s$-process,
in particular the neutron/seeds numeric ratio, hence the barium-peak production.
The inclusion of a stochastic process like rotation, where a range of initial angular velocities is possible, could explain the spread in $s$-process efficiencies,
observed in spectroscopic data and laboratory measurements of some isotopic-ratios in presolar grains \citep{herwig:03,herwig:05,battino:16}.
We are not going to present models including a self-consistent implementation of rotation,
yet, given the essential role rotational-induced mixing has in reproducing AGB observables,
we want to explore its possible impact in $s$-process nucleosynthesis.
For this reason, we apply a low constant mixing across the intershell during the interpulse period,
in order to mimic the effects of shear mixing, following a method very similar to \cite{herwig:03},
models with rotationally-induced mixing predict mixing coefficients around log(D[cm$^{2}$s$^{-1}$]) $\sim$ 2,
which eventually totally suppress the $s$-process production by an excessively large poisoning of the $^{13}$C-pocket.
\cite{cantiello:14} showed that models only accounting for angular momentum conservation \citep[as in][]{herwig:03} produce cores rotating about
10-1000 times faster than what has been found from asteroseismology, suggesting a missing angular-momentum transport process.
Since rotationally shear mixing coefficients depend on the square of the vertical velocity gradient \citep{zahn:92,maeder:00,mathis:04}
and the compact core rotates with velocity v$_{core}$,
faster than but similarly in order of magnitude to the expanded envelope which rotates with velocity v$_{env}$,
so that v$_{core}$ $\sim$ C$\times$v$_{env}$ \citep[typically 2$\lesssim$C$\lesssim$4, see][]{deheuvels:15}, we have:

\begin{equation}
D_{rot} \sim (K/N^{2})(dv/dr)^{2} \sim (K/N^{2})((v_{core}-v_{env})/(\delta\ (r)))^{2} \sim ((C-1)/C)^{2}(K/N^{2})((v_{core})/(\delta\ (r)))^{2}
\end{equation}

where K is the thermal diffusivity, N the Brunt-V{\"a}is{\"a}l{\"a} frequency and rotational velocity changes from v$_{core}$ to v$_{env}$ over a distance $\delta$($r$) along the stellar radius.
Therefore, if  v$_{core}$ from models is 10-1000 times faster than observed
(as suggested by asteroseismology),
then the expected mixing coefficients from rotationally-induced mixing should decrease from log(D$_{rot}$/(cm$^{2}$s$^{-1}$))$\sim$2 to
-4$<$log(D$_{rot}$)(cm$^{2}$s$^{-1}$)$<$0.
Hence, in order to mimic the effects of shear mixing and following a method very similar to what was done in \cite{herwig:03},
we present in Table \ref{tab:modrot} six additional models with an additional low constant mixing across the intershell,
consistent in the range we have just defined, during the interpulse period.
It is anyway important to notice the big assumption we are making here,
that is stellar rotation being an efficient extra-mixing source in AGB stars.
This is actually still a matter of debate \citep[see][]{straniero:15,deheuvels:15,herwig:05}.
On the other hand, any extra-mixing process able to satisfy the conditions described above could be considered.

\section{Post-processing nucleosynthesis calculations}
\label{sec:postpro}

\begin{figure}[htbp]
\centering
\resizebox{14.8cm}{!}{\rotatebox{0}{\includegraphics{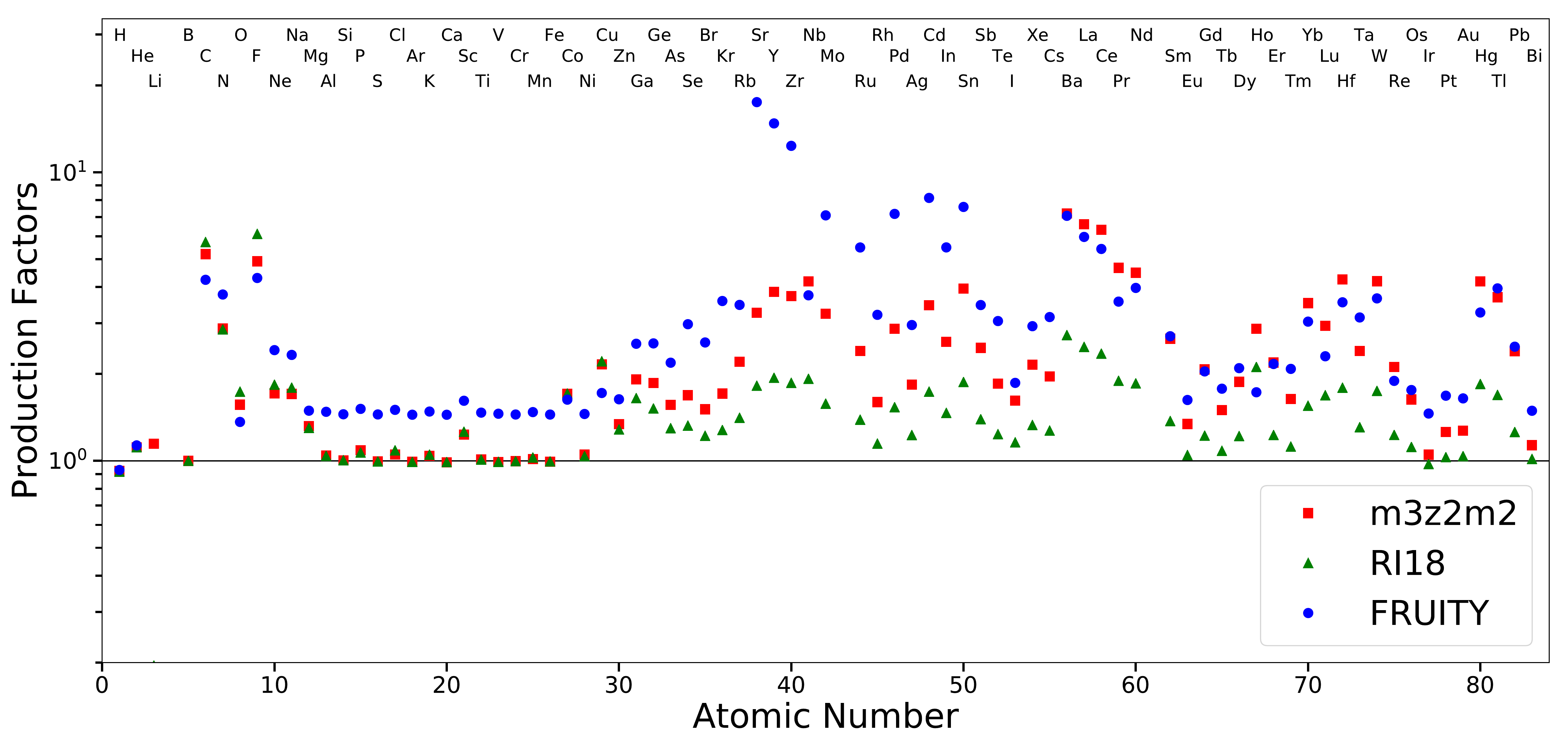}}}
\caption{Comparison of heavy elements production factors between m3z2m2, RI18 and FRUITY models.}
\label{eldistrib:compSet1}
\end{figure}

\begin{figure}[htbp]
\centering
\resizebox{13.8cm}{!}{\rotatebox{0}{\includegraphics{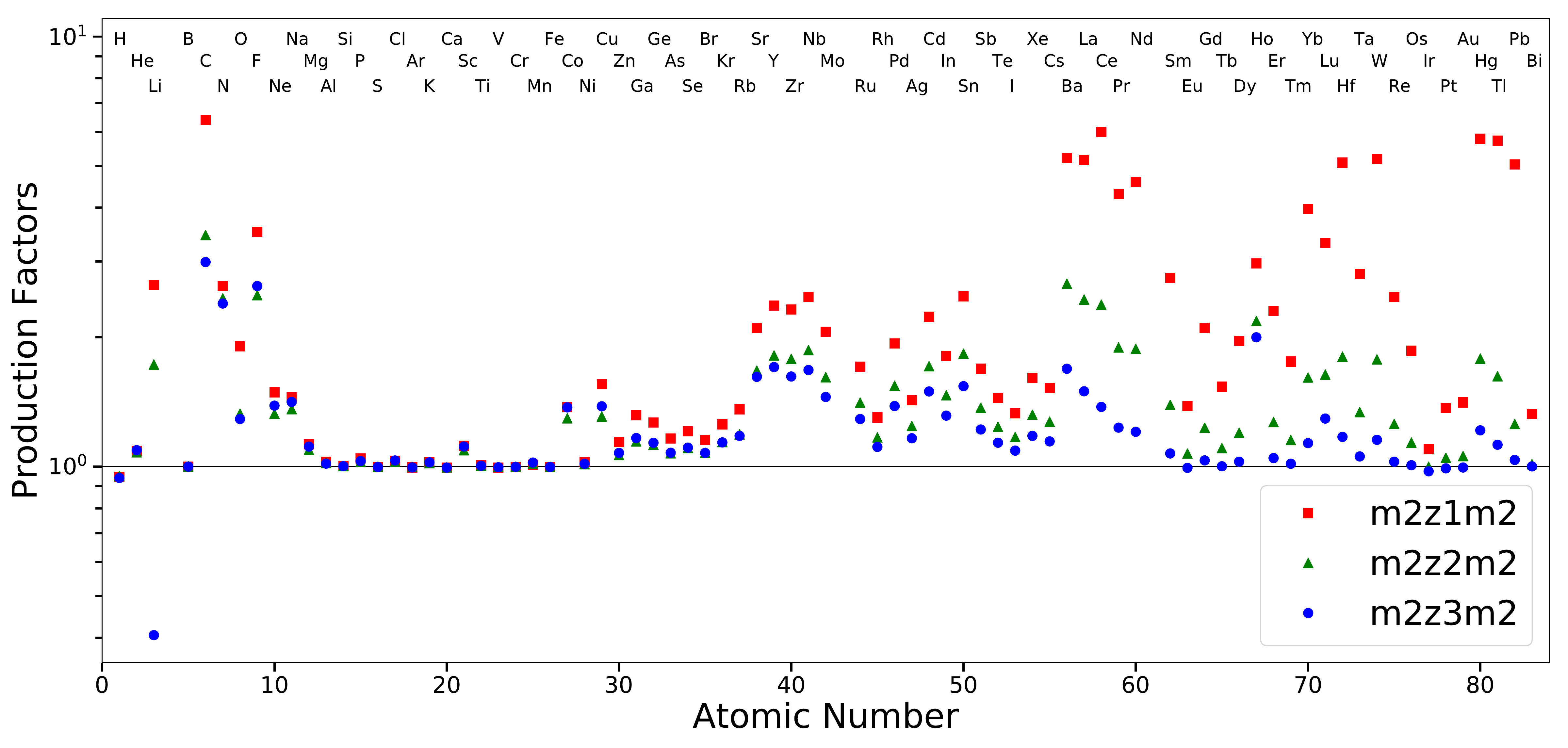}}}
\resizebox{13.8cm}{!}{\rotatebox{0}{\includegraphics{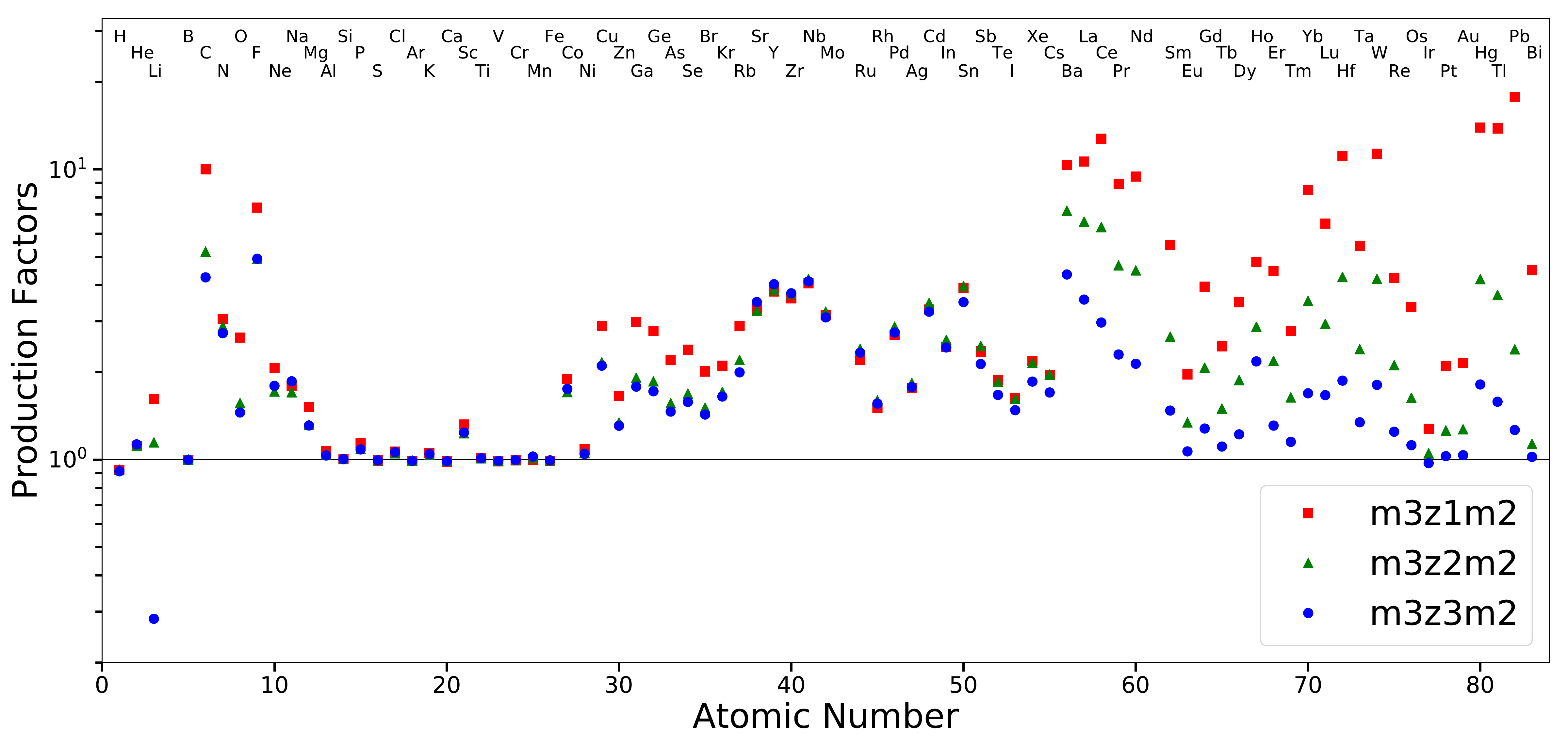}}}
\caption{Upper panel: Heavy elements production factors of 2 M$_\odot$ models listed in table \ref{tab:model_def}.
  Lower panel: Same as in the upper panel, but for 3 M$_\odot$ models.}
\label{eldistrib:all}
\end{figure}

\begin{figure}[htbp]
\centering
\resizebox{14.8cm}{!}{\rotatebox{0}{\includegraphics{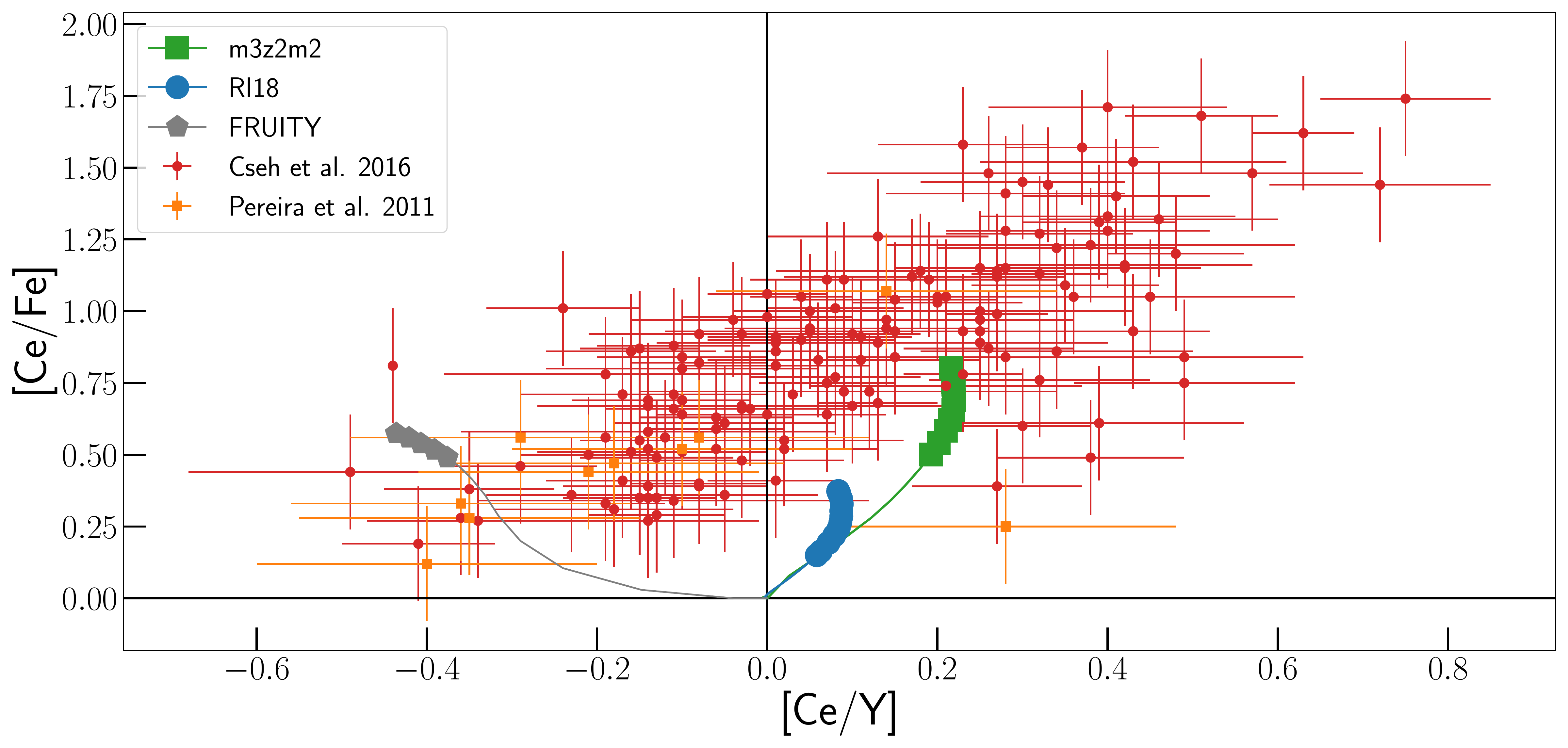}}}
\caption{Comparison of  [Ce/Fe] vs [Ce/Y] index between m3z2m2 and RI18. Also observational data of barium-stars from \cite{cseh:18} and \cite{pereira:11} are shown}
\label{hsfe:hslsSet1}
\end{figure}

The $s$-process nucleosynthesis in low-mass AGB stars heavily depends on the properties of the ${^{13}}$C-pocket. As already described, the stellar models presented in this work form a ${^{13}}$C-pocket that is about 3 times larger than in RI18. This has profound consequences on the resulting heavy element production, as shown in Fig. \ref{eldistrib:compSet1}. Models described in RI18 exhibit a low $s$-process production compared to what is inferred from spectroscopic observations. Changing the treatment of convective boundaries according to \cite{battino:16}, results in about 3 times larger $s$-process production factors in agreement with observations.
In the same figure, we also show the results from FRUITY calculations \citep{cristallo:11}.
The gap in barium production when comparing RI18 and FRUITY is not present anymore in this work,
while the difference persists when considering Sr peak abundances.
This is due to FRUITY models not including any CBM at the base of the PDCZ.
This leads to lower ${^{12}}$C abundances in the intershell in FRUITY models,
hence lower ${^{13}}$C abundances and a less efficient neutron flux in the ${^{13}}$C-pocket,
which favors Sr-peak over Ba-peak elements \citep[see][]{lugaro:03a}. 
The production factors of all our models are shown in Fig. \ref{eldistrib:all}: in particular, lower metallicity models show a stronger production of the second (Ba region) and third (Pb region) $s$-process peaks, while the first peak (strontium region) is favored in higher metallicity models. In Fig. \ref{hsfe:hslsSet1} we show the tracks of m3z2m2 and the models with the same initial mass and metallicity in RI18 and FRUITY. Since there are now large enough and internally consistent data sets of individual elements representing second-peak (hs) and first-peak (ls)elements, each symbol in the figure gives the surface [Ce/Y] and [Ce/Fe],
being an update to the classic [hs/ls] and [hs/Fe] indices respectively, as discussed by \cite{cseh:18}.
The theoretical tracks are compared to the largest homogeneous set of Ba giant star observations presented in \cite{cseh:18},
including data from \cite{pereira:11} to achieve a better statistic at super-solar metallicities.
As the star evolves, TDU events gradually enrich the envelope in carbon eventually resulting in surface C/O$>$1,
entering the carbon-rich phase that we indicate with bigger-size symbols.
The figure shows the larger $s$-process efficiency in the m3z2m2 and FRUITY model compared to RI18, demonstrated by the higher [Ce/Fe] value.
Additionally, since the intershell material in RI18 and m3z2m2 has the same [Ce/Y] the two tracks initially perfectly overlap, while FRUITY model evolves towards negative [Ce/Y] values, reflecting the absence of CBM under the PDCZ as previously discussed.
This is not surprising, since both RI18 and m3z2m2 models treat the CBM at the bottom of the intershell during TP in the same way,
resulting in very similar ${^{12}}$C intershell abundance (as seen in Section \ref{sec:poc_int}) and hence neutron exposures.
Because of the larger amount of $s$-process material brought to the surface at every TDU,
the m3z2m2 track is pushed further away from the origin towards a higher final [Ce/Y].

\subsection{Comparison with spectroscopic observations}
\label{sec:spectro}

\begin{figure}[htbp]
\centering
\resizebox{13.8cm}{!}{\rotatebox{0}{\includegraphics{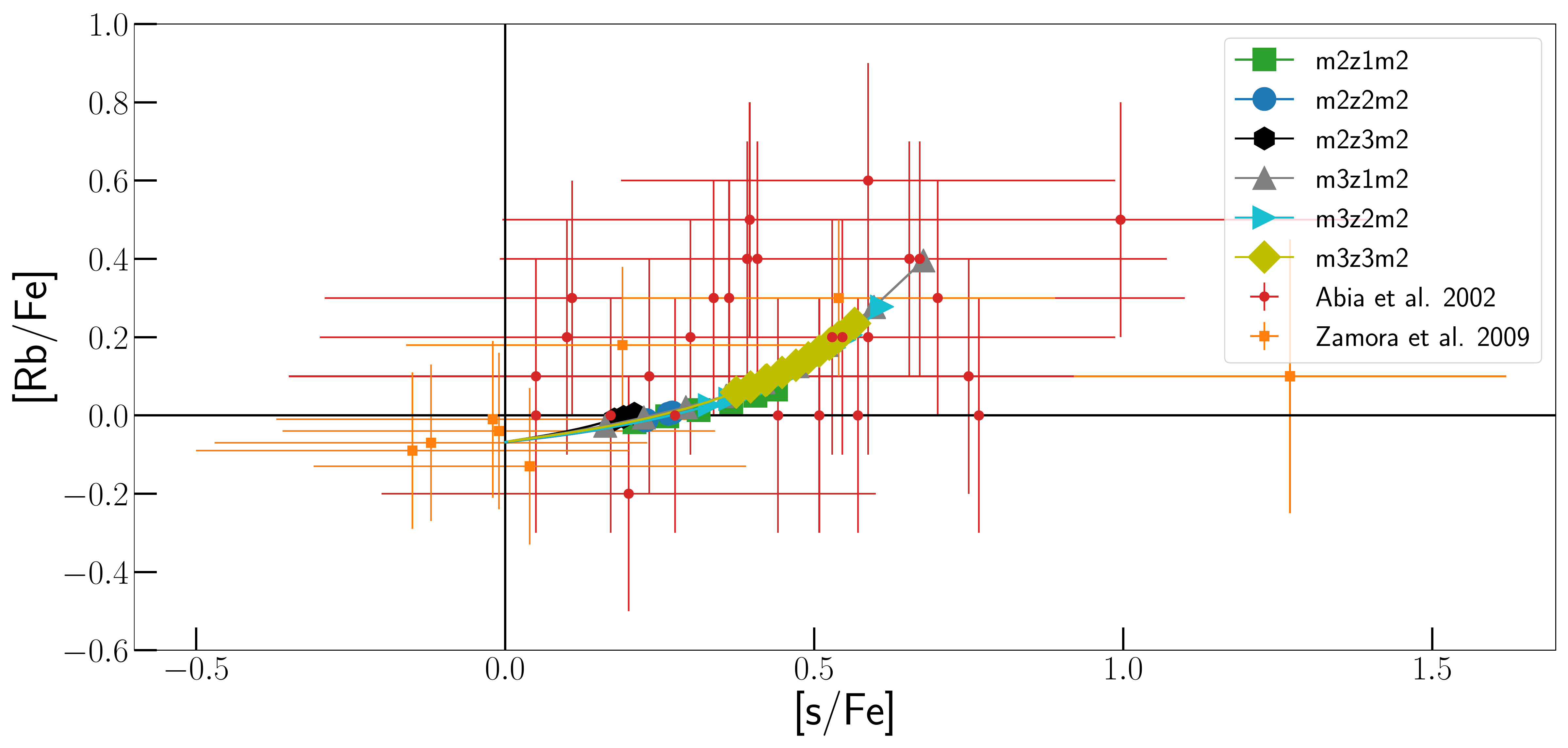}}}
\resizebox{13.8cm}{!}{\rotatebox{0}{\includegraphics{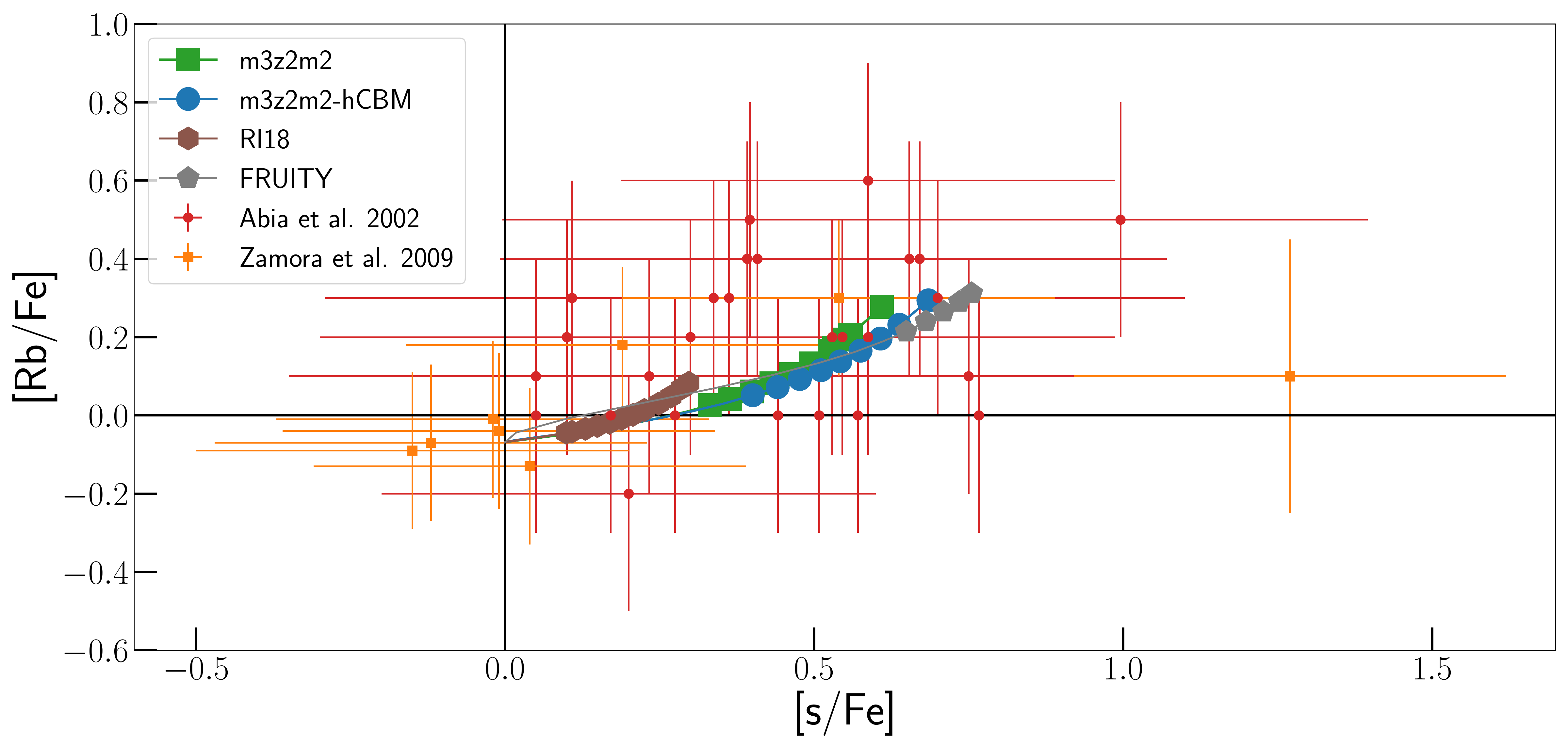}}}
\caption{Upper panel: Comparison of Rb abundance vs the total $s$-process production inferred from spectroscopy analysis of carbon stars by \cite{abia:02} and \cite{zamora:09} with the abundance predicted by our models in Table \ref{tab:model_def}. Lower panel: Same as in the upper panel, but comparing m3z2m2 with same mass and metallicity models from RI18 and the FRUITY database. We also show predictions from m3z2m2-hCBM, whose comparison with m3z2m2 shows the impact of the increased \cdr-pocket size due to higher CBM efficiency during TDUs. }
\label{rb:sfe}
\end{figure}

\begin{figure}[htbp]
\centering
\resizebox{14.8cm}{!}{\rotatebox{0}{\includegraphics{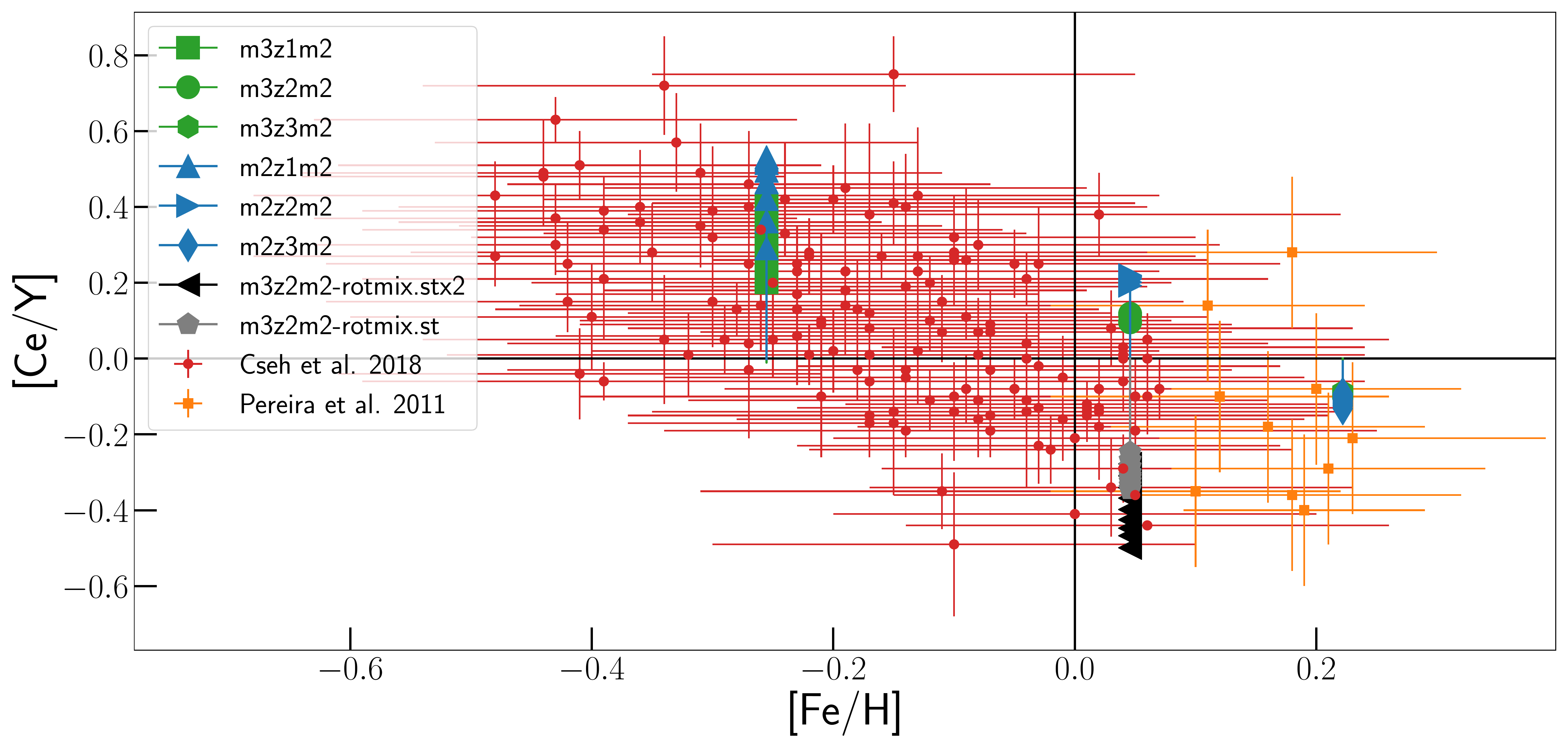}}}
\caption{Comparison of [Ce/Y] vs [Fe/H] results from the whole evolution of models listed in Table \ref{tab:model_def} adding also m3z3m2.rotmix.st and m3z2m2.rotmix.stx2, which include an artificial mixing to replicate stellar rotation effects. The values inferred from spectroscopy analysis of  barium-stars by \cite{cseh:18} and \cite{pereira:11} are also shown as comparison.}
\label{hsls:feh}
\end{figure}

\begin{figure}[htbp]
\centering
\resizebox{14.8cm}{!}{\rotatebox{0}{\includegraphics{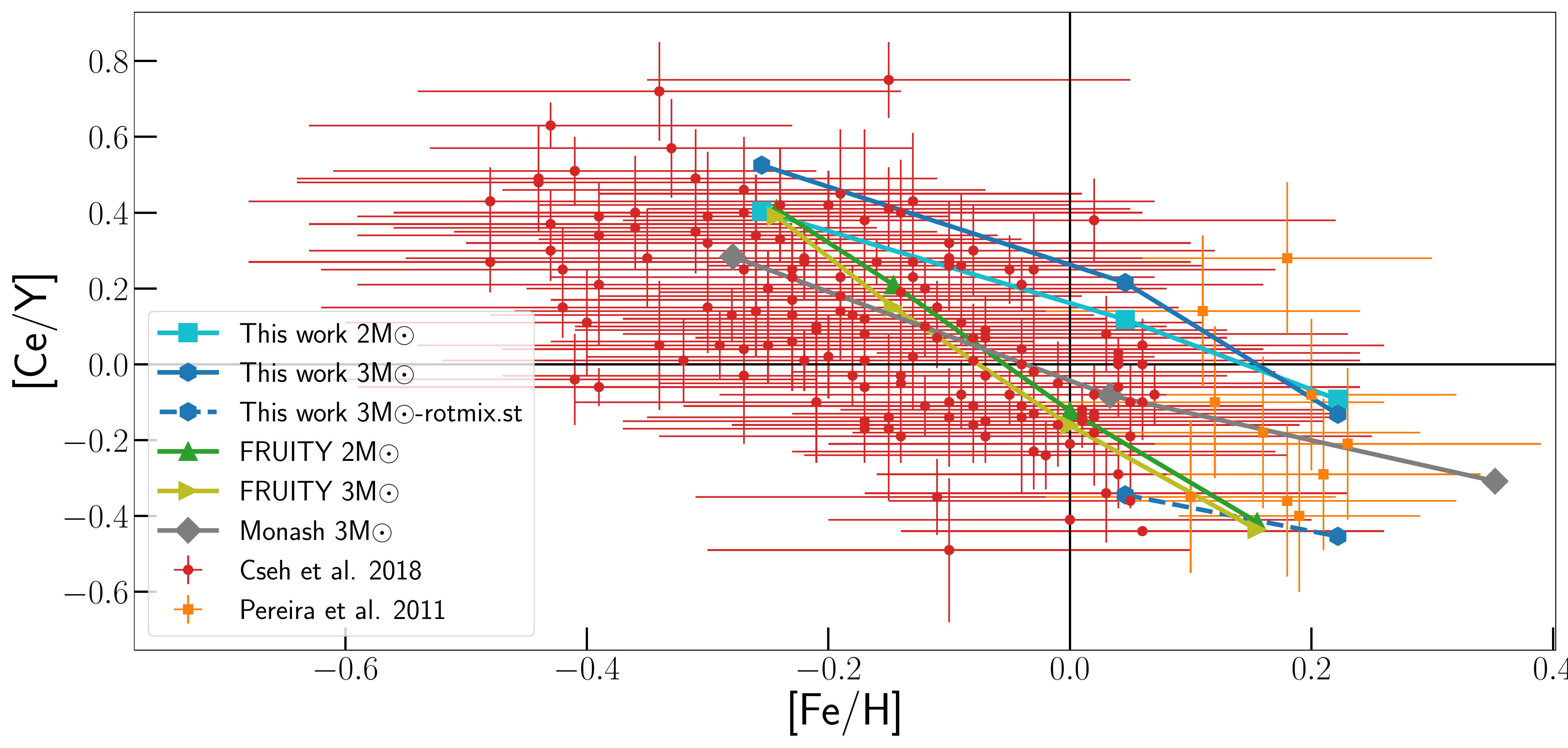}}}
\caption{Comparison of [Ce/Y] vs [Fe/H] results from the final surface abundances of models presented in Fig. \ref{hsls:feh} here we also include results from the FRUITY database and Monash models as a comparison.}
\label{cey:feh_comp}
\end{figure}

Low-mass AGB stars produce the bulk of the $s$-process material in the ${^{13}}$C-pocket, but a non-negligible amount of neutrons comes from ${^{22}}$Ne($\alpha$,n)${^{25}}$Mg activated during the TP. Additionally, some isotopes in proximity of branching points are efficiently produced only in the high neutron density conditions achieved during the TP \citep{raut:13}. One example is rubidium, whose neutron magic isotope ${^{87}}$Rb is produced only in high enough neutron density conditions to open the branching at ${^{86}}$Rb (18.642 days half-life).
Spectroscopic observables allow access to Rb abundances as well as abundances of other $s$-process elements produced entirely in the ${^{13}}$C-pocket.
Fig. \ref{rb:sfe} shows the rubidium abundance vs the total $s$-process production inferred from spectroscopy analysis of carbon stars
compared to the predictions by our models. The $s$-process production is described by the [s/Fe] index, expressed with formalism defined by Eq.\ref{square_formalism}, with the numerator being the averaged abundance between Sr and Ba peak elements. The slope of our models' tracks are in agreement with the observed relative contribution of the TP with respect to the ${^{13}}$C-pocket. Moreover, we are able to reproduce the highest observed $s$-process production within observational uncertainties.
As described in \cite{busso:01}, a range of ${^{13}}$C-pocket sizes is required to reproduce the spread of [hs/ls], and hence [Ce/Y], observed in stars for a given metallicity. Indeed, a stochastic process like rotation could produce this effect, as described in Section \ref{sec:poc_int}  \citep[see also][]{herwig:03,herwig:05}. Indeed, the higher the initial rotational velocity, the lower the final [Ce/Y].
As a consequence, stellar models not including rotation should reproduce the highest observed [Ce/Y],
and adding the effects of rotation should explain the lower [Ce/Y] values observed \citep{herwig:05}.
This is successfully reproduced by our models, as shown in Fig. \ref{hsls:feh}.
We plot the results from the whole evolution of the models listed in Table \ref{tab:model_def},
adding also two of the models described in Table \ref{tab:modrot},
which include an artificial mixing to replicate stellar rotation effects.
The theoretical predictions reproduce the [Ce/Y] vs [Fe/H] slope around solar metallicity,
as well as the observed spread in [Ce/Y] for specific metallicities.
Moreover, we present another similar comparison in Fig. \ref{cey:feh_comp},
where we show our final surface abundances and include the results from FRUITY and Monash group \citep{karakas:16} datasets.
Similarly to what previously discussed about FRUITY results,
Monash models do not include any CBM at the base of the PDCZ, hence a lower final [Ce/Y] compared to our models.
Additionally, our models including rotational mixing present a final [Ce/Y] on average 0.4 dex lower than our standard setting,
suggesting rotational mixing as a strong candidate to cover the whole observed range of $s$-process efficiencies.
However, it is important to notice the possibility that the necessary stochasticity to reproduce the observed spread in [Ce/Y]
may be present in CBM processes already.
Our understanding of convection, which is the physical process originating IGW and hence CBM in our models, is not in a satisfying state yet.
The picture gets even more complicated when considering additional physics like magnetic fields,
that have already been proposed to play a key role in the formation of the  ${^{13}}$C-pocket \citep[see][]{trippella:16},
whose interplay with IGW has not been investigated yet.
This may introduce a stochastic component in the CBM process,
resulting in the spread of neutron exposures and ${^{13}}$C-pocket sizes, possibly including the results obtained by RI18.

\subsection{Comparison with presolar grains measurements}
\label{sec:grains}

When the condition C/O$>$1 is met and a carbon-star is formed,
a sudden rise of the opacity occurs, making the atmosphere expand and cool \citep{marigo:02,kalirai:14}.
In these conditions silicon carbide (SiC) grains can form.
The vast majority of SiC grains ("mainstream" SiC, more than 90$\%$ of SiC grains)
form in the atmospheres around carbon-rich AGB stars \citep{ferrarotti:06,nanni:13,lugaro:18}.
Each specific grain formed in a single specific stellar source.

Recently, \cite{lugaro:18} compared predictions from AGB models computed with the Monash stellar structure code \citep{karakas:07}
with isotopic ratio measurements, focusing on Zr, Sr, and Ba isotopic ratios,
matching measurements from \cite{liu:14a} and \cite{liu:15}.
On the other hand, a number of limitations in the stellar models where also highlighted,
the most important of these being the absence of any CBM at the base of the TP-driven convective zone,
despite the indications from multi-D hydrodynamic simulations \citep{herwig:07}
and observations of H-deficient post-AGB stars as described in Section \ref{sec:poc_int}.
In the same Section, we also explained how the enhancement in $^{12}$C,
following the mixing at the bottom intershell convective boundary results in a more efficient neutron flux in the $^{13}$C-pocket,
hence favoring a higher production of Barium-peak isotopes and in a stronger activation of the $^{22}$Ne neutron source,
leaving a clear fingerprint in branching-point sensitive isotopic ratios like ${^{96}}$Zr/${^{94}}$Zr \citep[see][]{herwig:05,battino:16}).
It is then interesting to compare stellar models where such CBM processes are included,
like in the present work, to the stardust SiC data.

\subsubsection{Sr}
\label{sec:sr}

\begin{figure}[htbp]
\centering
\resizebox{9.8cm}{!}{\rotatebox{0}{\includegraphics{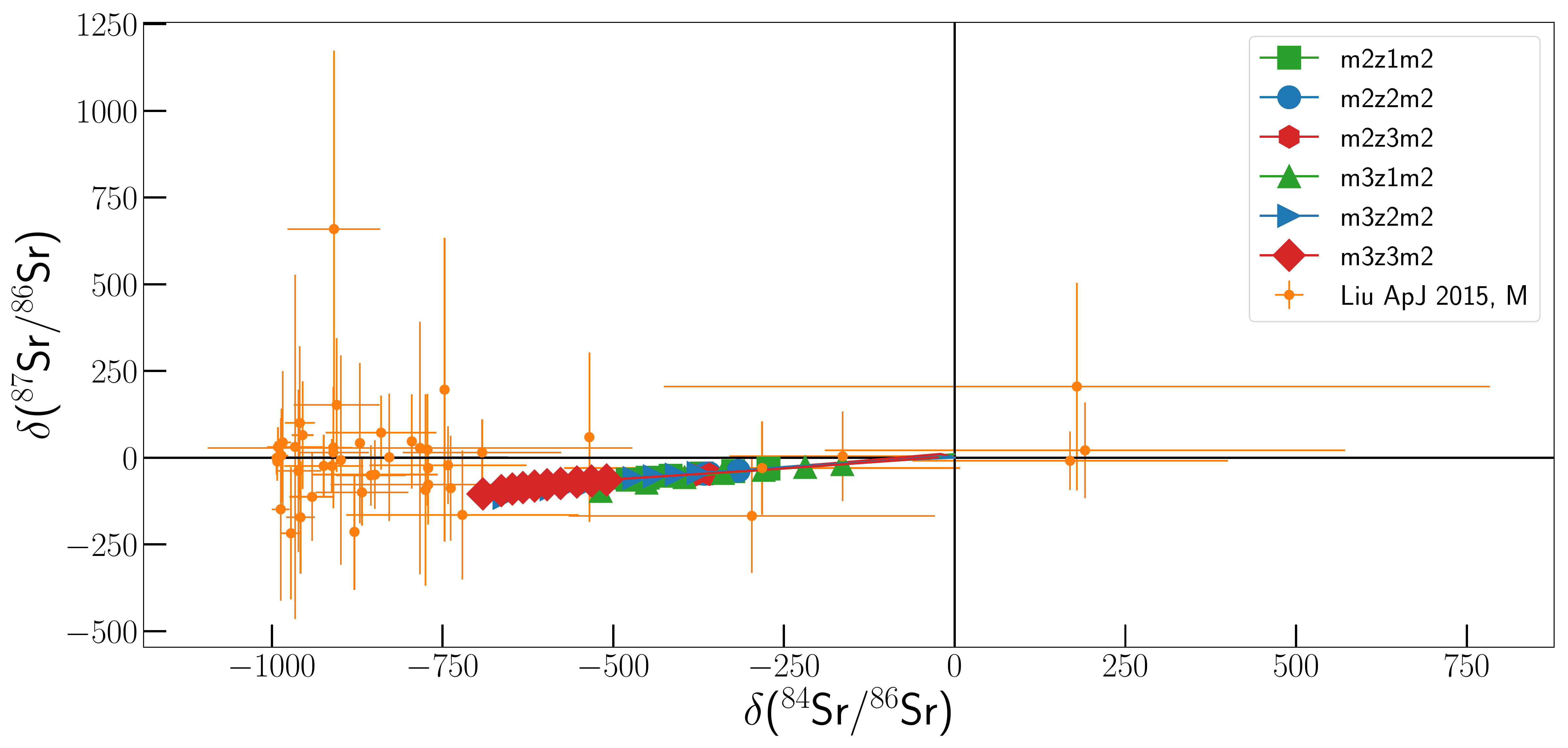}}}
\resizebox{9.8cm}{!}{\rotatebox{0}{\includegraphics{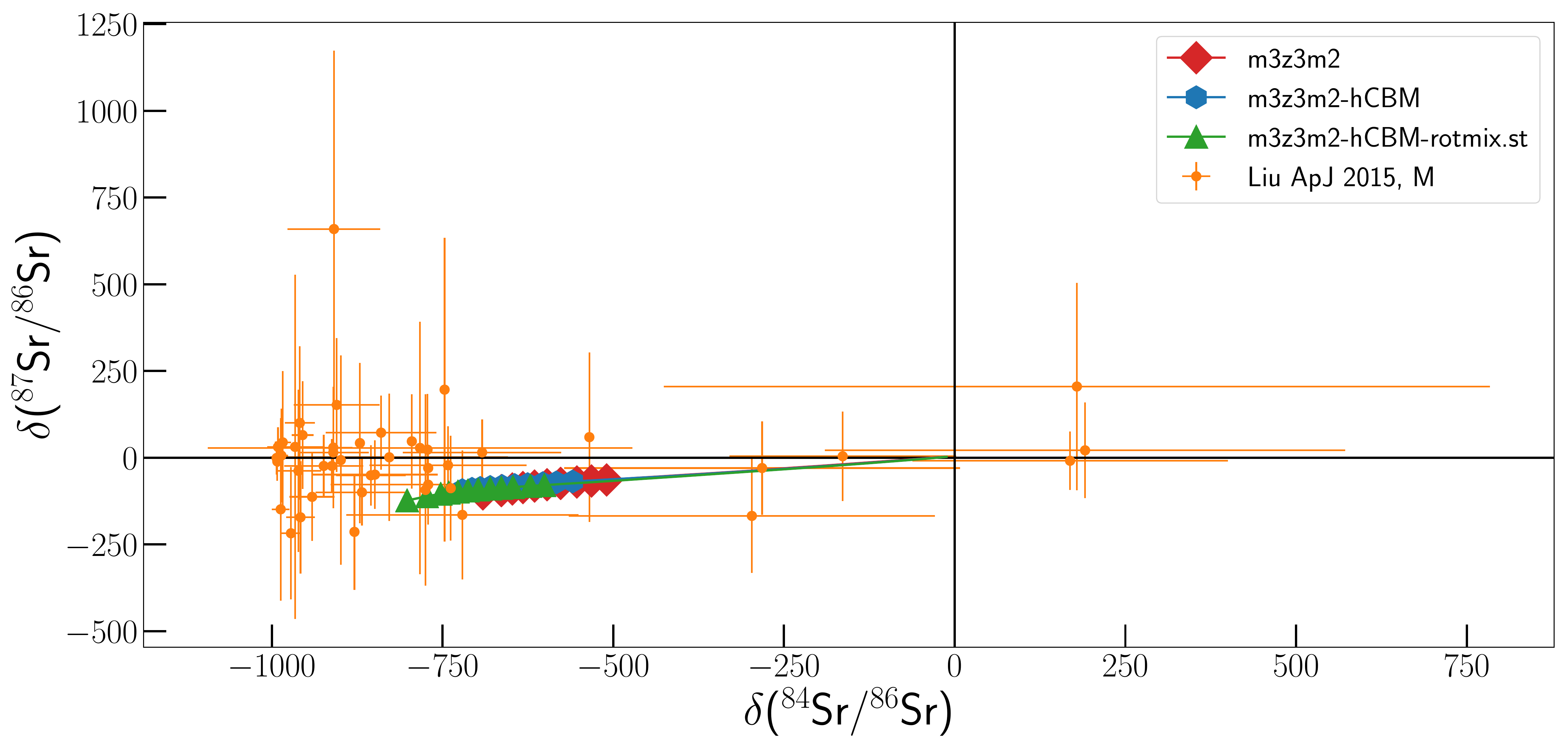}}}
\resizebox{9.8cm}{!}{\rotatebox{0}{\includegraphics{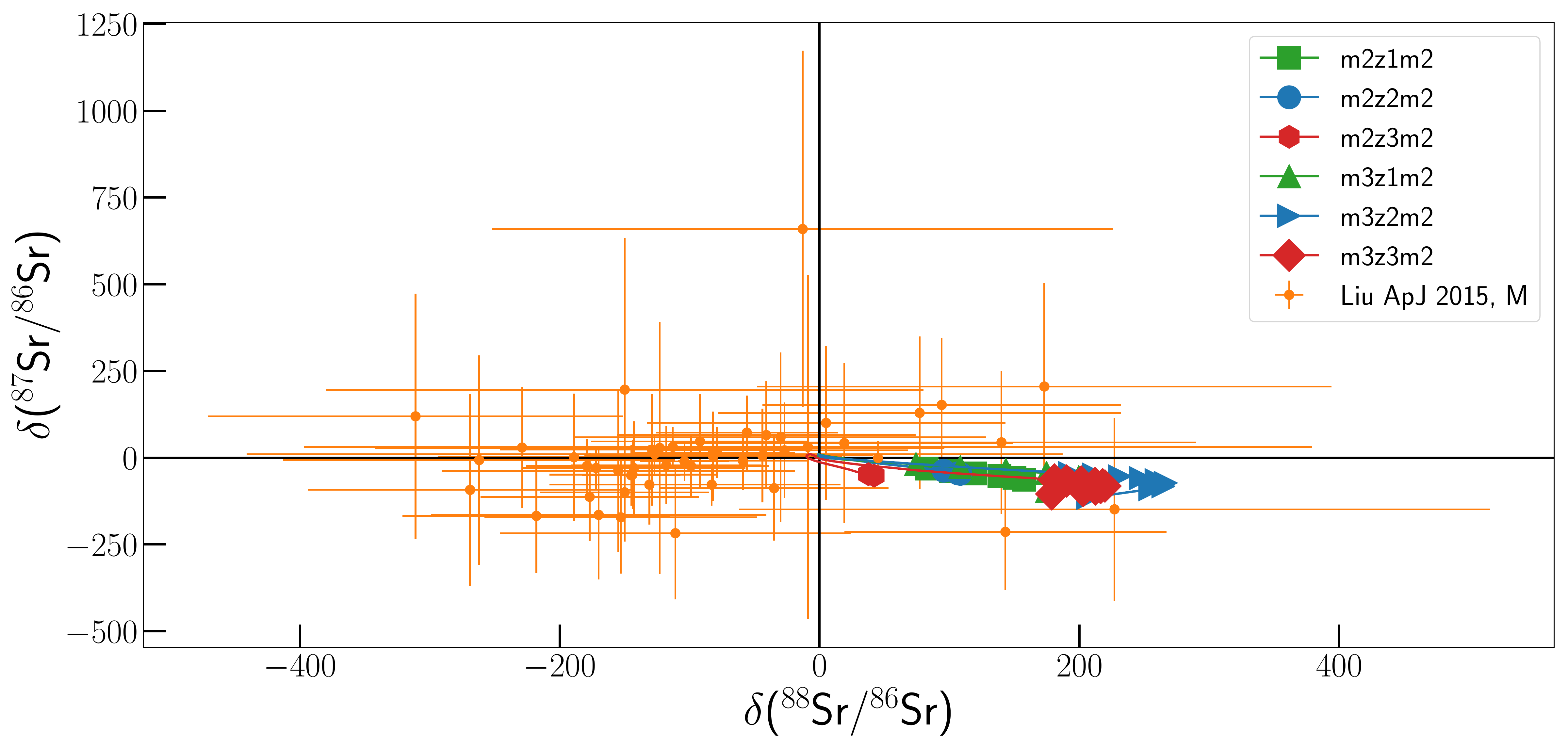}}}
\resizebox{9.8cm}{!}{\rotatebox{0}{\includegraphics{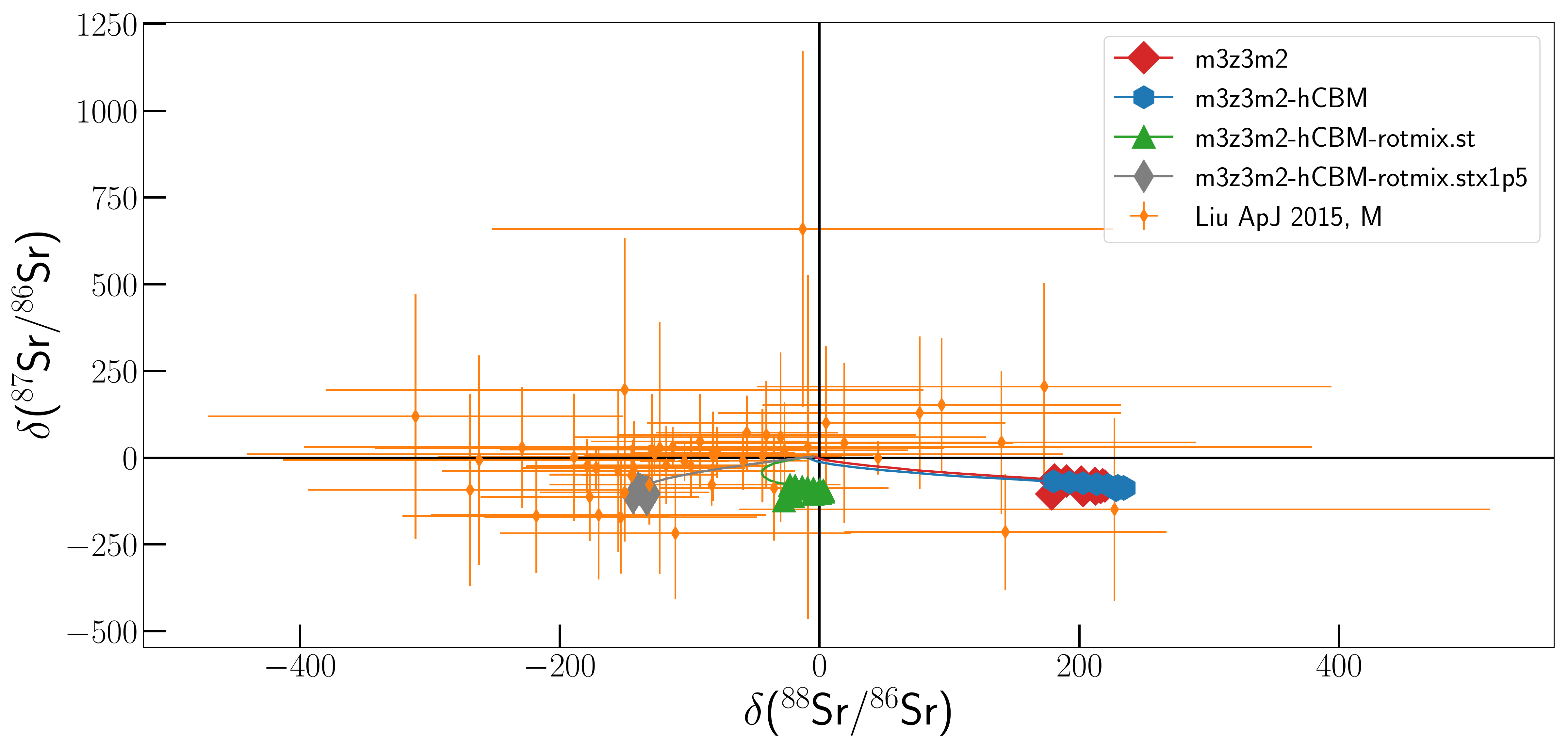}}}
\caption{Comparison of stellar models presented in this work with measured Sr isotopic ratios from presolar SiC grains. Each symbol marking a theoretical prediction corresponds to an interpulse-period, bigger-size symbols corresponding to the carbon-rich phase. Is visible how rotation-induced mixing may help self-consistently cover the whole observed range, in particular in ${^{88}}$Sr/${^{86}}$Sr. Error bars accounts for a two $\sigma$ uncertainty.}
\label{sr:iso}
\end{figure}

In Fig. \ref{sr:iso} we compare our models with measured Sr isotopic ratios.
Plotted values are given in $\delta$-value notation to represent the isotopic ratios,
i.e. the permil variation with respect to the solar ratio (for which  $\delta$=0),
so that $\delta$=(($model$ $ratio$/$solar$ $ ratio$)-1)$\times$1000.
Each symbol marking a theoretical predictions corresponds to an interpulse-period,
with bigger-size symbols corresponding to the carbon-rich phase, which is a necessary condition for grains to form.
As visible in panels 2 and 4, we tested both a larger TDU efficiency and rotation-induced mixing to consistently cover the whole observed range, with the latter having the largest impact.
This is particularly important for ${^{88}}$Sr/${^{86}}$Sr,
where the neutron-magic ${^{88}}$Sr is depleted more and more by higher diffusion of ${^{14}}$N inside the ${^{13}}$C-pocket,
as it would occur in faster rotating models.
More precisely, Fig. \ref{sr:iso} shows how a range of initial rotation velocity,
able to produce the additional intershell mixing between zero and the value inserted in m3z2m2-rotmix.stx1p5,
would be able to cover the bulk of the observed values.
Rotation also improves the comparison to measured ${^{84}}$Sr/${^{86}}$Sr ratios,
pushing the tracks towards the bulk of data which have typically values lower than $~$800,
due to a lower destruction of ${^{86}}$Sr as a consequence of the lower neutron-exposure,
while ${^{84}}$Sr is unaffected being a $p$-only isotope.
On the other hand, this is still not enough to reproduce the typical ${^{84}}$Sr/${^{86}}$Sr measured from most of the grains,
possibly suggesting a too weak depletion of ${^{84}}$Sr.

\subsubsection{Ba}
\label{sec:ba}

\begin{figure}[htbp]
\centering
\resizebox{10.8cm}{!}{\rotatebox{0}{\includegraphics{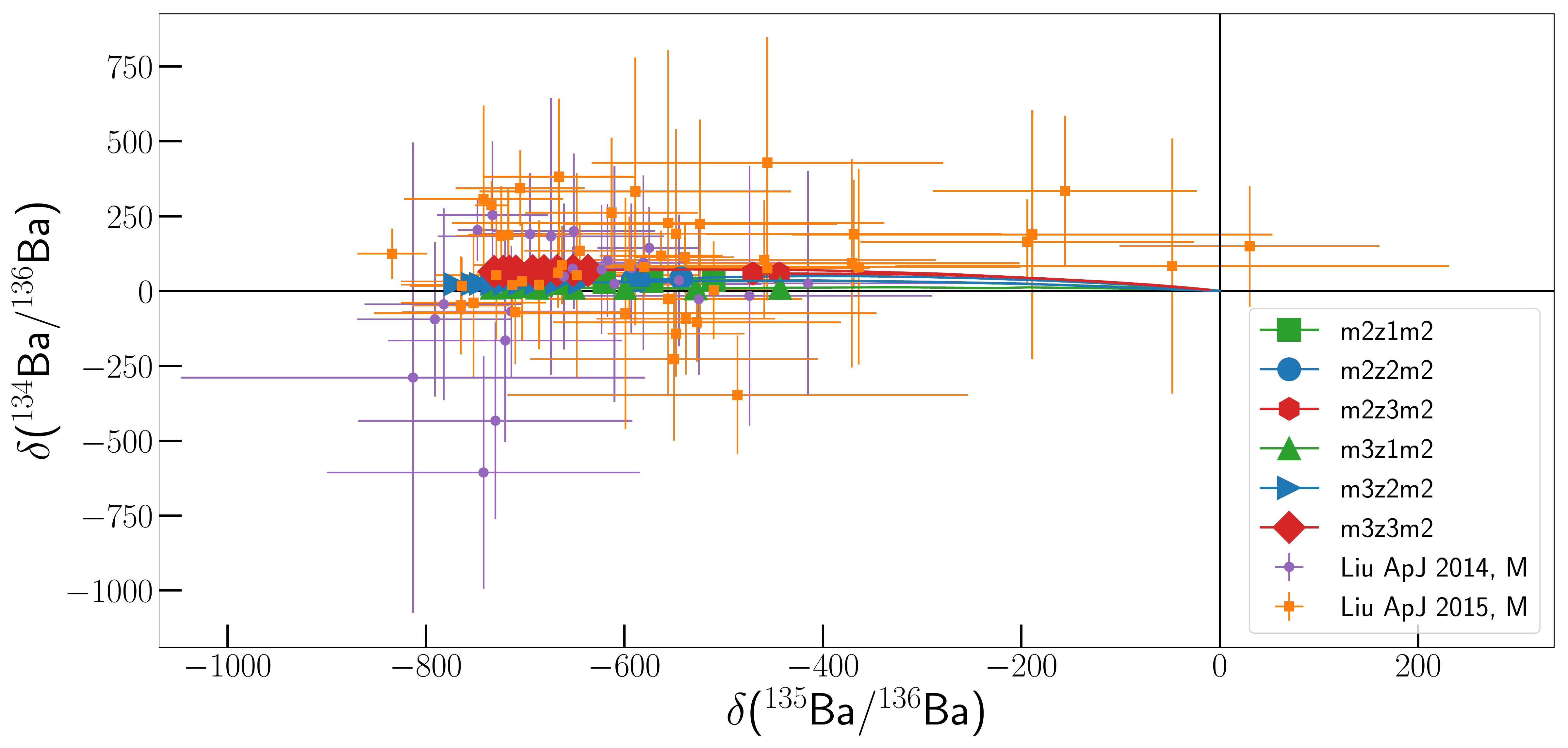}}}
\resizebox{10.8cm}{!}{\rotatebox{0}{\includegraphics{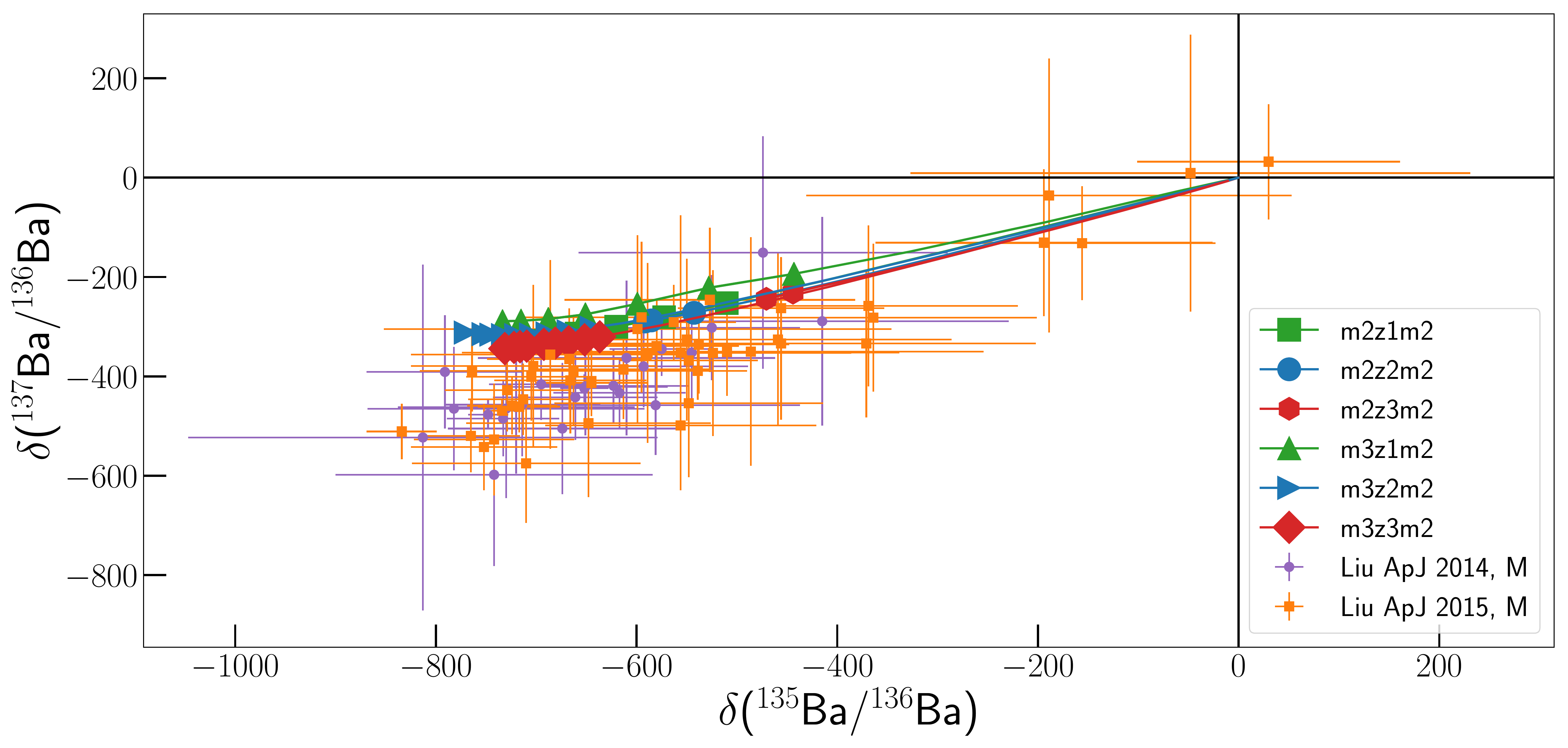}}}
\resizebox{10.8cm}{!}{\rotatebox{0}{\includegraphics{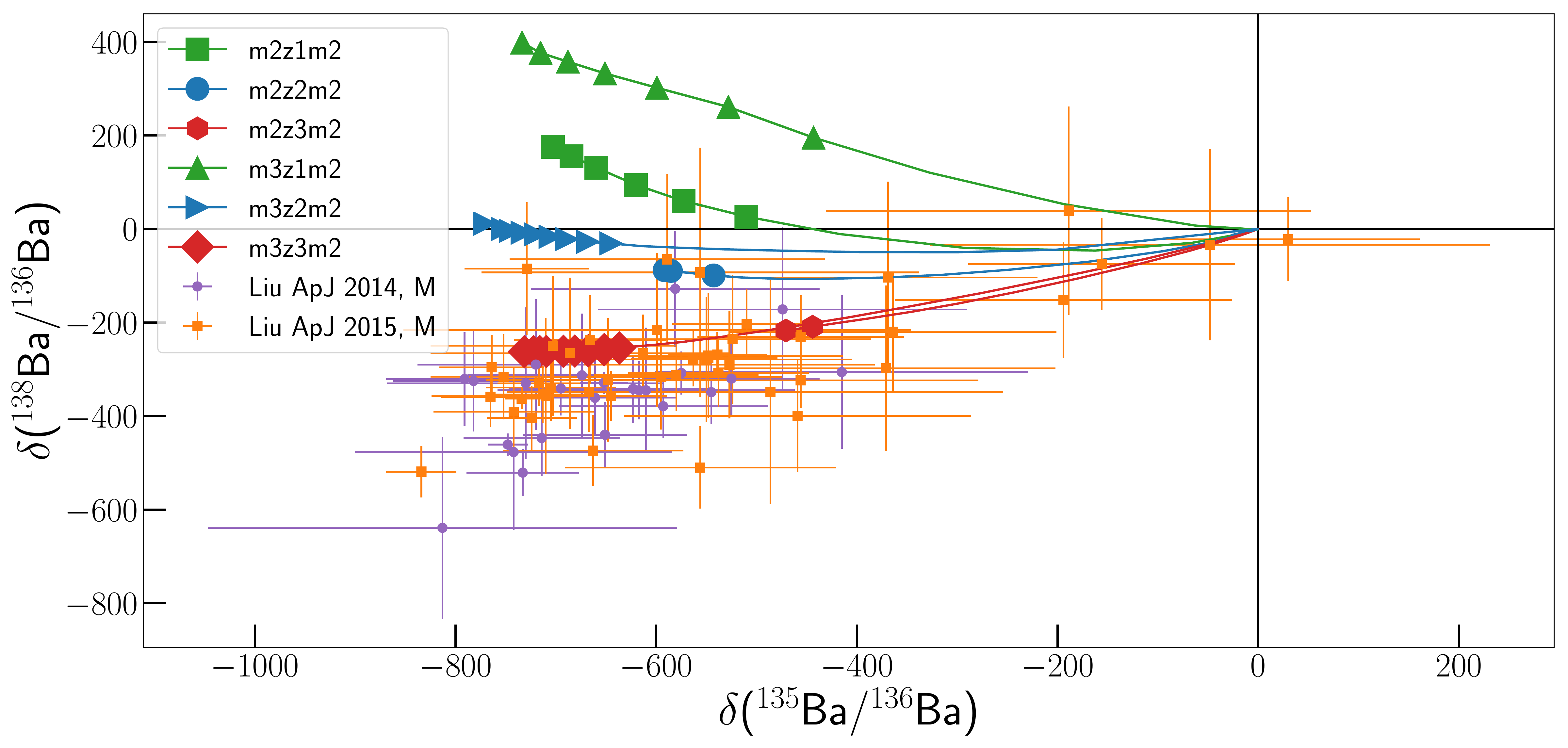}}}
\caption{Comparison of stellar models presented in this work with measured Ba isotopic ratios from presolar SiC grains.}
\label{ba:iso}
\end{figure}

\begin{figure}[htbp]
\centering
\resizebox{14.8cm}{!}{\rotatebox{0}{\includegraphics{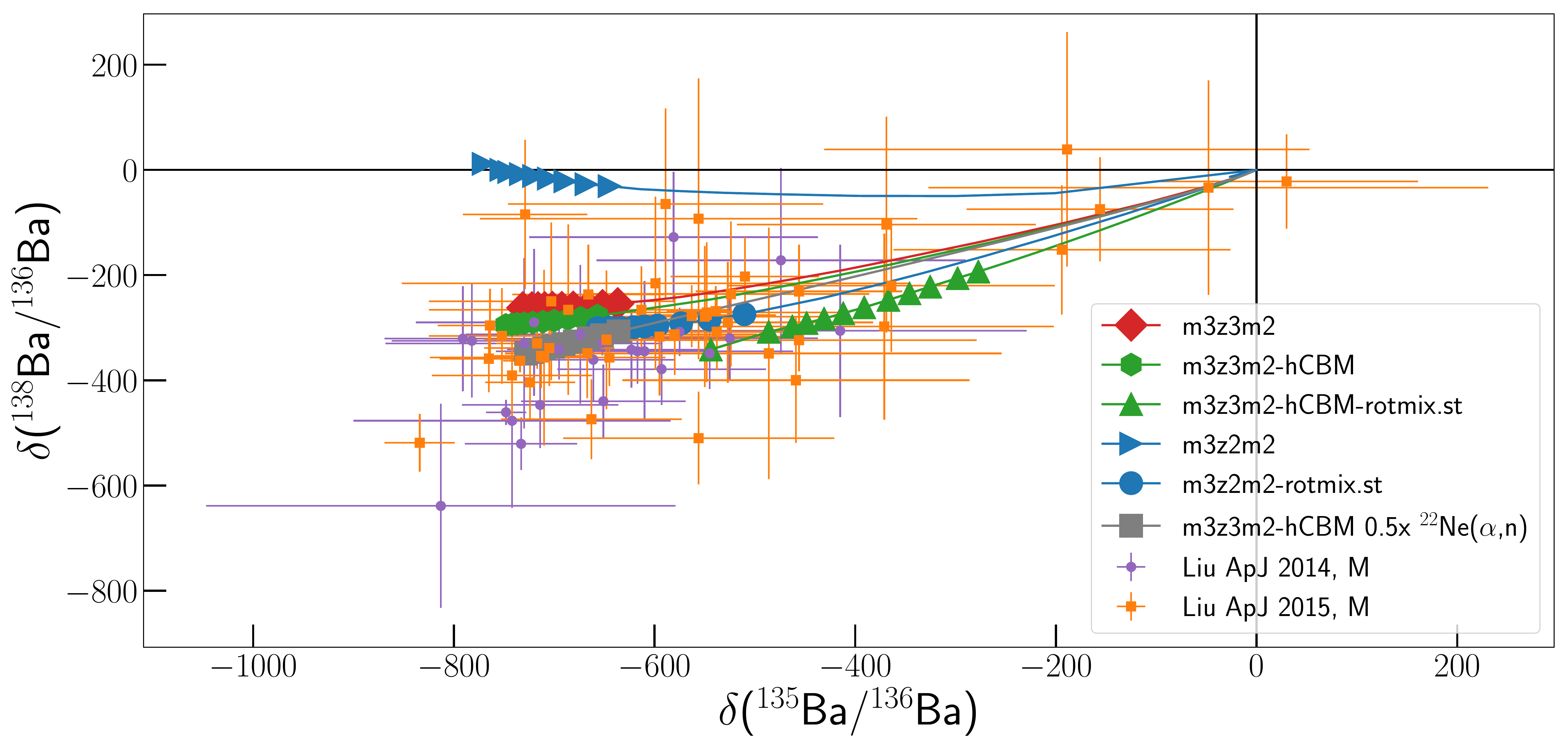}}}
\caption{Comparison of $\delta$(${^{138}}$Ba/${^{136}}$Ba) vs $\delta$(${^{135}}$Ba/${^{136}}$Ba) from m2z2m2, m2z3m2, m3z2m2 and m3z3m3.
  We also included two models including artificial rotation-induced mixing, m3z2m2.rotmix.st and m3z2m2.rotmix.std2.}
\label{ba:isorot}
\end{figure}

In Fig. \ref{ba:iso} we compare our models with measured Ba isotopic ratios. As shown in fig. \ref{ba:isorot},
the artificial inclusion of rotation-induced mixing pushes our stellar tracks down to lower $\delta$(${^{138}}$Ba/${^{136}}$Ba),
as the neutron-magic ${^{138}}$Ba decreases with decreasing neutron-exposure,
closer to the experimental data from \cite{liu:14a} and \cite{liu:15}.
In the same figure, m3z3m2-hCBM m3z3m2 are also presented,
showing how models with a larger \cdr\ -pocket perform better in reproducing laboratory measurements.
Additionally, we tested the lower limit of our adopted $^{22}$Ne($\alpha$,n)$^{25}$Mg rate, the main neutron source at He-flash temperatures, by dividing it by a factor of two, consistently with a 2$\sigma$ variation according to the Monte Carlo calculation by \cite{iliadis:10}, whose recommended rate is very consistent with \cite{jaeger:01}.
Interestingly, the stellar track is pushed down right into the experimental data-point,
indicating how our models are actually in good agreement with laboratory data, within nuclear uncertainties.

\begin{figure}[htbp]
\centering
\resizebox{14.8cm}{!}{\rotatebox{0}{\includegraphics{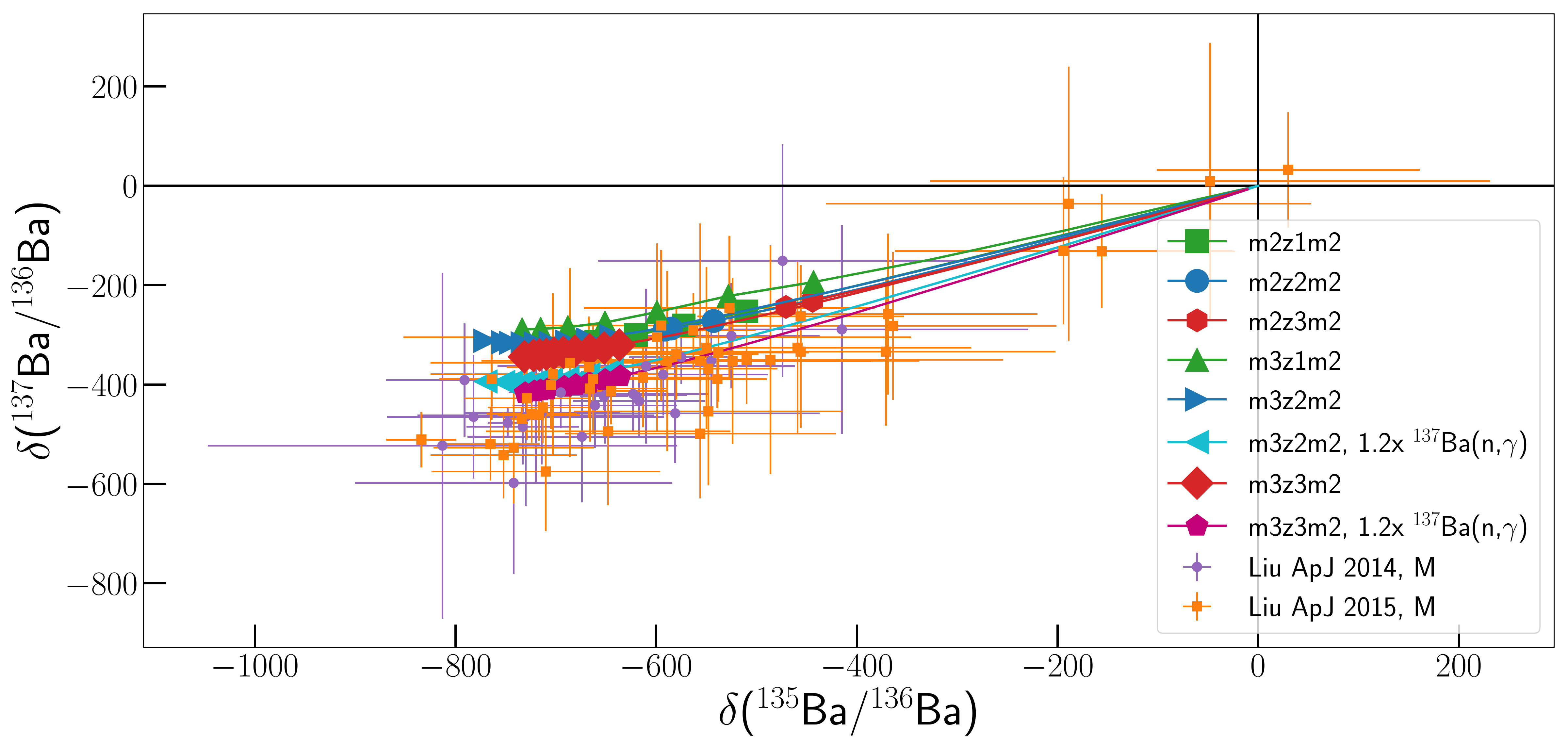}}}
\caption{Comparison of m3z2m2 and m3z3m2 $\delta$(${^{137}}$Ba/${^{136}}$Ba) vs $\delta$(${^{135}}$Ba/${^{136}}$Ba) with measurements from SiC grains: we show results obtained when adopting the ${^{137}}$Ba(n,$\gamma$)${^{138}}$Ba given by Kadonis 0.3 (that we used as standard)
  to what is recommended in Kadonis 1.0 (i.e. a factor of 1.2 higher than Kadonis 0.3}.
\label{ba137:nuctest}
\end{figure}

Rotationally induced mixing does not impact $\delta$(${^{134}}$Ba/${^{136}}$Ba) and $\delta$(${^{137}}$Ba/${^{136}}$Ba),
since they mainly depend on the neutron capture cross-sections: the models reproduce well the observed $\delta$(${^{134}}$Ba/${^{136}}$Ba),
on the other hand it looks like it is not the case when comparing them with measured $\delta$(${^{137}}$Ba/${^{136}}$Ba).
The ${^{137}}$Ba(n,$\gamma$)${^{138}}$Ba reaction rate has been considerably increased (by a factor of 1.2) from Kadonis 0.3 (that we adopt) to Kadonis 1.0.
Therefore, we tested this new rate in m3z2m2 and m3z3m2.
Indeed, fig. \ref{ba137:nuctest} shows how pre-solar grains and our models prediction are in better agreement when using the newer  ${^{137}}$Ba(n,$\gamma$)${^{138}}$Ba rate.

\begin{figure}[htbp]
\centering
\resizebox{13.8cm}{!}{\rotatebox{0}{\includegraphics{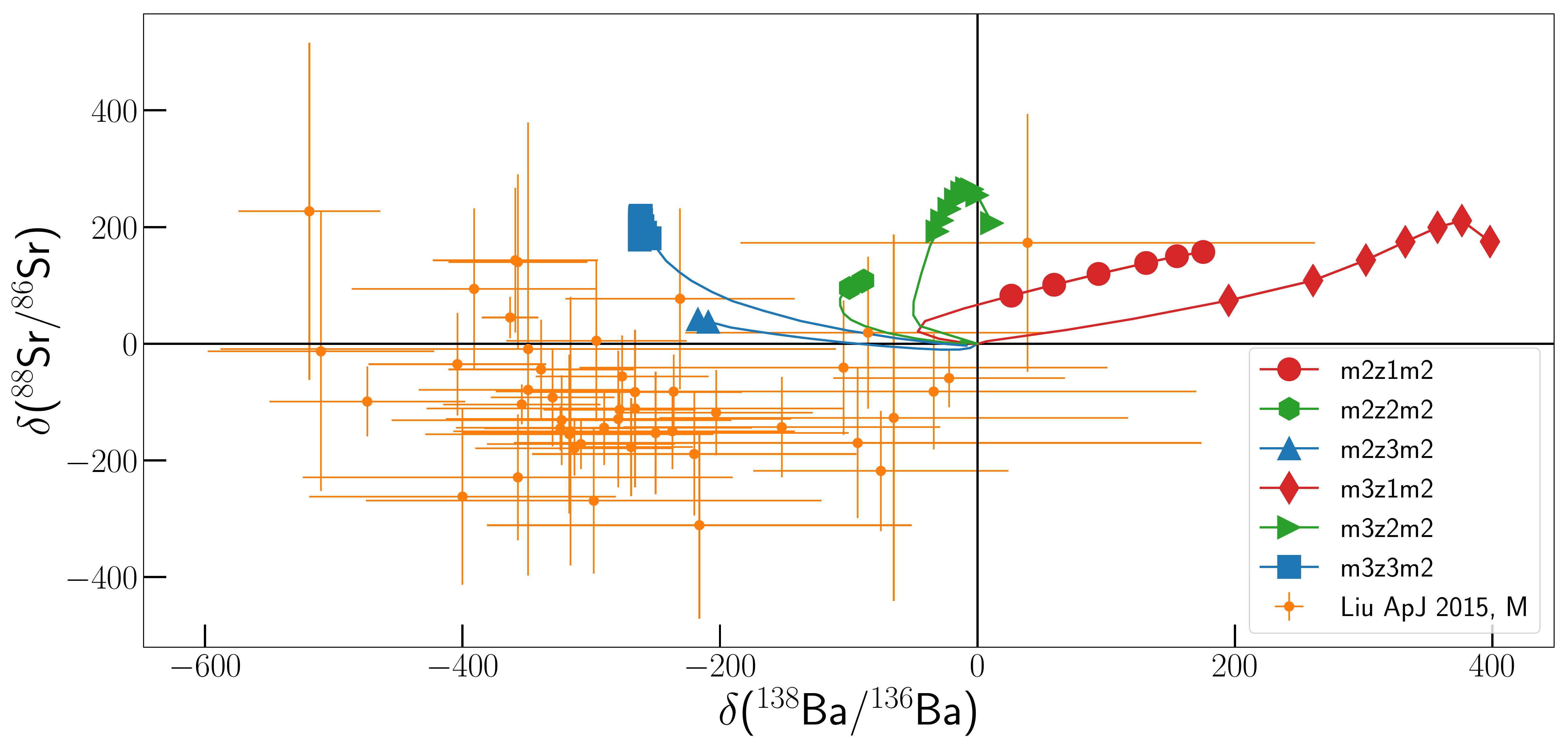}}}
\resizebox{13.8cm}{!}{\rotatebox{0}{\includegraphics{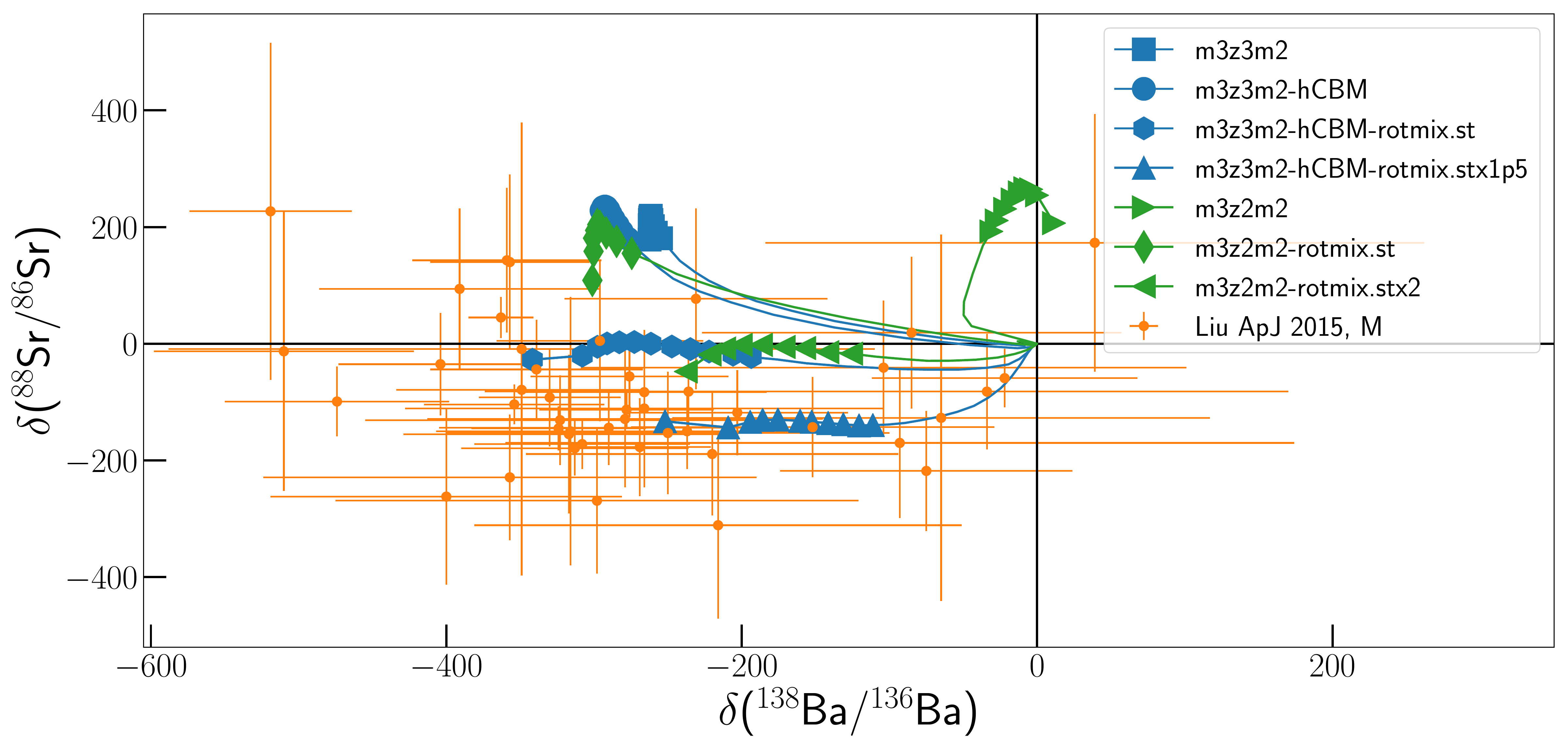}}}
\caption{Correlated measurements of Sr and Ba of \cite{liu:15}, compared to our standard set in the upper panel and to models including artificial rotation-induced mixing in the lower one.}
\label{sr:ba}
\end{figure}

Since \cite{liu:15} provides Ba and Sr data coming from the same grain,
we perform an additional comparison using the observed correlation between ${^{138}}$Ba/${^{136}}$Ba versus ${^{88}}$Sr/${^{86}}$Sr.
Fig. \ref{sr:ba} shows that most of the grains present
-200 $<$ $\delta$(${^{88}}$Sr/${^{86}}$Sr) $<$ 0 and -400 $<$ $\delta$(${^{138}}$Ba/${^{136}}$Ba) $<$ -200. In the lower panel,
m3z3m2-hCBM-rotmix.st, m3z3m2-hCBM-rotmix.stx1p5 and m3z2m2-rotmix.stx2 tracks successfully enter this specific area in the diagram,
hence being able to explain the bulk of grains data. It is important to notice how all these three models are *rotmix* models,
having all a slow constant extra-mixing active into the intershell,
suggesting rotation-induced mixing as an strong candidate to explain the range of observed values in presolar grains.
In particular, m3z3m2-hCBM-rotmix.st nicely reproduce the range of observed ${^{138}}$Ba/${^{136}}$Ba values.
At the same time, a range of stronger stronger extra-mixing, as in m3z3m2-hCBM-rotmix.stx1p5,
may effectively reproduce the observed ${^{88}}$Sr/${^{86}}$Sr range.

\subsubsection{Zr}
\label{sec:zr}

\begin{figure}[htbp]
\centering
\resizebox{10.8cm}{!}{\rotatebox{0}{\includegraphics{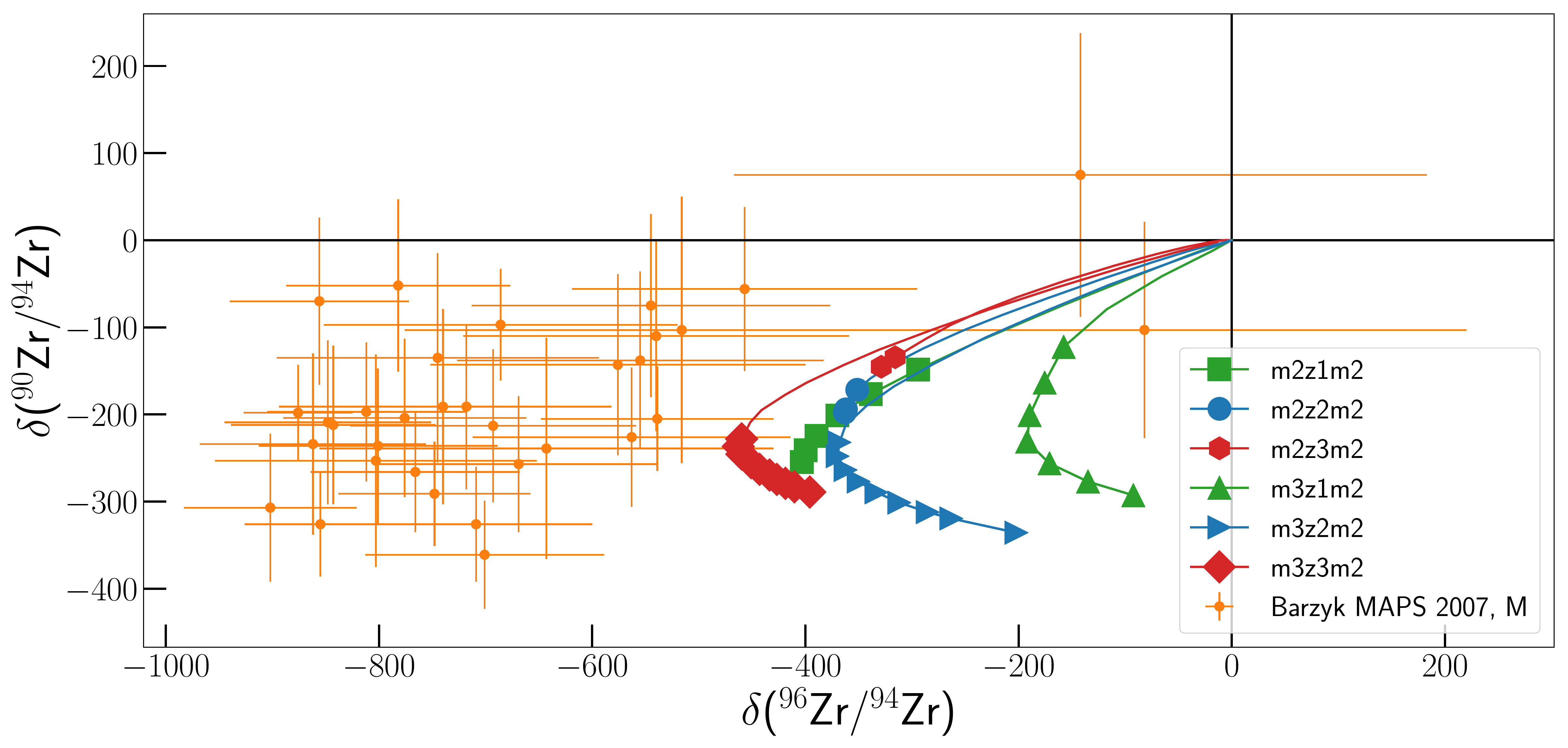}}}
\resizebox{10.8cm}{!}{\rotatebox{0}{\includegraphics{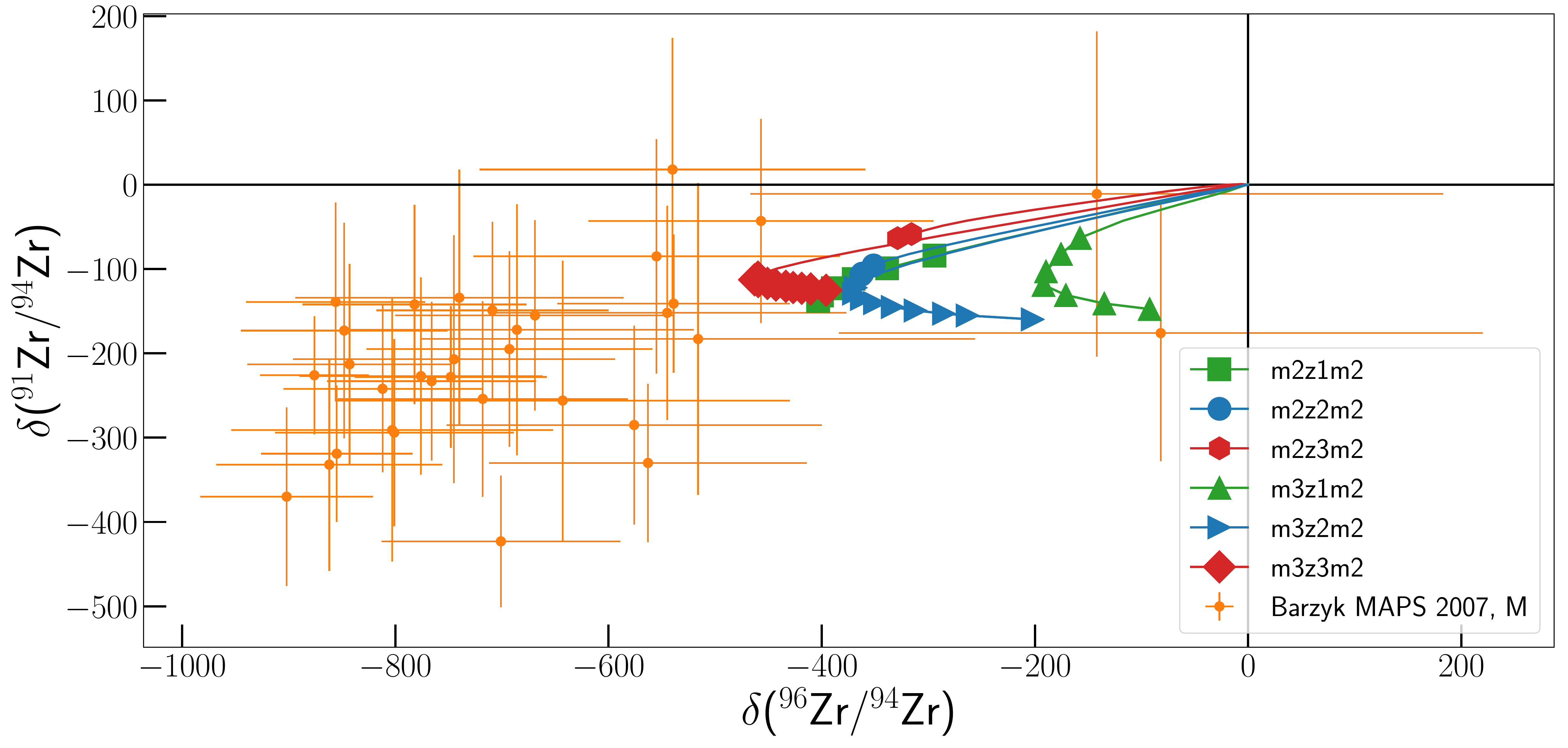}}}
\resizebox{10.8cm}{!}{\rotatebox{0}{\includegraphics{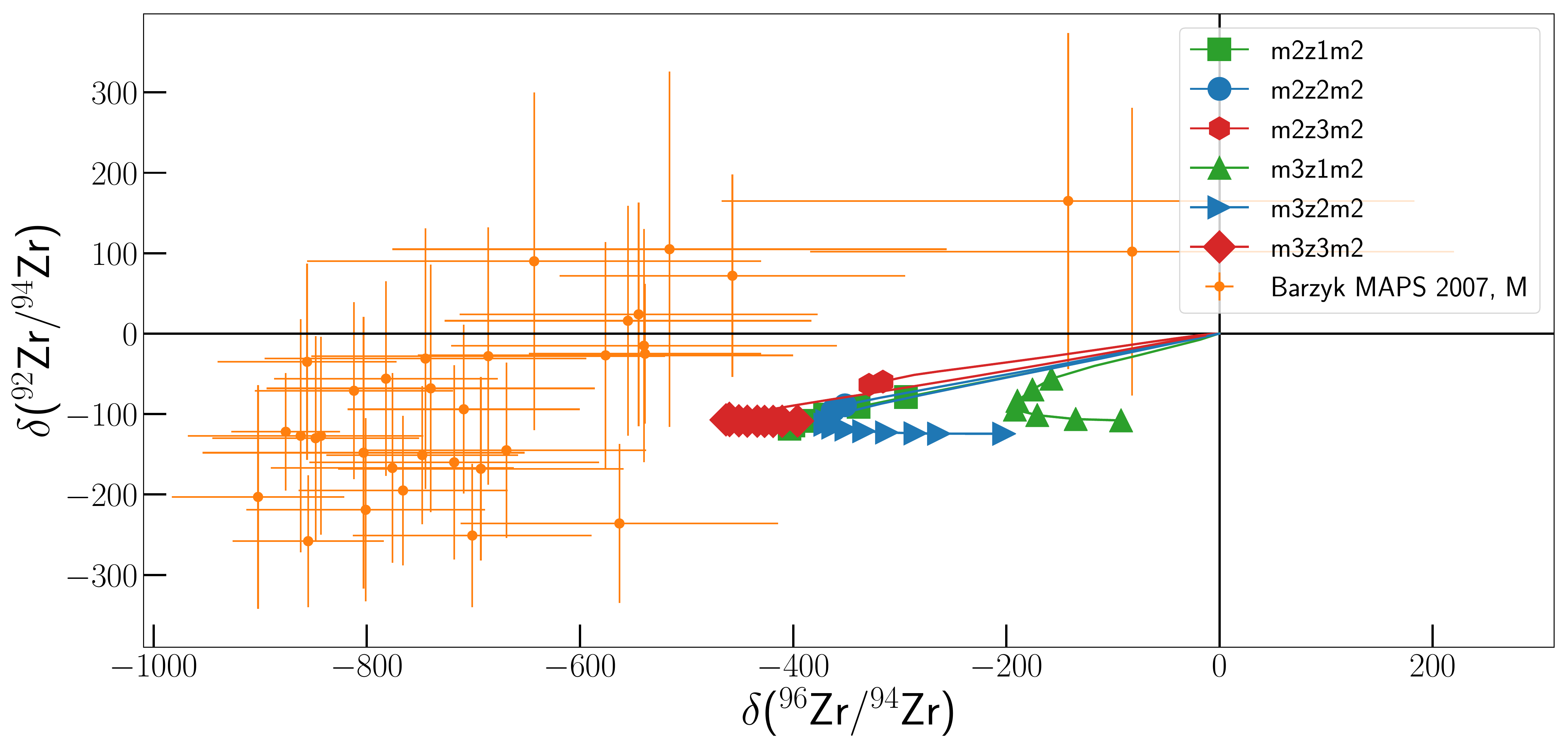}}}
\caption{Comparison of stellar models presented in this work with \cite{barzyk:07} measurements of Zr isotopic ratios.}
\label{zr:iso}
\end{figure}

\begin{figure}[htbp]
\centering
\resizebox{10.8cm}{!}{\rotatebox{0}{\includegraphics{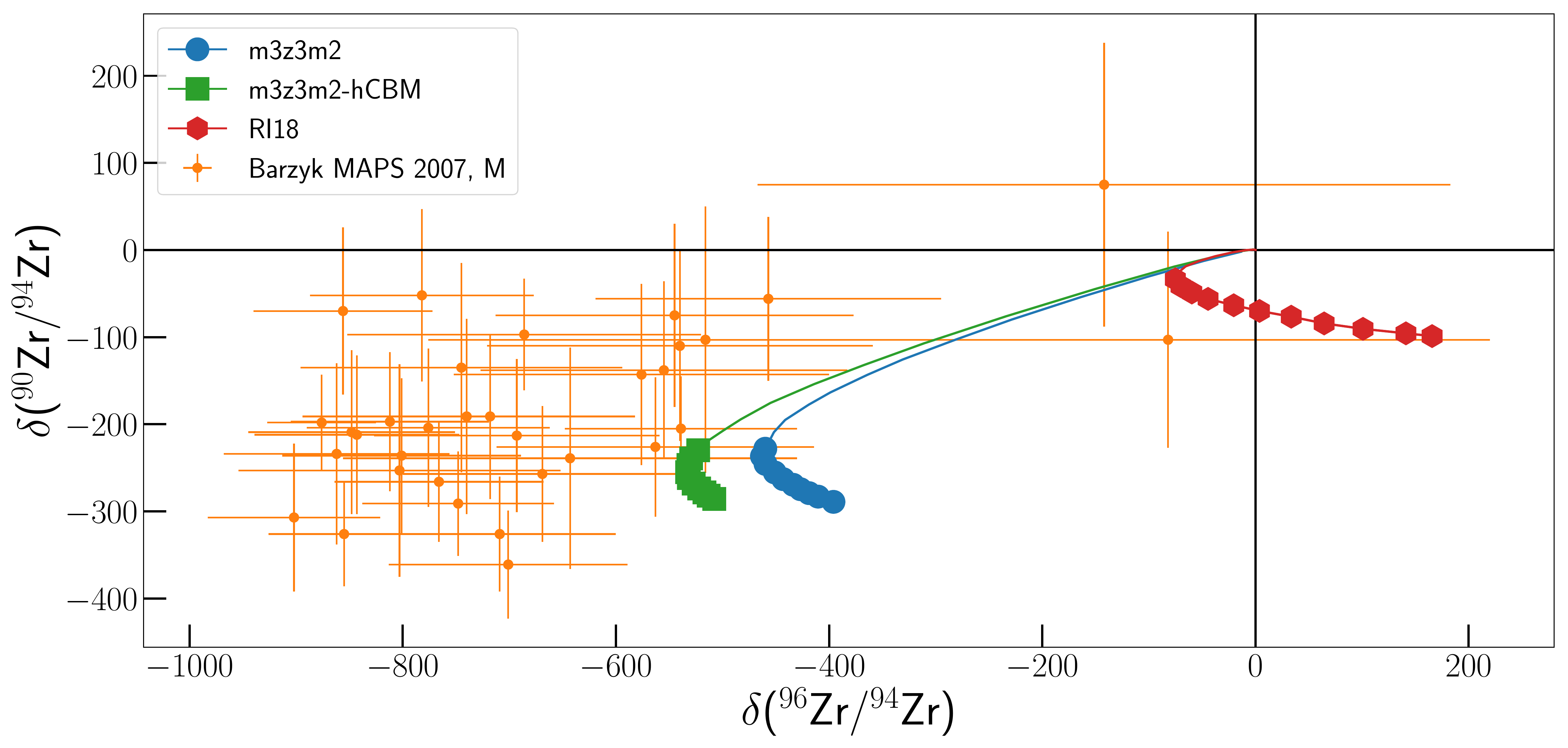}}}
\resizebox{10.8cm}{!}{\rotatebox{0}{\includegraphics{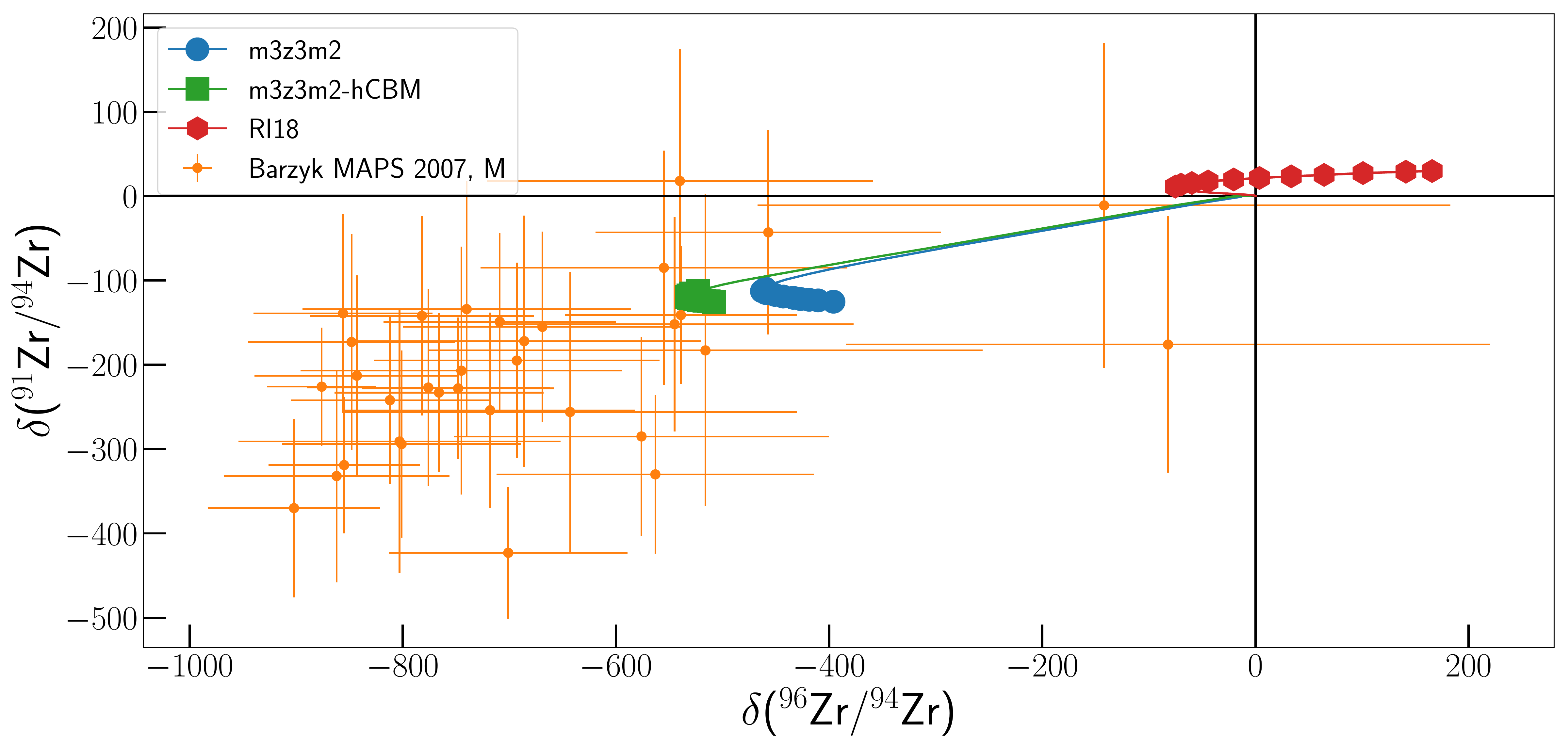}}}
\resizebox{10.8cm}{!}{\rotatebox{0}{\includegraphics{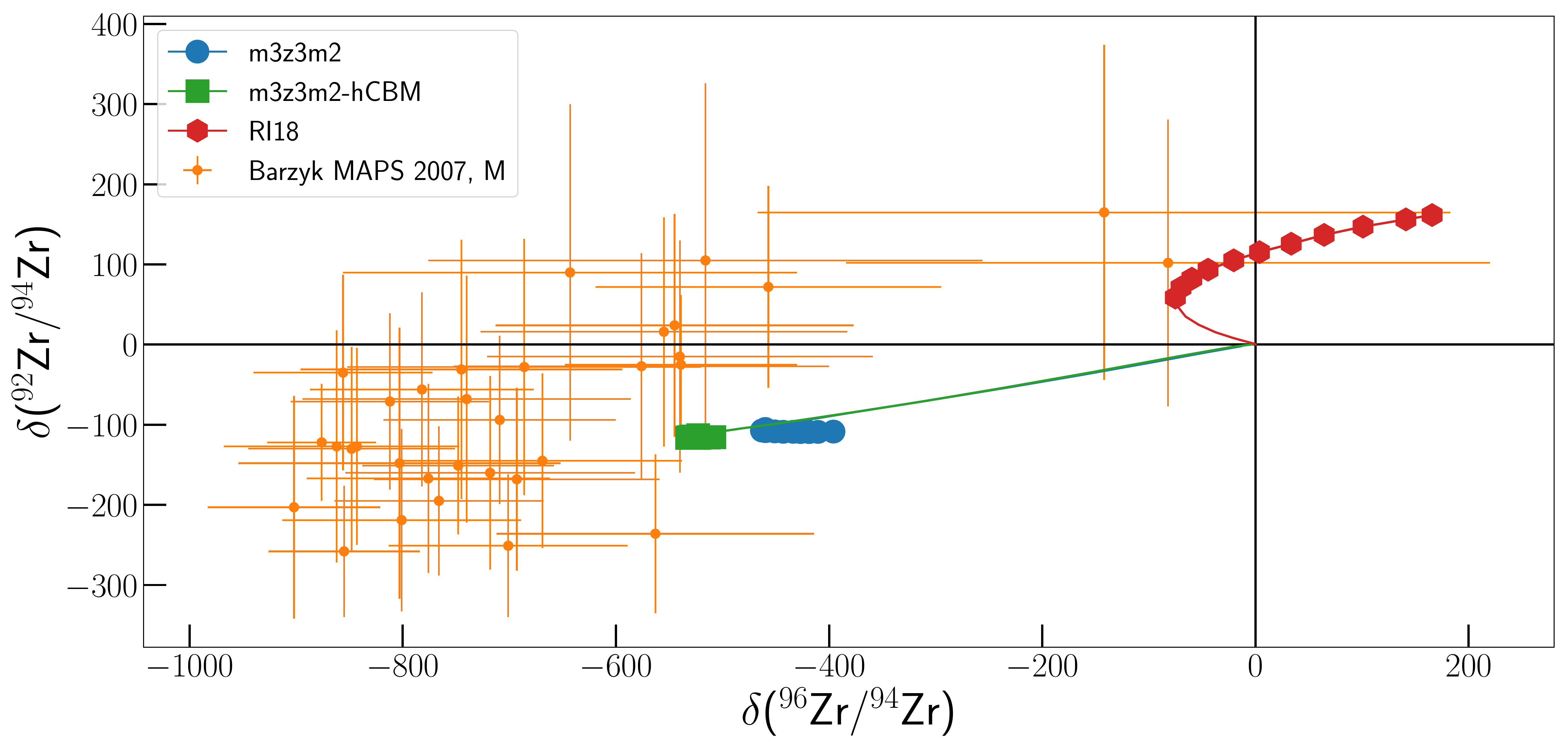}}}
\caption{Same as in Fig. \ref{zr:iso}, but the results are shown for models m3z3m2 and m3z3m2-hCBM.
  The larger $s$-process production in m3z3m2-hCBM is a consequence of a $^{13}$C-pocket 50$\%$ larger in mass-coordinate compared to m3z3m2,
leading to a stronger production of $^{94}$Zr and hence decreasing the $^{94}$Zr/ $^{96}$Zr isotopic ratio.}
\label{zr:poctest}
\end{figure}

\begin{figure}[htbp]
\centering
\resizebox{10.8cm}{!}{\rotatebox{0}{\includegraphics{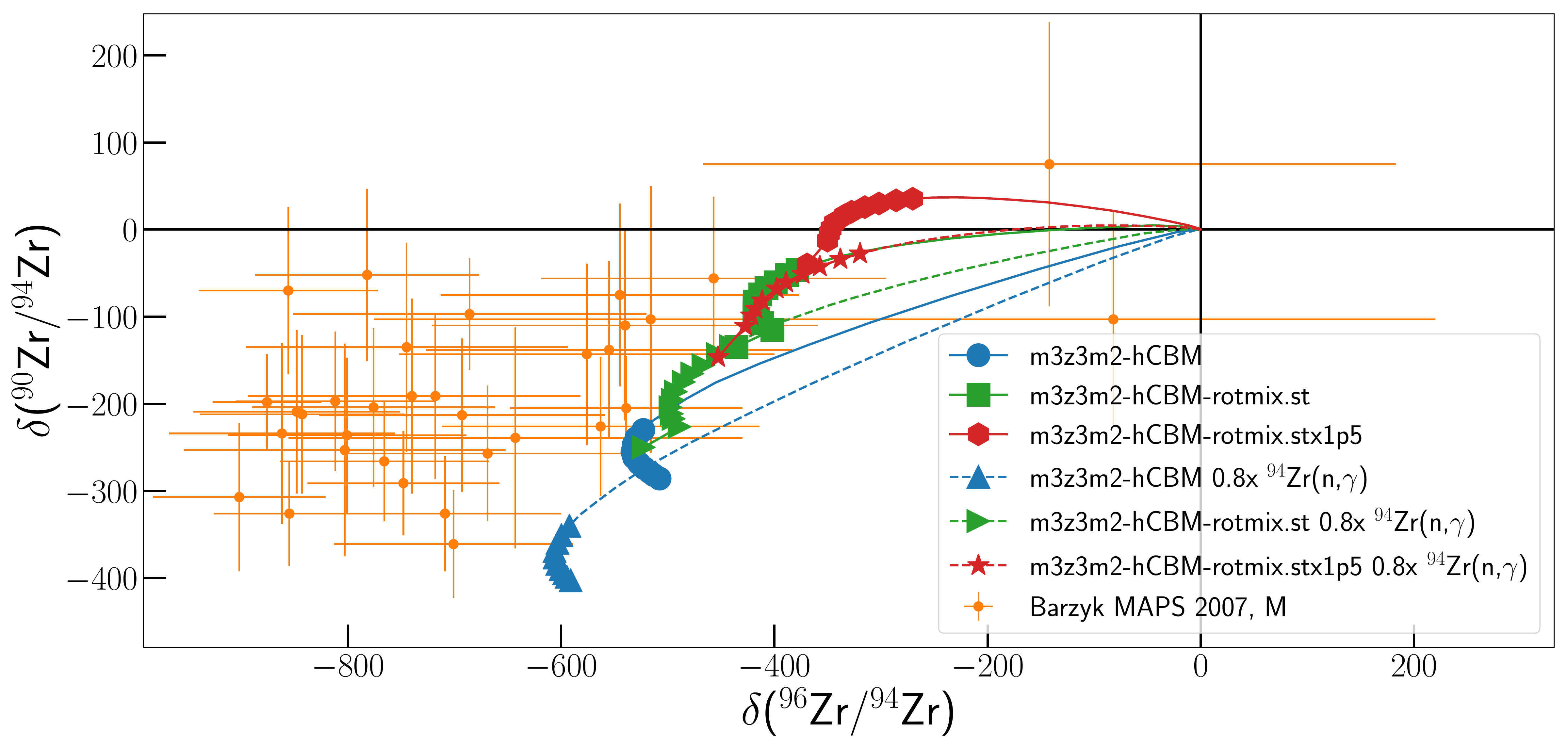}}}
\resizebox{10.8cm}{!}{\rotatebox{0}{\includegraphics{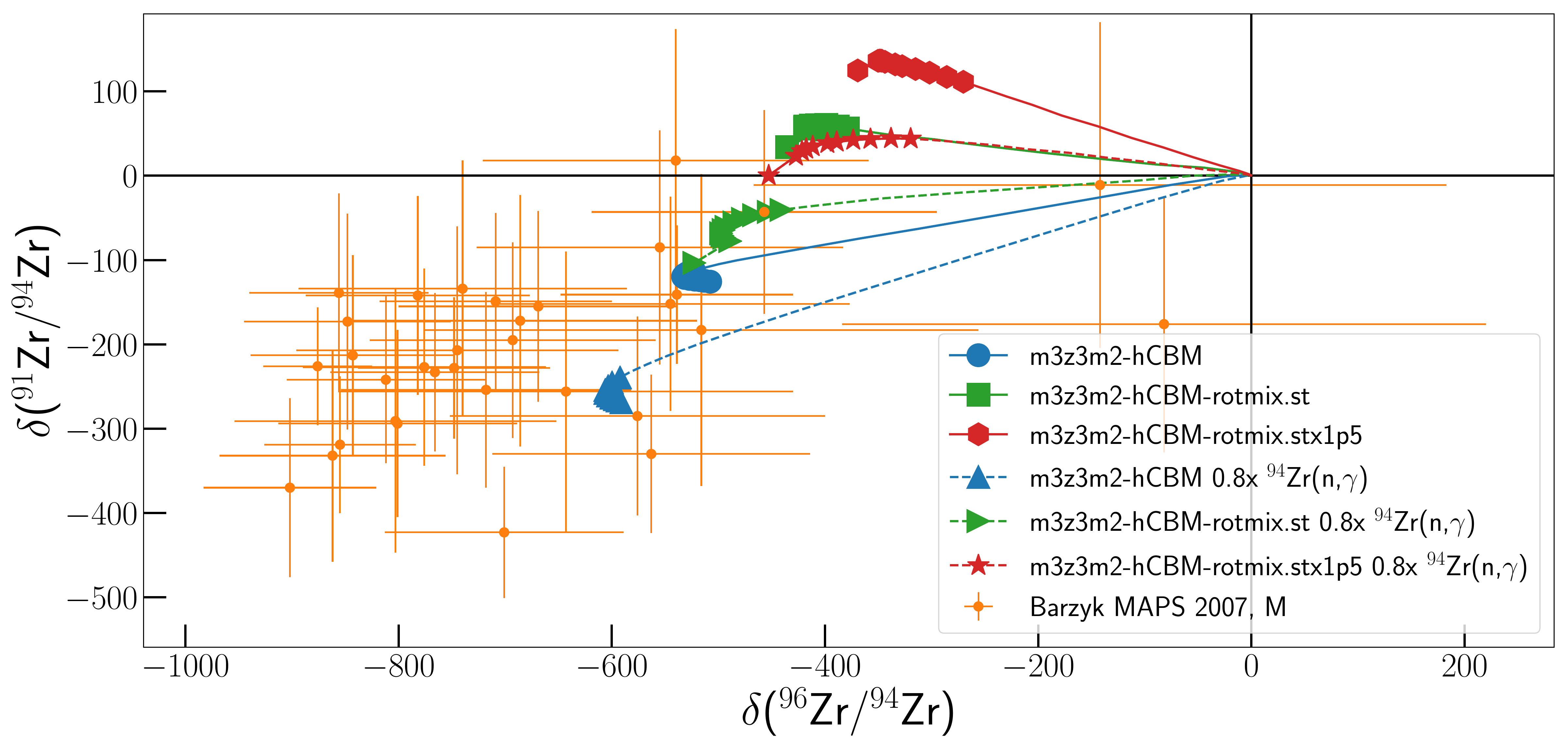}}}
\resizebox{10.8cm}{!}{\rotatebox{0}{\includegraphics{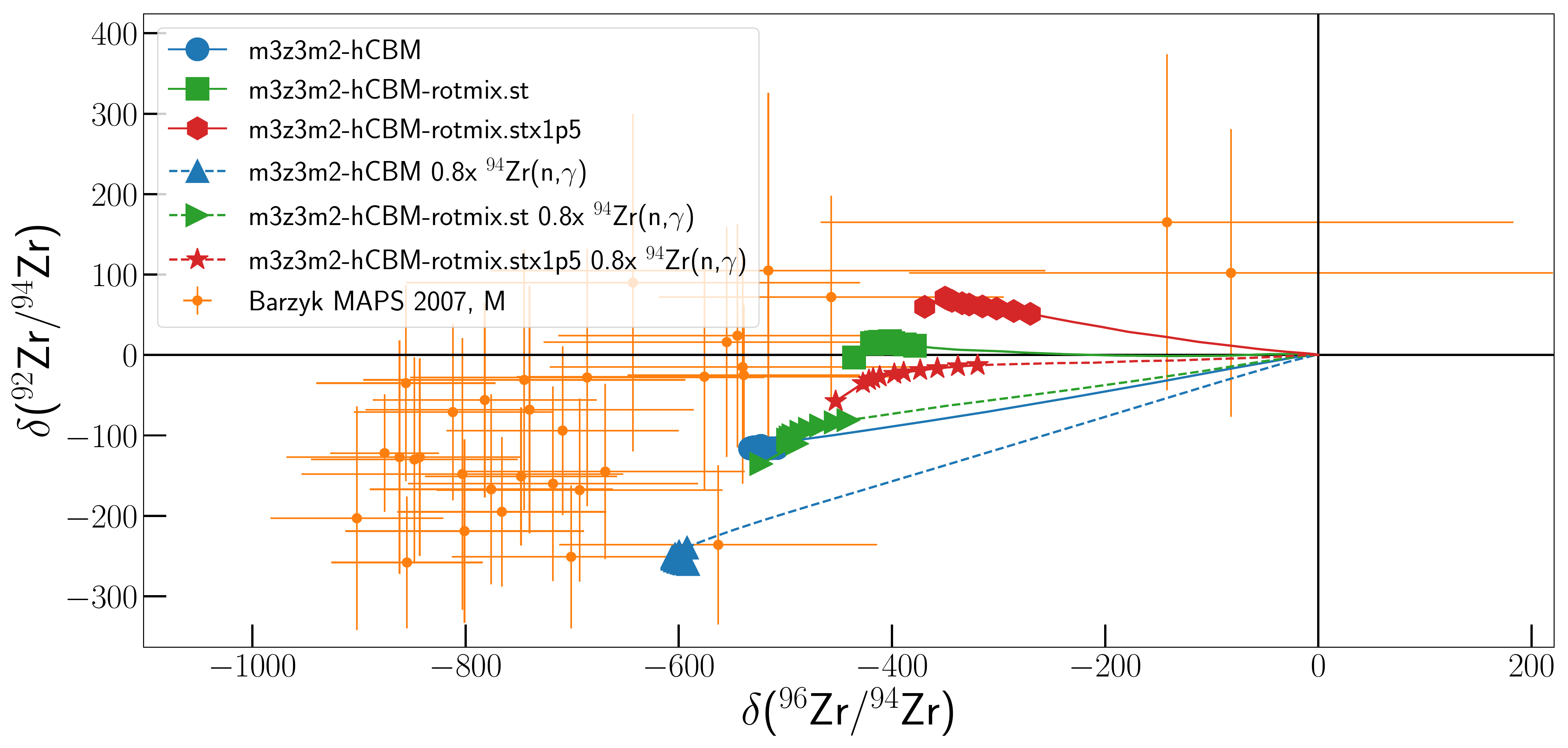}}}
\caption{Same as in figure \ref{zr:iso}, but here we show the impact of the neutron capture rate on $^{94}$Zr on our theoretical predictions. In particular, we apply a factor of 0.8 to the $^{94}$Zr(n,$\gamma$)$^{95}$Zr reaction rate to test the value recommended in Kadonis 0.3, since it is 20$\%$ lower than the \cite{lugaro:14} recommended rate that we adopted. We also show the effect of rotation-induced mixing which, combined to neutron capture reaction rate uncertainties, effectively reproduce the whole range of measured $^{90}$Zr/ $^{94}$Zr (already reproduced by our standard set as shown in figure \ref{zr:iso}), $^{91}$Zr/ $^{94}$Zr and $^{92}$Zr/ $^{94}$Zr values.}
\label{zr:zr94test}
\end{figure}

Fig. \ref{zr:iso} shows the predictions of zirconium isotopic ratios for our models compared to \cite{barzyk:07} measurements.
${^{90,91,92}}$Zr/${^{94}}$Zr depend on the nucleosynthesis taking place in the ${^{13}}$C-pocket, while ${^{96}}$Zr/${^{94}}$Zr also depends on TP conditions, since it is affected by the 64 days half-life branching point at ${^{95}}$Zr which can only be opened in high neutron density conditions. Models with higher metallicities produce lower ${^{96}}$Zr/${^{94}}$Zr values for two reasons: 1) the higher the initial metallicity, the more first peak elements are favored compared to second peak ones; 2) a higher metallicity affects stellar opacities and structure, resulting in lower TP temperatures and lower $^{22}$Ne($\alpha$,n)$^{25}$Mg activation. Anyway, our standard setting apparently do not perform well when compared to observations, with the exception of the range of ${^{90}}$Zr/${^{94}}$Zr values.
In Fig. \ref{zr:poctest} we test m3z3m2 and m3z3m2-hCBM models with grains measurements, showing a the model of RI18 as comparison:
m3z3m2-hCBM performs better than m3z3m2, as it reproduces lower ${^{96}}$Zr/${^{94}}$Zr achieving a better agreement with observations,
even if still not good enough.
The big difference between m3z3m2 and RI18 is mainly due to a different adopted ${^{95}}$Zr neutron capture cross section,
that is in our case more than a factor of two lower than the rate used by RI18.
${^{13}}$C-pocket sizes in m3z3m2-hCBM are around 10$^{-4}$ M$_\odot$ large in mass coordinate, a factor of about 1.5 larger than the typical pocket size in m3z3m2. This allows a higher ${^{94}}$Zr production and hence lower ${^{96}}$Zr/${^{94}}$Zr after every TDU.
We noticed
In addition to the $^{94}$Zr(n,$\gamma$)$^{95}$Zr, \cite{cescutti:18} also indicated ${^{95}}$Zr(n,$\gamma$)${^{96}}$Zr as a key reaction rate for ${^{96}}$Zr. We therefore considered nuclear uncertainties with a potential impact on zirconium isotopes.
In Fig. \ref{zr:zr94test} we the impact of the  $^{94}$Zr(n,$\gamma$)$^{95}$Zr rate on our predictions.
In particular, we apply a factor of 0.8 to the $^{94}$Zr(n,$\gamma$)$^{95}$Zr reaction rate to test the value recommended in Kadonis 0.3,
since it is 20$\%$ lower than the \cite{lugaro:14} recommended rate that we adopted. In this case, the lowest measured values of  ${^{91}}$Zr/${^{94}}$Zr and  ${^{92}}$Zr/${^{94}}$Zr are now reproduced. Additionally, when the effects of rotation-induced mixing are included,
also the highest values are explained. In summary, the effect of rotation-induced mixing, combined to neutron capture reaction rate uncertainties, effectively reproduce the whole range of measured $^{90}$Zr/ $^{94}$Zr (already reproduced by our standard set as shown in figure \ref{zr:iso}), $^{91}$Zr/ $^{94}$Zr and $^{92}$Zr/ $^{94}$Zr values.

\begin{figure}[htbp]
\centering
\resizebox{10.8cm}{!}{\rotatebox{0}{\includegraphics{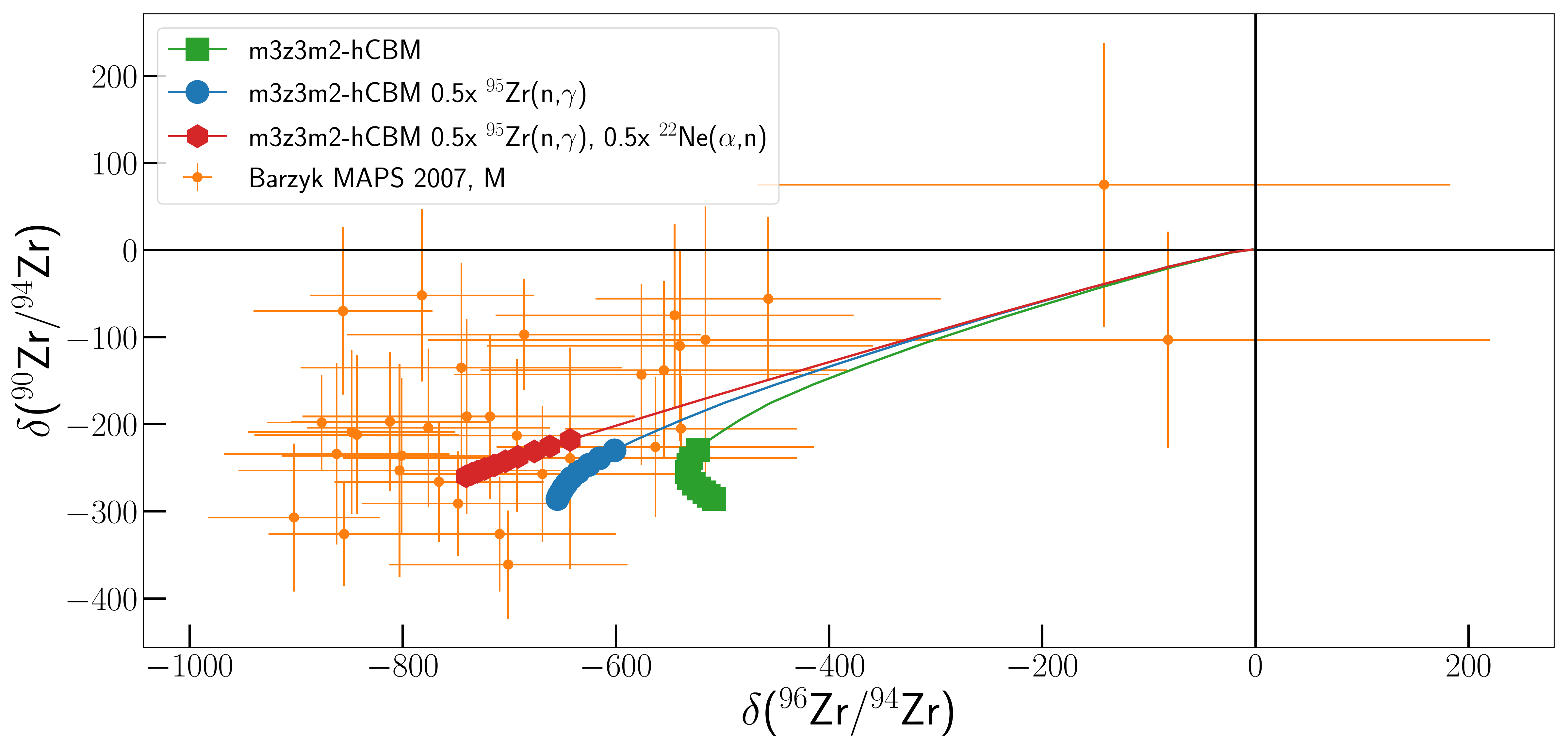}}}
\resizebox{10.8cm}{!}{\rotatebox{0}{\includegraphics{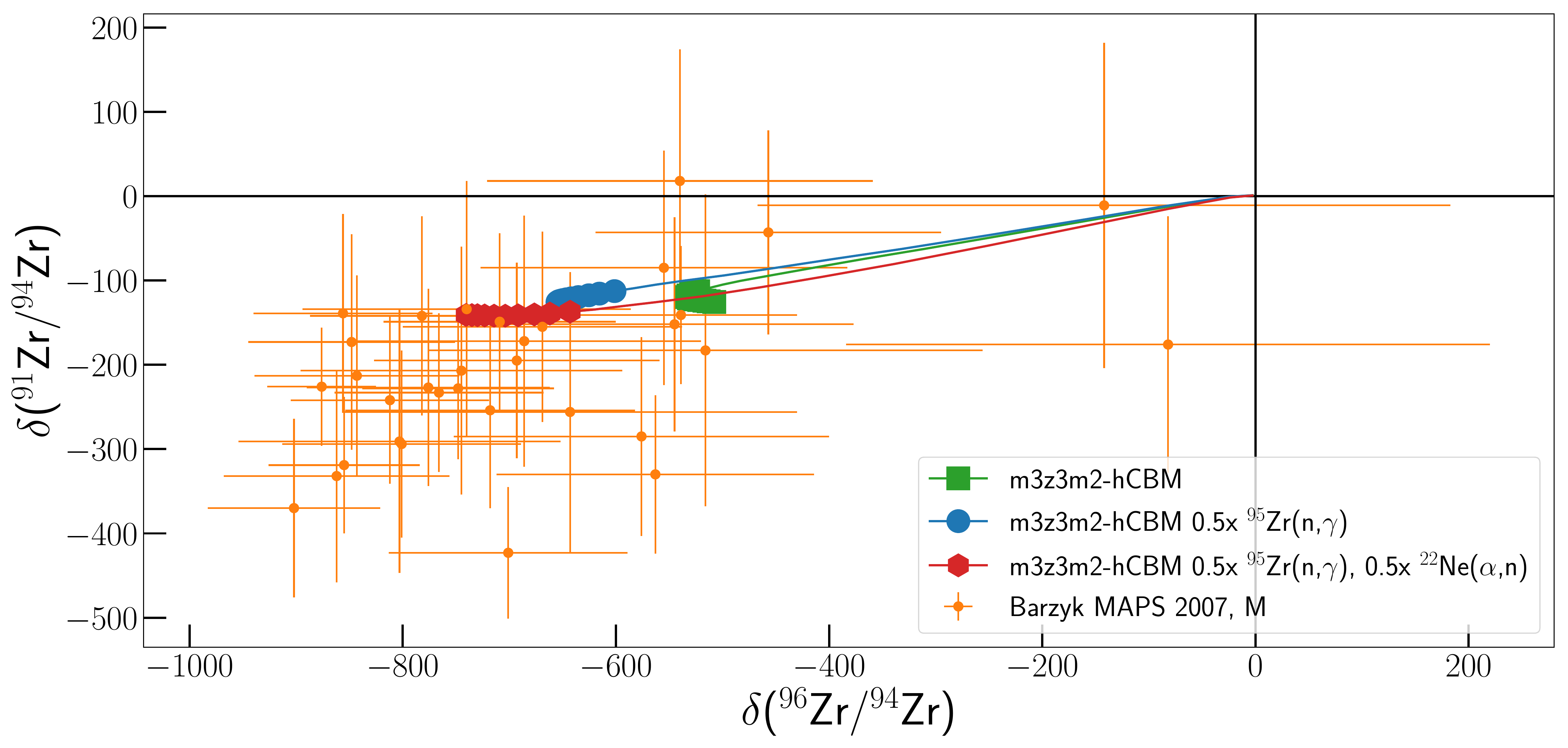}}}
\resizebox{10.8cm}{!}{\rotatebox{0}{\includegraphics{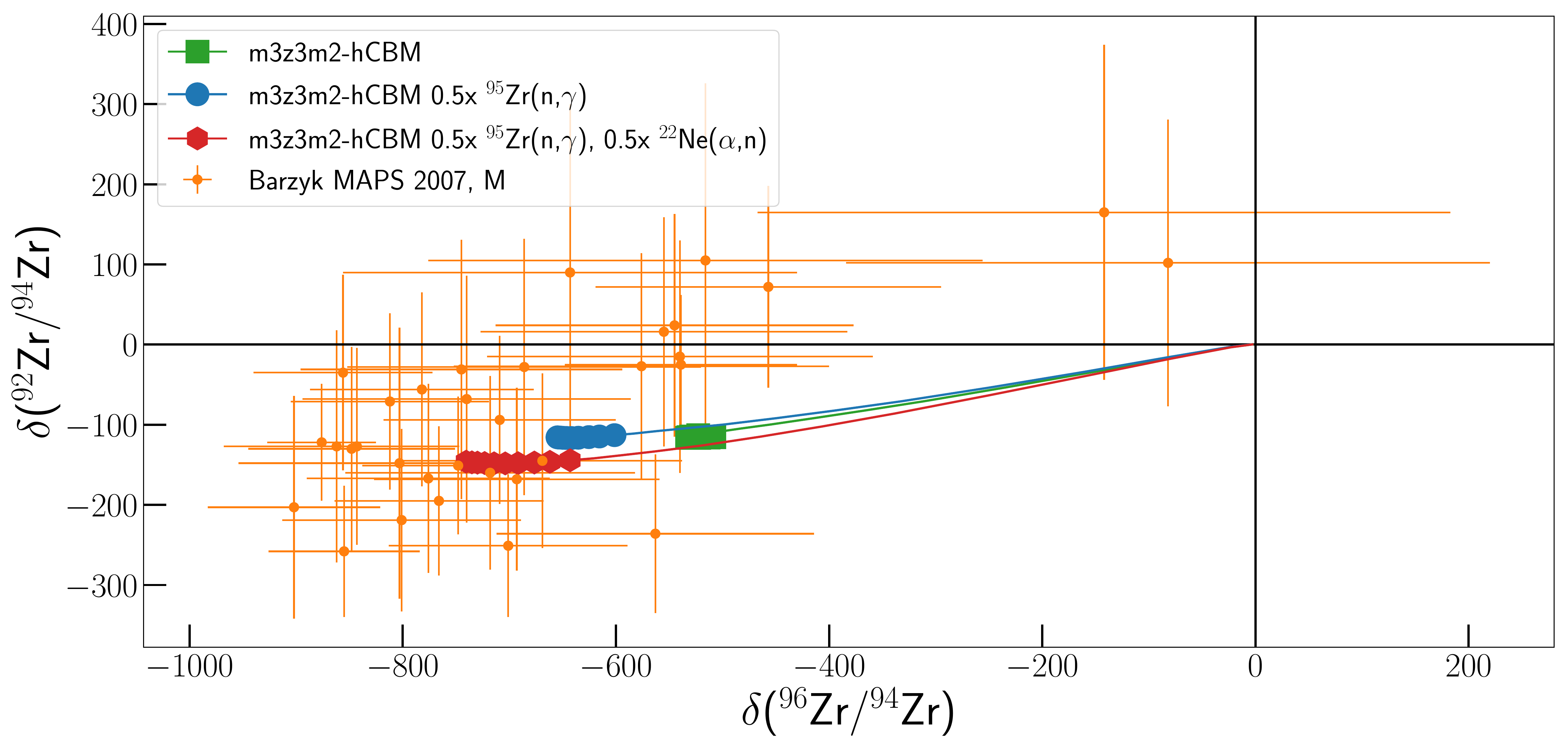}}}
\caption{Same as in figure \ref{zr:iso}, but the results are shown for the models calculated varying key-reaction rates that impact the observed isotopic ratios (see text for details).}
\label{zr:nuctest}
\end{figure}

Fig. \ref{zr:nuctest} shows that our prediction are in better agreement with laboratory measurements when considering stellar modelling and nuclear physics uncertainties: \cite{lugaro:14} gives a factor of 2 uncertainty for ${^{95}}$Zr(n,$\gamma$)${^{96}}$Zr, additionally we we apply the same factor to test the lower limit of our adopted $^{22}$Ne($\alpha$,n)$^{25}$Mg, in the same way as discussed in section \ref{sec:ba}.
The majority of grains data have -800$<$ $\delta$(${^{96}}$Zr/${^{94}}$Zr) $<$-600 and m3z3m2-hCBM track successfully reproduce data in this interval, ranging between -750 and -650 in delta values during the carbon-rich phase.
On the other hand it is not possible for our models to explain those grains with $\delta$(${^{96}}$Zr/${^{94}}$Zr) $<$-800,
failing to reproduce the whole ${^{96}}$Zr/${^{94}}$Zr observed range.

\subsubsection{Mo}
\label{sec:mo}

\begin{figure}[htbp]
\centering
\resizebox{10.8cm}{!}{\rotatebox{0}{\includegraphics{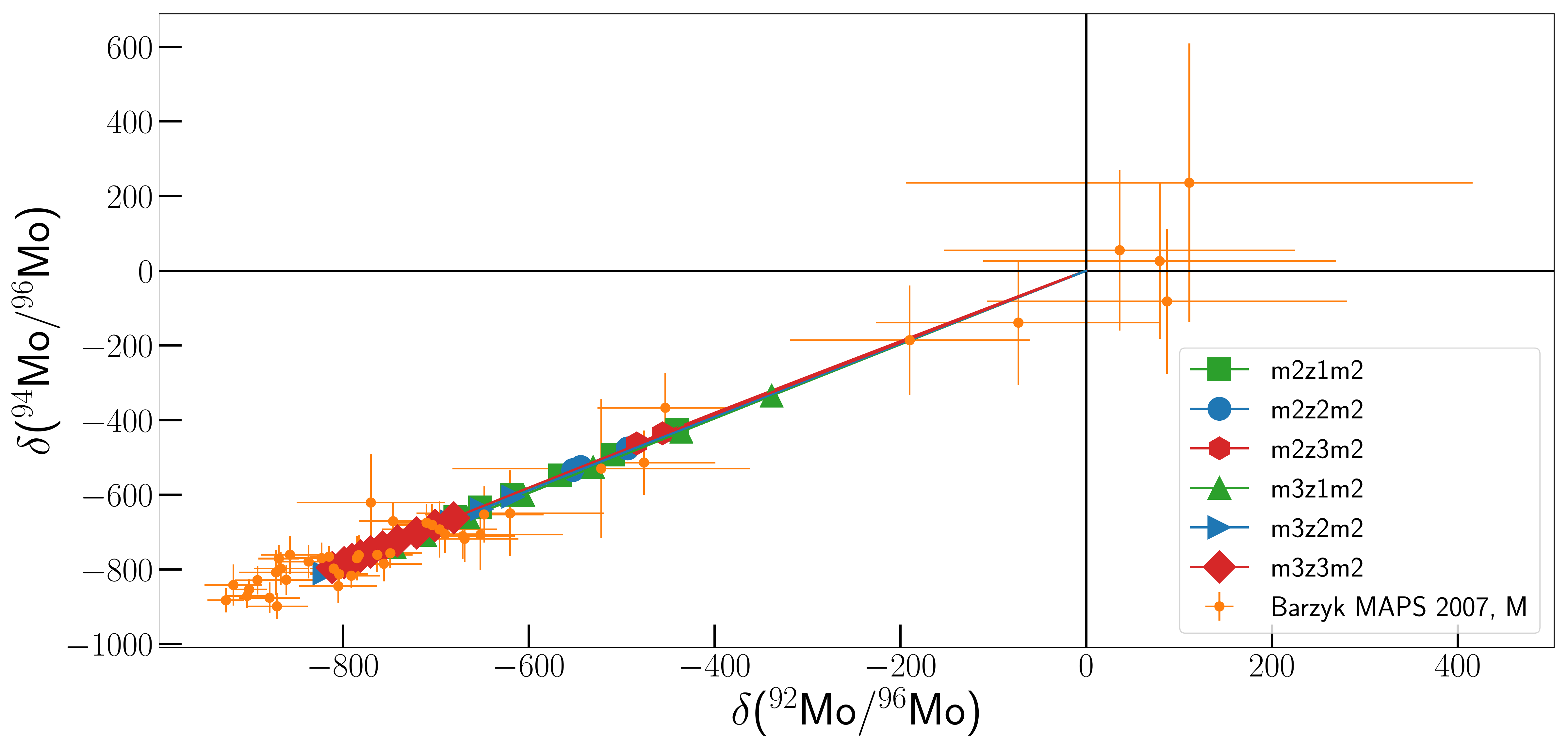}}}
\resizebox{10.8cm}{!}{\rotatebox{0}{\includegraphics{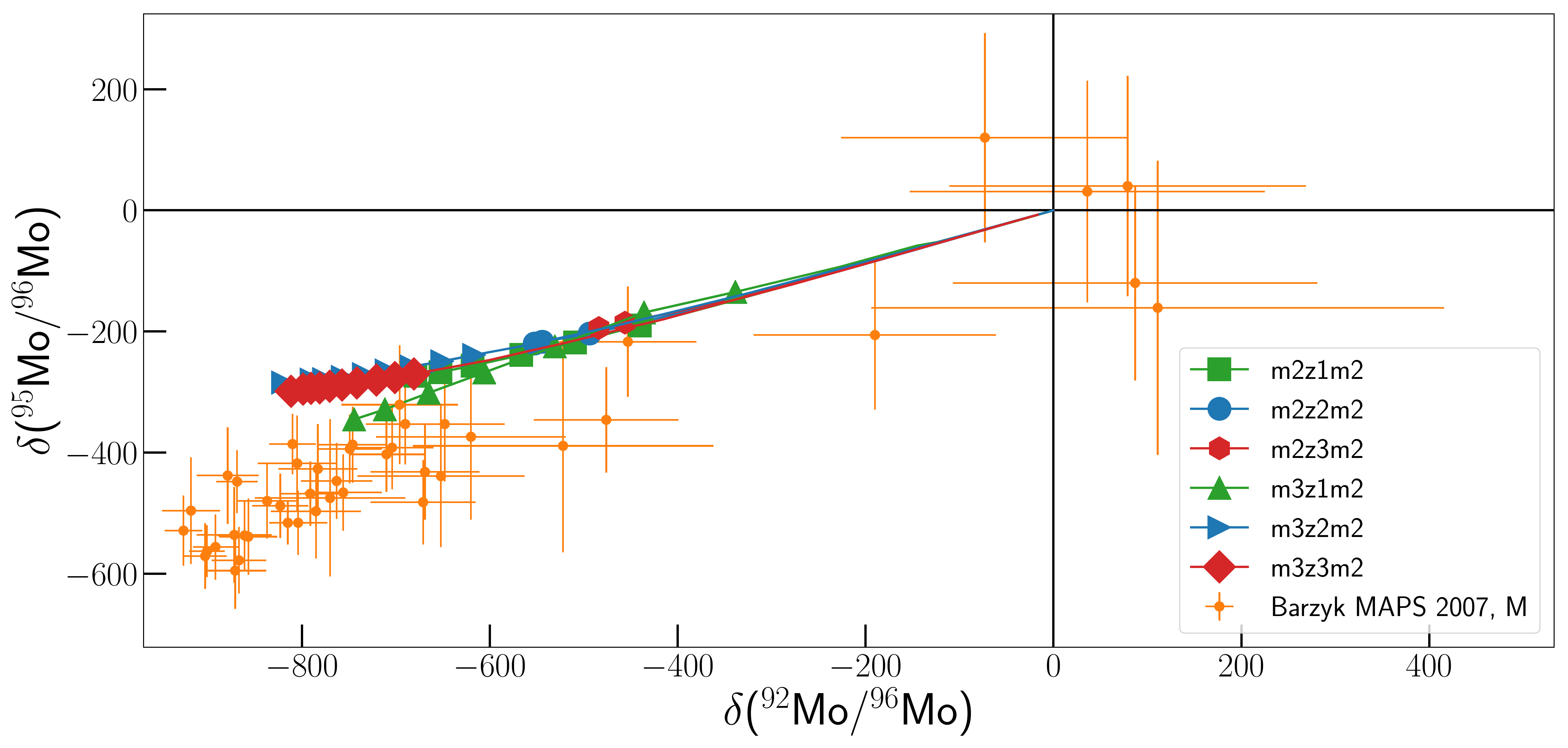}}}
\resizebox{10.8cm}{!}{\rotatebox{0}{\includegraphics{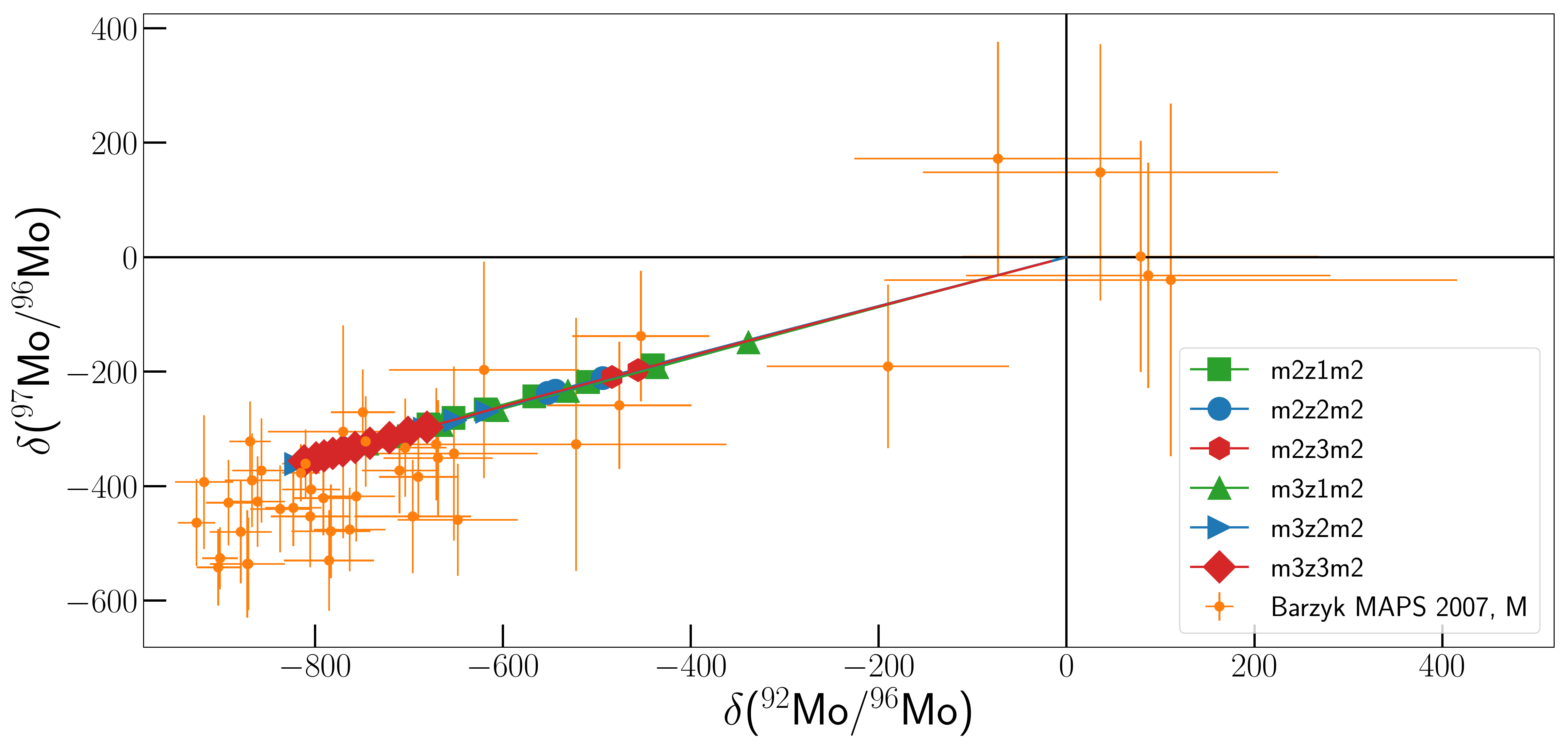}}}
\caption{Comparison of stellar models presented in this work with \cite{barzyk:07} measurements of Mo isotopic ratios.}
\label{mo:iso}
\end{figure}

\begin{figure}[htbp]
\centering
\resizebox{10.8cm}{!}{\rotatebox{0}{\includegraphics{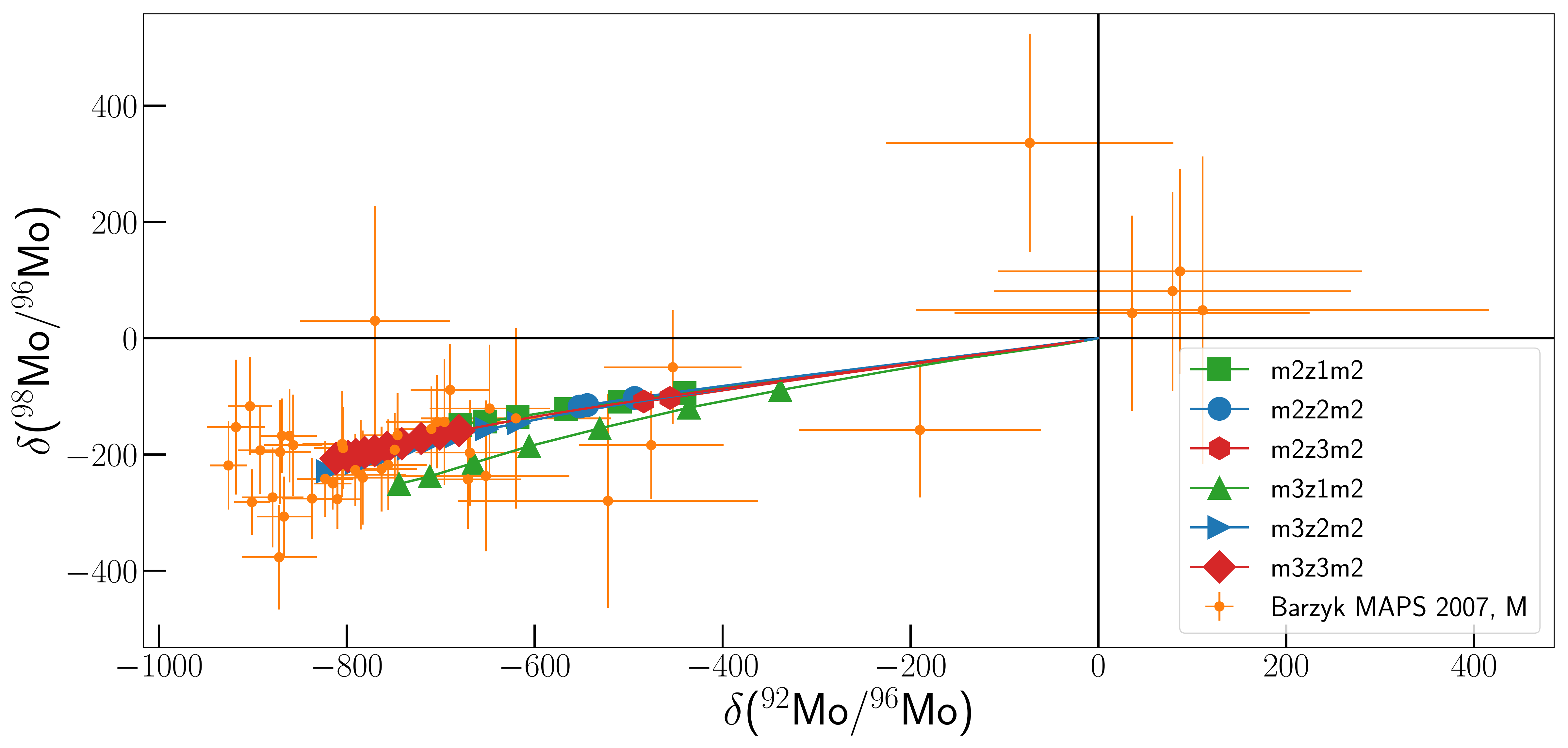}}}
\resizebox{10.8cm}{!}{\rotatebox{0}{\includegraphics{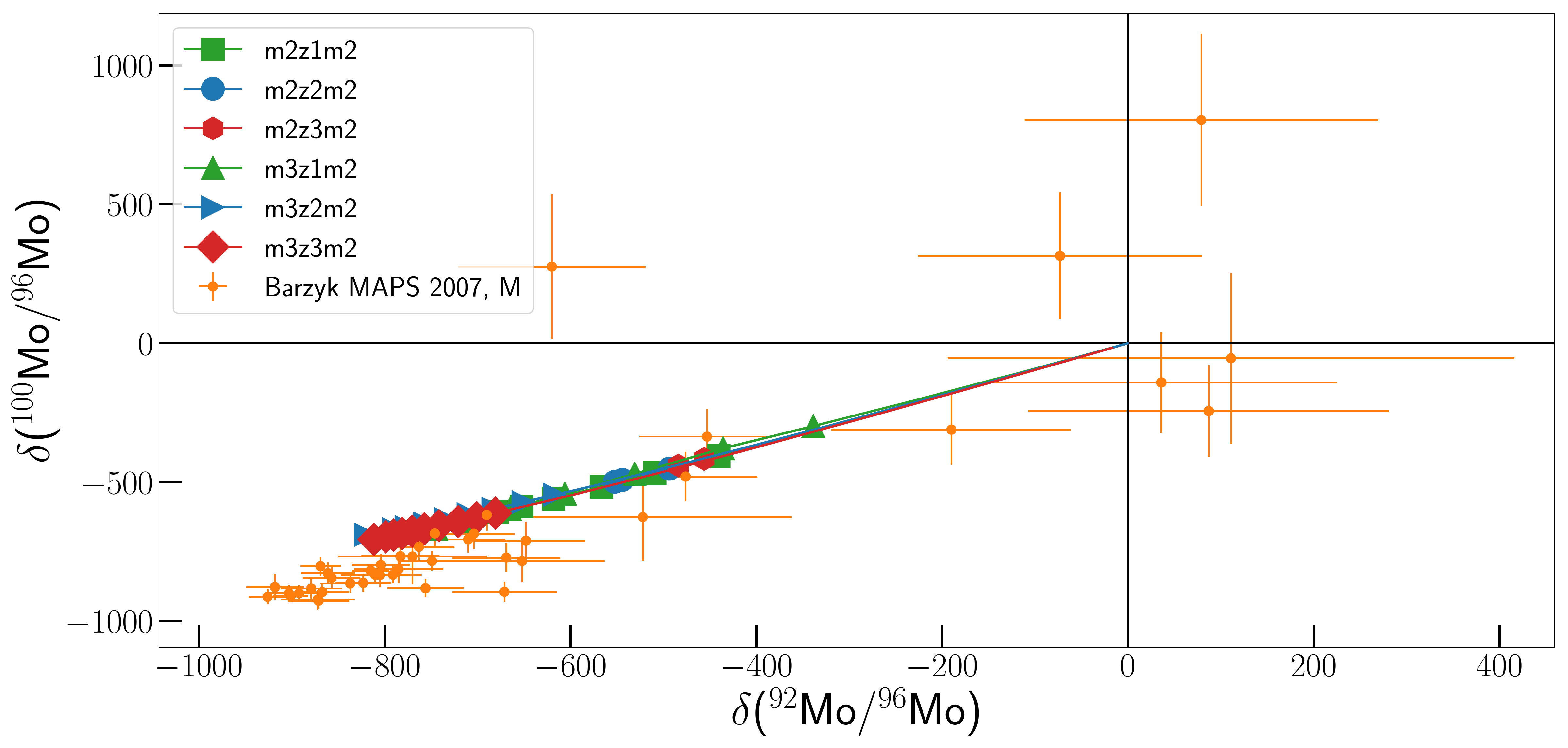}}}
\caption{Continuing figure \ref{mo:iso}.}
\label{mo:iso2}
\end{figure}

\begin{figure}[htbp]
\centering
\resizebox{10.8cm}{!}{\rotatebox{0}{\includegraphics{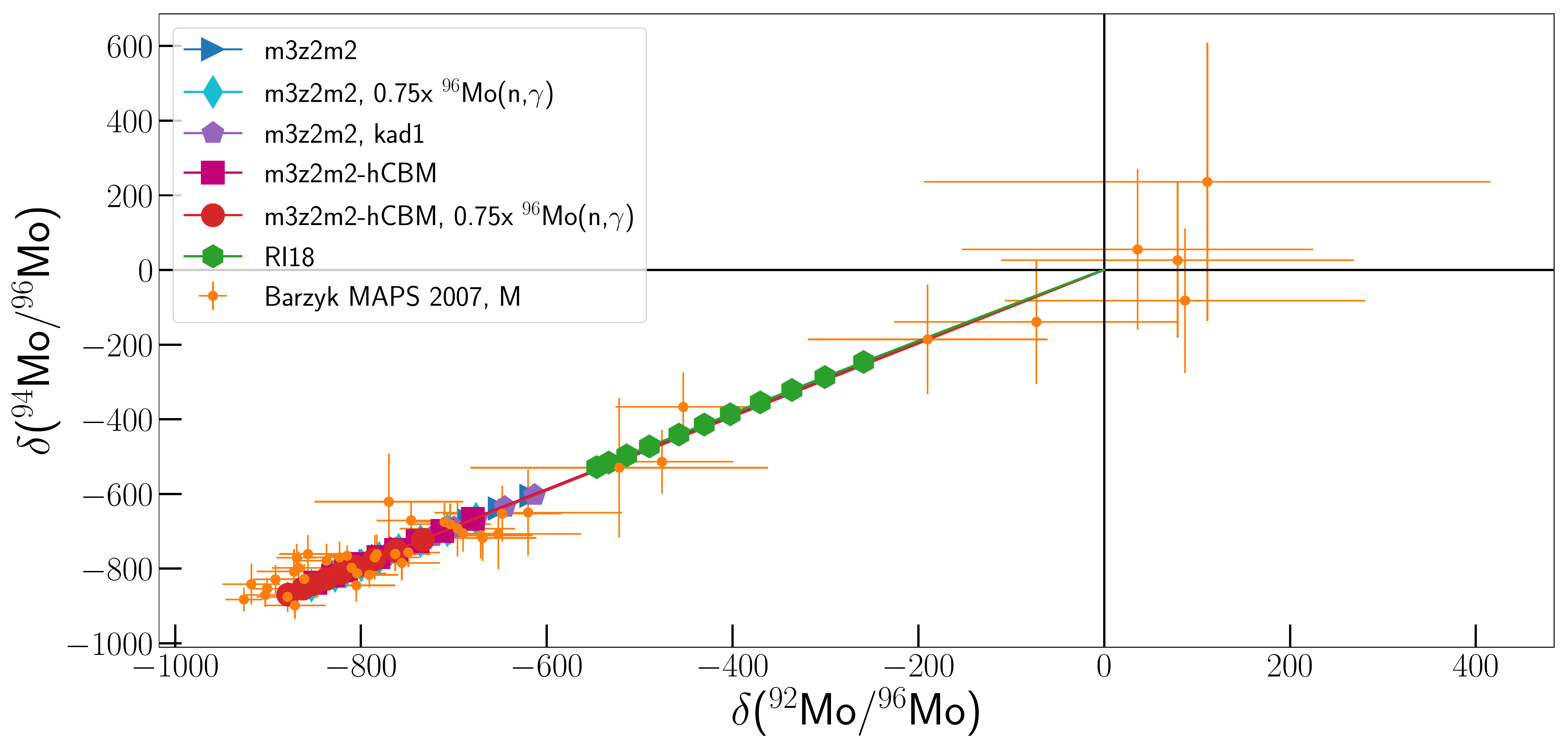}}}
\resizebox{10.8cm}{!}{\rotatebox{0}{\includegraphics{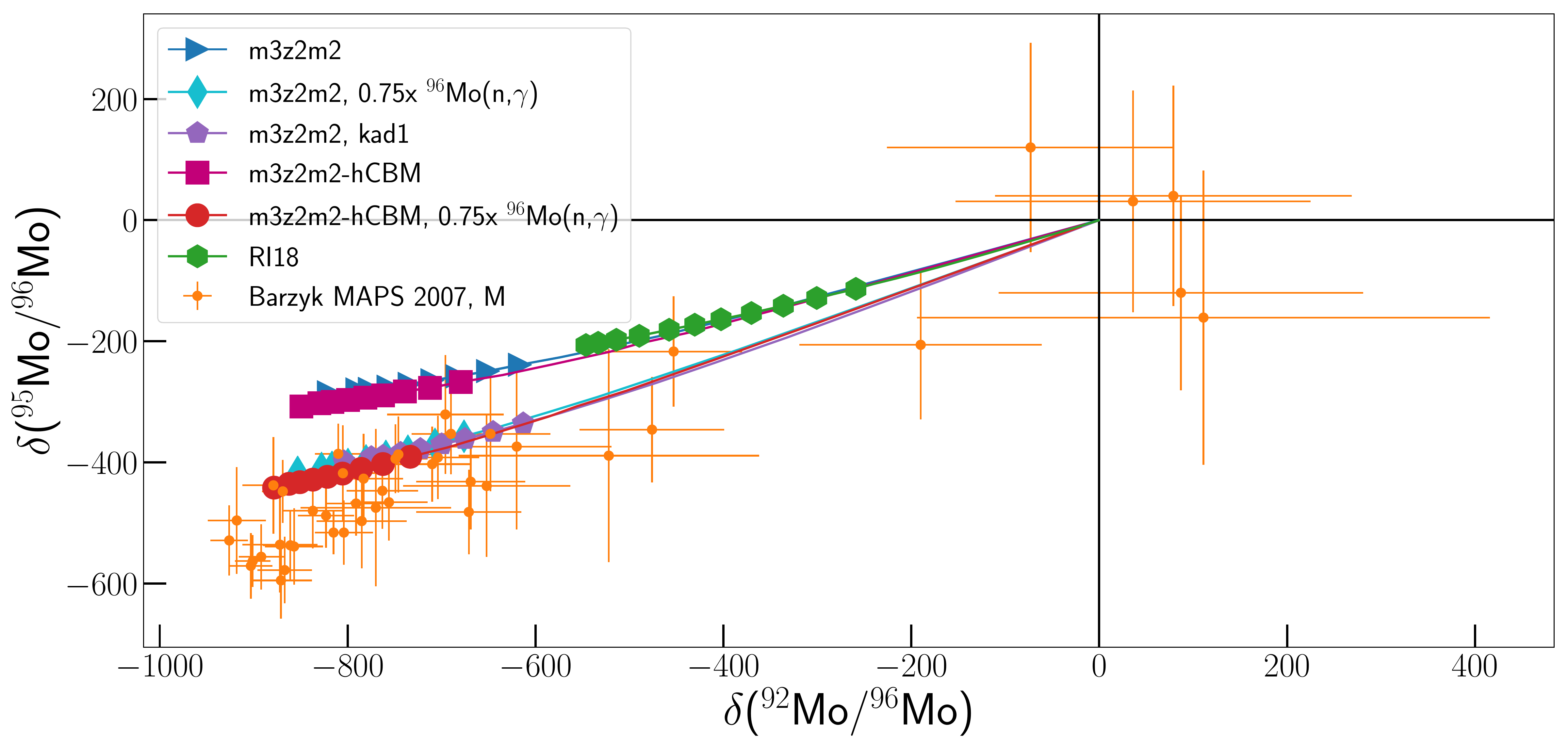}}}
\resizebox{10.8cm}{!}{\rotatebox{0}{\includegraphics{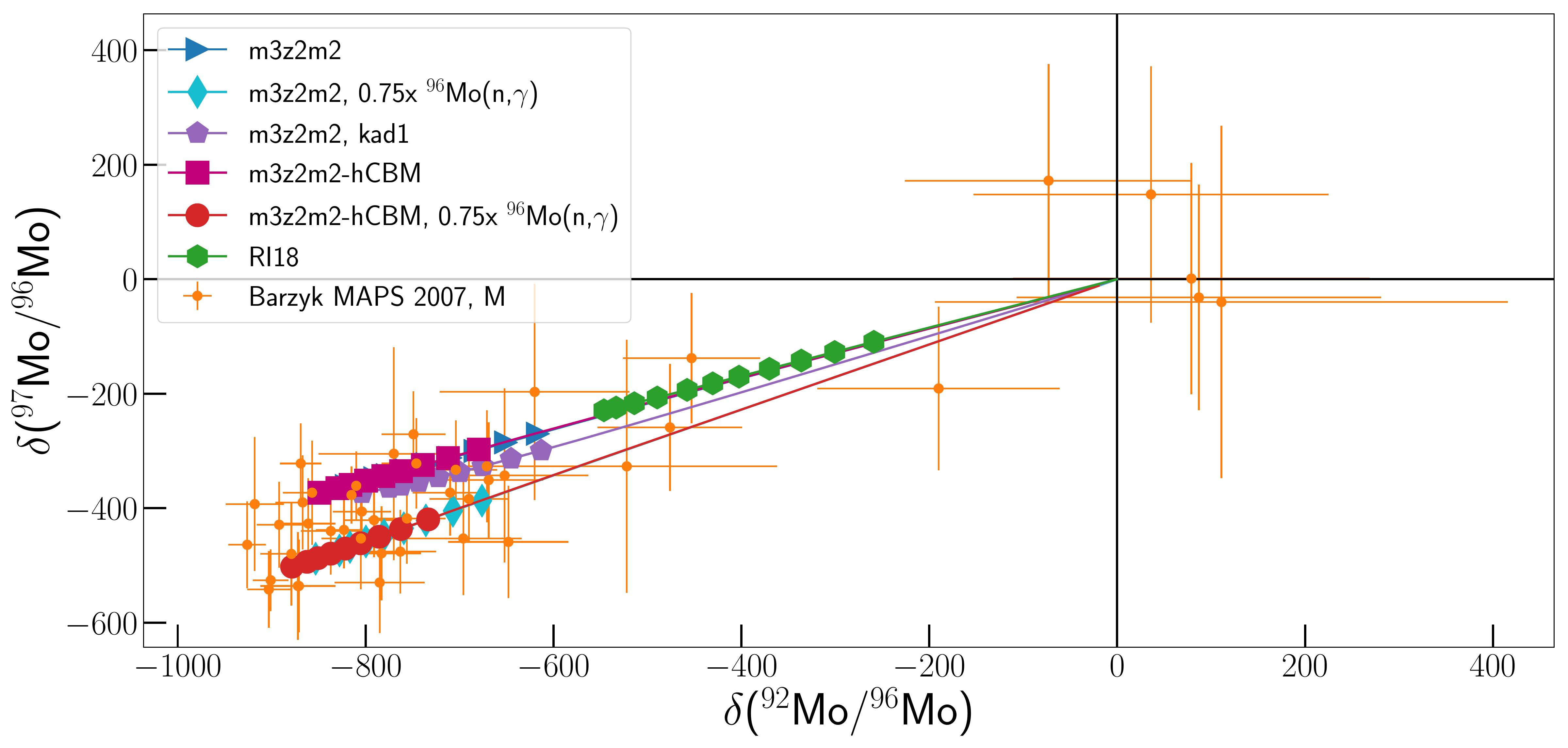}}}
\caption{Predictions from m3z2m2 model calculated with Kadonis 0.3,
  Kadonis 1.0 and Kadonis 0.3 with the ${^{96}}$Mo(n,$\gamma$)${^{97}}$Mo set to its lower limit (i.e. multiplied by a factor 0.75).}
\label{mo:nuctest}
\end{figure}

\begin{figure}[htbp]
\centering
\resizebox{10.8cm}{!}{\rotatebox{0}{\includegraphics{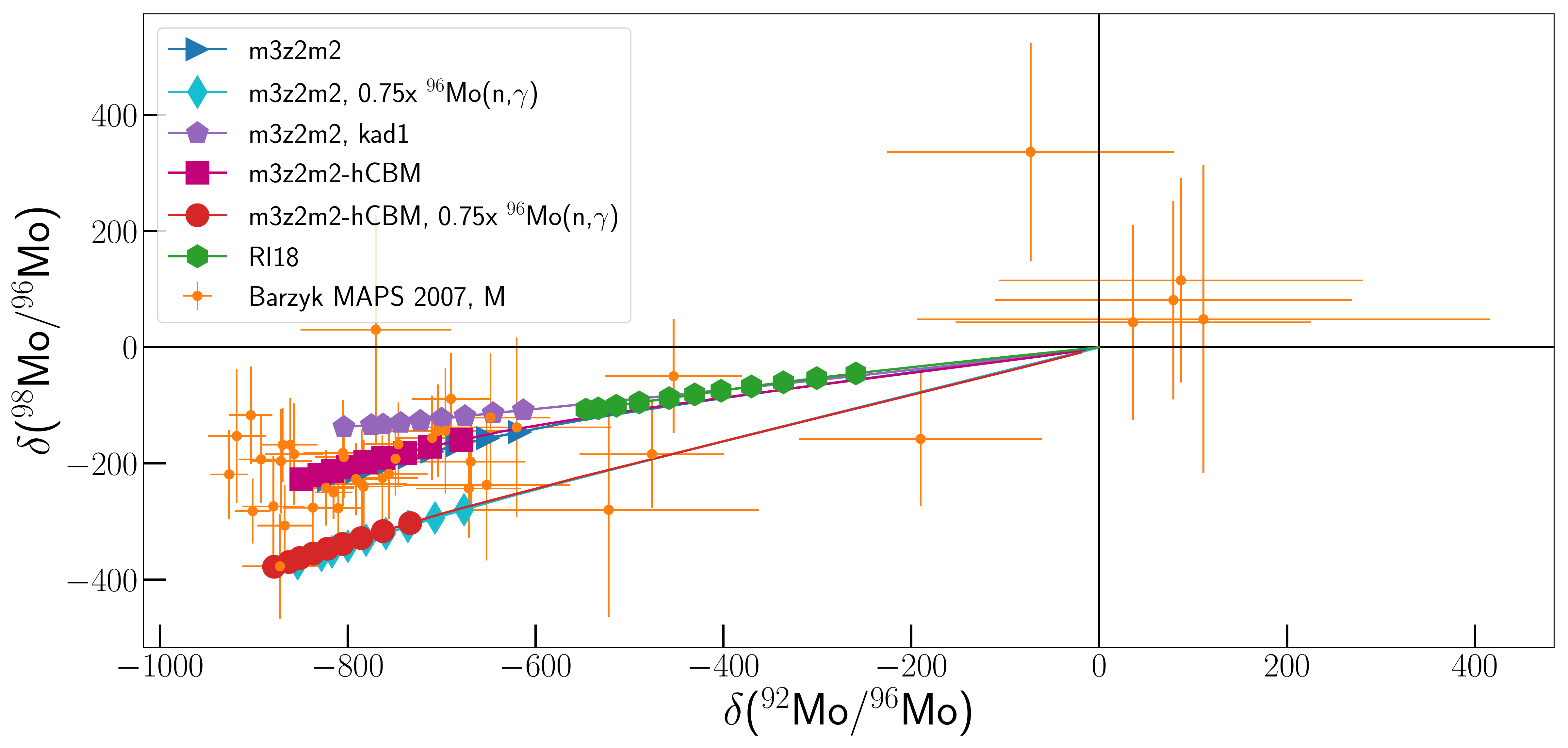}}}
\resizebox{10.8cm}{!}{\rotatebox{0}{\includegraphics{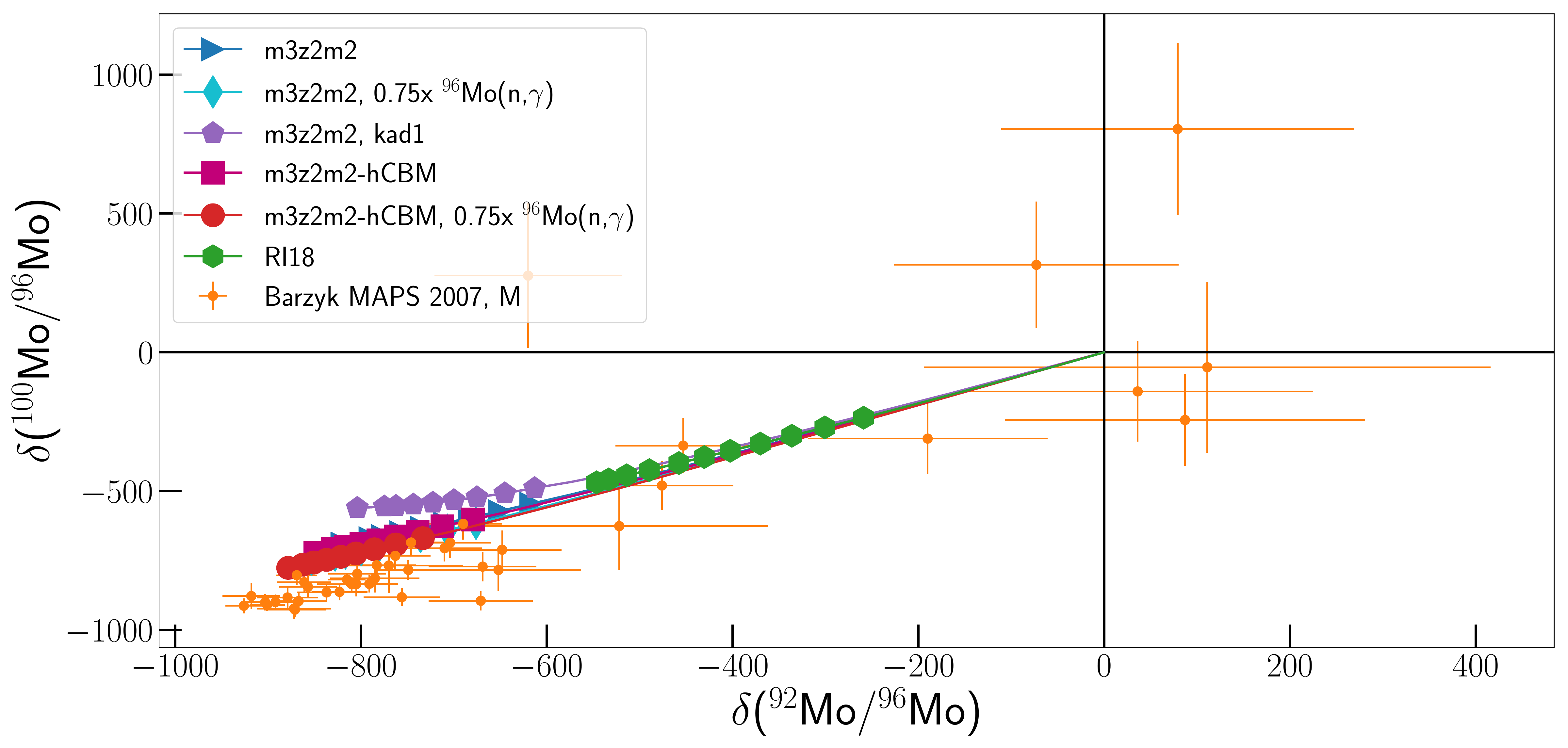}}}
\caption{Continuing figure \ref{mo:nuctest}.}
\label{mo:nuctest2}
\end{figure}

Figs. \ref{mo:iso} and \ref{mo:iso2} show predictions of isotopic-ratios compared to \cite{barzyk:07} measurements.
The agreement is not satisfactory good for ${^{92}}$Mo/${^{96}}$Mo, ${^{95}}$Mo/${^{96}}$Mo, ${^{97}}$Mo/${^{96}}$Mo and ${^{100}}$Mo/${^{96}}$Mo.
In particular, in both figures ${^{92}}$Mo looks like it is not burned enough.
Neutron captures on Mo isotopes are considerably different between Kadonis 0.3 and Kadonis 1.0,
with ${^{96}}$Mo(n,$\gamma$)${^{97}}$Mo also having an uncertainty around 20$\%$ at  ${^{13}}$C-pocket temperatures.
In figs. \ref{mo:nuctest} and \ref{mo:nuctest2} we show predictions from m3z2m2 and m3z3m2 calculated with Kadonis 0.3 and
m3z2m2 computed with Kadonis 1.0. We also show the predictions from m3z2m2 and m3z3m2-hCBM when Kadonis 0.3 is adopted,
but with the ${^{96}}$Mo(n,$\gamma$)${^{97}}$Mo from Kadonis 1.0 set to its lower limit (i.e. multiplied by a factor 0.8).
We also show the results from the M=3\msun\ , Z=0.02 model from RI18 as a comparison. First of all,
it is evident how the comparison with pre-solar grains is definitely improved compared to RI18 due to the larger \cdr-pocket.
The second aspect is that the observed ${^{97}}$Mo/${^{96}}$Mo range can actually be explained within nuclear-physics uncertainties.
Finally, considering m3z2m2-hCBM,
we notice how the bunch of grains with the lowest ${^{92}}$Mo/${^{96}}$Mo observed could be reproduced with one more TDU event,
which is well inside model uncertainties.
On the other hand, even considering both nuclear and model uncertainties, our models are not able to reproduce the measured spread of  ${^{95}}$Mo/${^{96}}$Mo and ${^{100}}$Mo/${^{96}}$Mo.


\subsection{Key reaction rates}
\label{sec:keyreac} 

Table \ref{tab:keyreac} shows the reaction rates we found important when comparing our results to observations.
We identified six key reactions, five of them being neutron captures.
Additionally, four out of five of these (n,$\gamma$) reactions have been classified as '$Level$ $1$' key-rates by \cite{cescutti:18},
which means they showed a strong correlation to the abundances of specific $s$-process isotopes listed in the second column.
We hence agree with \cite{cescutti:18} and propose them as candidates for improved measurement by future experiments,
since more precise measurement of these rates will allow significantly more precise nucleosynthesis calculations.
We also highlight the importance of ${^{95}}$Zr(n,$\gamma$)${^{96}}$Zr,
which is classified as '$Level$ $2$' key rate by \cite{cescutti:18}, hence less correlated to final abundances than '$Level$ $1$' key-rates.
Despite this fact, we notice that this rate is actually the main source of the still significant uncertainty affecting ${^{96}}$Zr.
Finally, we include in the list also the main neutron source during TP events, the  $^{22}$Ne($\alpha$,n)$^{25}$Mg,
affecting branching-point sensitive isotopes like ${^{87}}$Rb and ${^{96}}$Zr.

\subsection{Ejected yields}
\label{sec:yields}

We have calculated full yields for all our models. These are available in Tables online at the CADC
(The Canadian Astronomical Data Center, \url{http://www.cadc-ccda.hia-iha.nrc-cnrc.gc.ca})
and can be analyzed interactively through the web interface WENDI at \url{wendi.nugridstars.org}.
Table \ref{tab:yields} shows a comparison between
the yields presented in this work for $m3z2m2$ and the yields presented by
\cite{karakas:10a} and \cite{cristallo:11} for their models with same initial mass and metallicity.
The final ejected masses of $^{12}$C, $^{14}$N, and $^{16}$O for $m3z2m2$ are 0.0340, 0.0070, and 0.0316 M$_{\odot}$ respectively.
For the same isotopes and the same star, RI18 provides 0.0445, 0.0077 and 0.0383, \cite{karakas:10a} 0.0207, 0.0056,
and 0.0211, and \cite{cristallo:11} 0.0186, 0.0066 and 0.0211.
For $^{12}$C we obtain an abundance that is factor of 1.83 and 1.64 higher than \cite{cristallo:11} and \cite{karakas:10a}.
A higher $^{12}$C enrichment in our models is due to the CBM activated at the bottom of the PDCZ.
The consistent amount of ejected $^{12}$C with RI18 is a consequence of the same CBM scheme adopted at the bottom of the intershell.
The $^{14}$N yields are consistent within 20$\%$.
Concerning $^{16}$O, m3z2m2 show a larger production, up to 60$\%$ compared to \cite{cristallo:11} and \cite{karakas:10a},
while it is consistent with RI18. As for $^{12}$C,
this higher production corresponds to the CBM scheme we applied during the TP,
that both \cite{cristallo:11} and \cite{karakas:10a} do not include.
Concerning the $s$-process nucleosynthesis, the final ejected masses from m3z2m2 of $^{88}$Sr, $^{138}$Ba and $^{208}$Pb
are 3.34$\times$10$^{-7}$, 1.96$\times$10$^{-7}$ and 3.28$\times$10$^{-8}$ M$_{\odot}$ respectively.

\cite{cristallo:11} predicts a much higher production of $^{88}$Sr,
which is 1.63$\times$10$^{-6}$ M$_{\odot}$, due to the smaller $^{13}$C-pockets obtained in our models.
On the other hand, the calculated ejected amount of $^{138}$Ba and $^{208}$Pb are consistent with our results being 1.69$\times$10$^{-7}$ and
4.82$\times$10$^{-8}$  M$_{\odot}$ respectively,
which is explained by the higher neutron-exposure of our models originating from the  higher $^{12}$C enrichment in the intershell.

RI18 predicts on average a factor of two lower production,
with 1.98$\times$10$^{-7}$, 7.58$\times$10$^{-8}$, 2.24$\times$10$^{-8}$ M$_{\odot}$ of  $^{88}$Sr, $^{138}$Ba and $^{208}$Pb respectively.
This is due to the $^{13}$C-pockets being between two and three times larger in our models compared to RI18,
but with the same peak abundance of $^{13}$C in them which give similar [hs/ls] values,
again because of the same CBM scheme adopted at the bottom of the intershell.

\section{Discussion and conclusions}
\label{sec:conclusions}

In this work we presented a significant update of low-mass AGB star models and nucleosynthesis calculations presented in RI18.
In that work, the $s$-process production was low compared to observations.
We tackled this by re-calculating the low-mass AGB models with the same stellar code,
general input physics  parameters with the only difference being describing the convective boundaries during TDU events according to the scheme described by \cite{battino:16}, which was based on the IGW-mixing scenario described by \cite{denissenkov:03}.
The direct consequence of this is a $^{13}$C-pocket up to three times larger in mass-coordinate than in RI18,
with the final $s$-process production increasing by almost a factor of three and now in much better agreement with observations.
One additional difference compared to RI18 is the inclusion of the additional metallicity Z=0.03,
since its contribution to dust production and hence presolar grains can be very significant \citep{lugaro:18}.
Moreover, we compute two additional models (labelled '-hCBM') with metallicities Z=0.02 and 0.03 with an increased CBM under the convective envelope during TDUs.
This increased CBM-efficiency is well inside the uncertainties characterizing the IGW-mixing parameterization of \cite{denissenkov:03},
and produces a $^{13}$C-pocket about 50$\%$ larger compared to when the standard setting is adopted.

We validated our results by comparing them with a large set of observables,
including carbon-stars and barium-stars surface abundances inferred from spectroscopy and isotopic-ratios from presolar grains.
We noticed how '-hCBM' models, forming a larger $^{13}$C-pocket, generally performs better when compared to observations.
This indicates how uncertainties affecting CBM impacts our \spr\ results,
motivating us to do a future dedicated study before completing re-computing the whole metallicity grid of RI18.
Within all uncertainties (stellar modeling, nuclear physics and observations) our models agree with most of observational data.
The most difficult observables to be reproduced are the full ranges of some isotopic-ratios,
precisely ${^{84}}$Sr/${^{86}}$Sr, ${^{137}}$Ba/${^{136}}$Ba, ${^{96}}$Zr/${^{94}}$Zr, ${^{95,100}}$Mo/${^{96}}$Mo.

We explored the role of rotation-induced mixing adopting a simple parametric approach,
confirming it as a strong candidate to explain the range of observed values in presolar grains.
It is anyway important to notice our assumption about stellar rotation being an efficient extra-mixing source in AGB stars,
even though this is actually still a matter of debate \citep[see][]{straniero:15,deheuvels:15,herwig:05}.


We identified a number of reaction-rates that impact our results, some of which already been classified as key-reaction
rate for AGB nucleosynthesis by \cite{cescutti:18}. We therefore want to suggest them as priority candidates for future measurements.

Finally, we provide the final ejected yields from our models that can be used as inputs for galactic-chemical evolution simulations.
Full tables are available online as described in the text.

This research was enabled in part by support provided by WestGrid (www.westgrid.ca) and Compute Canada Calcul Canada (www.computecanada.ca).
NuGrid data is served by Canfar CADC. This work received support from the Science and Technology Facilities Council UK (ST/M006085/1),
and the European Research Council ERC-2015-STG Nr. 677497 and ERC-2016-CO Grant 724560. Part of this work was performed under the auspices of the U.S. Department of Energy by Lawrence Livermore National Laboratory under Contract DE-AC52-07NA27344 and was supported by the LLNL-LDRD Program under Project No. 19-LW-033. LLNL-JRNL-765023. This work used the DiRAC Complexity system, operated by the University of Leicester IT Services, which forms part of the STFC DiRAC HPC Facility (www.dirac.ac.uk ). This equipment is funded by BIS National E-Infrastructure capital grant ST/K000373/1 and  STFC DiRAC Operations grant ST/K0003259/1. DiRAC is part of the National E-Infrastructure. RH acknowledges support from the World Premier International Research Centre Initiative (WPI Initiative), MEXT,Japan. This article is based upon work from the ChETEC COST Action (CA16117), supported by COST (European Cooperation  in  Science  and  Technology).

\bibliography{astro}

\clearpage

\begin{sidewaystable}
\begin{center}
  \caption{Main properties of the low-mass AGB models: initial mass, initial metallicity, H-free core mass at the beginning and the end of the AGB phase and total lifetime are given.
    Core masses and total lifetimes obtained in RI18 for models with same mass/metallicity combinations are also presented. 
}
\begin{tabular}{lcccccccc}
\hline
name & mass [M$_{\odot}$] & Z &  H-free M$_{ini}$ [M$_{\odot}$]  & H-free M$_{end}$ [M$_{\odot}$]  & $\tau$ $_{tot}$ [yrs] & H-free M$_{ini}$ RI18 & H-free M$_{end}$ RI18 [M$_{\odot}$] & $\tau$ $_{tot}$ RI18\\
\hline
\hline
m2z1m2 & 2  & 0.01 & 0.502 & 0.629 & 1.28$\times$10$^{9}$ & 0.498 & 0.617 & 1.28$\times$10$^{9}$  \\ 		
m3z1m2 & 3  & 0.01 & 0.644 & 0.669 & 4.13$\times$10$^{8}$ & 0.646 & 0.659 & 4.13$\times$10$^{8}$ \\ 	
m2z2m2 & 2  & 0.02 & 0.508 & 0.646 & 1.40$\times$10$^{9}$ & 0.510 & 0.620 & 1.42$\times$10$^{9}$ \\ 		
m3z2m2 & 3  & 0.02 & 0.599 & 0.659 & 4.85$\times$10$^{8}$ & 0.596 & 0.642 & 4.82$\times$10$^{9}$ \\ 	
m2z3m2 & 2  & 0.03 & 0.511 & 0.643 & 1.71$\times$10$^{9}$ & - & - & - \\ 		
m3z3m2 & 3  & 0.03 & 0.562 & 0.650 & 6.03$\times$10$^{8}$ & - & - & - \\ 	

\noalign{\smallskip}
\hline
\end{tabular}
\label{tab:model_def}
\end{center}
\end{sidewaystable}

\begin{table}
\begin{center}
  \caption{The CBM parameters adopted during TDU events are given (see text for details) for the models shown in Table \ref{tab:model_def}. The 3 M$_{\odot}$ , Z=0.02 model from RI18 is shown as a comparison to model m3z2m2. We added two additional model with Z=0.02 and Z=0.03 to test the impact of more efficient TDUs.
}
\begin{tabular}{lccc}
\hline
Name & $f_{1}$ & D$_{2}$[cm$^{2}$s$^{-1}$] & $f_{2}$ \\
\hline
\hline
m2z1m2 & 0.014 & 10$^{11}$ &  0.27 \\ 		
m3z1m2 & 0.014 & 10$^{11}$ &  0.27 \\ 	
\hline
m2z2m2 & 0.014 & 10$^{11}$ &  0.27 \\ 		
m3z2m2 & 0.014 & 10$^{11}$ &  0.27 \\
RI18   & 0.126 & - & -  \\ 	
\hline
m2z3m2 & 0.014 & 10$^{11}$ &  0.27 \\ 		
m3z3m2 & 0.014 & 10$^{11}$ &  0.27 \\
\hline
m3z2m2-hCBM & 0.014 & 4.3$\times$10$^{11}$ &  0.27 \\ 	
m3z3m2-hCBM & 0.014 & 4.3$\times$10$^{11}$ &  0.27 \\ 	

\noalign{\smallskip}
\hline
\end{tabular}
\label{tab:cbm}
\end{center}
\end{table}

\begin{table}
\begin{center}
  \caption{Total number of TPs and number of TPs occurring during the AGB oxygen-rich phase for the same models shown in Table \ref{tab:cbm}. The 3 M$_{\odot}$, Z=0.02 model from RI18 is shown as a comparison to models m3z2m2 and m3z2m2-hCBM.
}
\begin{tabular}{lcc}
\hline
Name & Total TPs & O-rich TPs \\
\hline
\hline
m2z1m2 & 25 & 18 \\ 		
m3z1m2 & 16 &  8 \\ 	
\hline
m2z2m2 & 30 & 27 \\ 		
m3z2m2 & 24 & 15 \\
m3z2m2-hCBM & 23 & 15 \\ 	
RI18   & 21 & 13 \\ 	
\hline
m2z3m2 & 30 & 29 \\ 		
m3z3m2 & 31 & 20 \\
m3z3m2-hCBM & 30 & 19 \\ 	

\noalign{\smallskip}
\hline
\end{tabular}
\label{tab:model_agb}
\end{center}
\end{table}

\begin{sidewaystable}
\begin{center}
  \caption{List of AGB models with additional internal constant mixing to mimic the effects of rotation: initial mass, initial metallicity, initial and final H-free core mass, total lifetime and logarithmic value of constant mixing coefficient are given. 
   For comparison, also m3z2m2 and m3z3m2 models already introduced in Table \ref{tab:model_def} are presented.}
\begin{tabular}{lcccccccc}
\hline
Name & Ini mass [M$_{\odot}$] & Ini Z &  H-Free M$_{Ini}$ [M$_{\odot}$]  & H-Free M$_{end}$ [M$_{\odot}$]  & $\tau$ $_{tot}$ [yrs] & Total TPs & TPs O-rich & Log$_{10}$(D$_{mix}$) \\
\hline
\hline
m3z2m2 & 3  & 0.02 & 0.599 & 0.659 & 4.85$\times$10$^{8}$ & 24 & 15 & - \\
m3z2m2-rotmix.stx2 & 3  & 0.02 & 0.599 & 0.657 & 4.83$\times$10$^{8}$ & 24 & 17 & -0.4 \\
m3z2m2-rotmix.st & 3  & 0.02 & 0.599 & 0.659 & 4.83$\times$10$^{8}$ & 25 & 16 & -0.7 \\
m3z2m2-rotmix.std2 & 3  & 0.02 & 0.599 & 0.656 & 4.83$\times$10$^{8}$ & 25 & 16 & -1 \\ 		
\hline
m3z3m2 & 3  & 0.03 & 0.562 & 0.650 & 6.03$\times$10$^{8}$ & 31 & 20 & - \\
m3z3m2-rotmix.st & 3  & 0.03 & 0.562 & 0.645 & 6.05$\times$10$^{8}$ & 30 & 21 & -0.7 \\
m3z3m2-hCBM-rotmix.stx1p5 & 3  & 0.03 & 0.562 & 0.639 & 6.05$\times$10$^{8}$ & 30 & 19 & -0.5 \\
m3z3m2-hCBM-rotmix.st & 3  & 0.03 & 0.562 & 0.640 & 6.05$\times$10$^{8}$ & 31 & 20 & -0.7 \\

\noalign{\smallskip}
\hline
\end{tabular}
\label{tab:modrot}
\end{center}
\end{sidewaystable}

\begin{table}
\begin{center}
  \caption{List of the reaction rates with the highest impact on the observables we considered in this work. Also the most affected affected isotope by each reaction is shown.
  For neutron capture rates, we add the classification and the main nuclide affected given by \cite{cescutti:18}.}
\begin{tabular}{lcc}
\hline
Reaction rate & Affected observable & Cescutti et al. 2018 classification \\
\hline
\hline
${^{137}}$Ba(n,$\gamma$)${^{138}}$Ba & ${^{137}}$Ba/${^{136}}$Ba  & Level 1 for ${^{137}}$Ba \\
${^{95}}$Mo(n,$\gamma$)${^{96}}$Mo & ${^{95}}$Mo/${^{96}}$Mo & Level 1 for ${^{95}}$Mo \\
${^{96}}$Mo(n,$\gamma$)${^{97}}$Mo & ${^{94,95,97,98,100}}$Mo/${^{96}}$Mo & Level 1 for ${^{96}}$Mo \\
${^{94}}$Zr(n,$\gamma$)${^{95}}$Zr & ${^{90,91,92,96}}$Zr/${^{94}}$Zr & Level 1 for ${^{96}}$Zr\\ 	
${^{95}}$Zr(n,$\gamma$)${^{96}}$Zr & ${^{96}}$Zr/${^{94}}$Zr & Level 2 for ${^{96}}$Zr\\ 	
${^{22}}$Ne($\alpha$,n)${^{25}}$Mg & ${^{96}}$Zr/${^{94}}$Zr & - \\

\noalign{\smallskip}
\hline
\end{tabular}
\label{tab:keyreac}
\end{center}
\end{table}

\begin{table}
\begin{center}
  \caption{Comparison between the yields in solar masses for $m3z2m2$ and the yields from
    \cite{karakas:10a} (Ka10), \cite{cristallo:11} (Cr11) and RI18 for their model with same initial mass and metallicity.}
\begin{tabular}{lccccc}
\hline
Isotope  &   m3z2m2   & m3z2m2-hCBM &    RI18   &  Cr11     &   Ka10 \\
\hline
C 12     &  3.4035341E-02  &  3.43558877e-02   & 4.448E-02  &     1.86110E-02     &   2.0739544E-02\\
C 13     &  2.2231664E-04  &  2.18517069e-04   & 2.252E-04  &     2.20200E-04     &   1.9436399E-04\\
N 14     &  7.0151267E-03  &  6.90119758e-03   & 7.685E-03  &     6.64840E-03     &   5.6565693E-03\\
N 15     &  4.4046608e-06  &  4.25324193e-06   & 4.207E-06  &     4.29400E-06     &   5.0818235E-06\\
O 16     &  3.1633632E-02  &  3.22309748e-02   & 3.828E-02  &     1.94360E-02     &   2.1144016E-02\\
O 17     &  6.2956708e-05  &  6.20965699e-05   & 5.194E-05  &     7.91850E-05     &   5.5763638E-05\\
O 18     &  3.4374702e-05  &  3.34363411e-05   & 3.364E-05  &     3.12110E-05     &   3.6596495E-05\\
F 19     &  4.9649986e-06  &  4.92340864e-06   & 7.655E-06  &     3.68770E-06     &   4.3487280E-06\\
NE 20    &  4.2225594E-03  &  4.13259733e-03   & 4.356E-03  &     3.63520E-03     &   3.7571993E-03\\
NE 21    &  1.2744558e-05  &  1.24662448e-05   & 1.270E-05  &     9.90460E-06     &   1.0039988E-05\\
NE 22    &  2.7187524E-03  &  2.82343479e-03   & 3.937E-03  &     2.32210E-03     &   2.1113991E-03\\
NA 23    &  1.4976682E-04  &  1.49833381e-04   & 1.772E-04  &     1.87730E-04     &   1.2845088E-04\\
MG 24    &  1.3712457E-03  &  1.34172081e-03   & 1.421E-03  &     1.84710E-03     &   1.1949923E-03\\
MG 25    &  2.9955618E-04  &  2.92761999e-04   & 2.915E-04  &     2.43210E-04     &   1.6784266E-04\\
MG 26    &  4.5746367E-04  &  4.54130815e-04   & 4.726E-04  &     2.88120E-04     &   1.9374024E-04\\
AL 27    &  1.5420979E-04  &  1.50610193e-04   & 1.585E-04  &     2.08100E-04     &   1.3861095E-04\\
SI 28    &  1.7175335E-03  &  1.67961680e-03   & 1.770E-03  &     2.36270E-03     &   1.5164100E-03\\
SI 29    &  9.2162458e-05  &  9.01472831e-05   & 9.501E-05  &     1.24570E-04     &   7.9920115E-05\\
SI 30    &  6.7840128e-05  &  6.65214218e-05   & 6.975E-05  &     8.60130E-05     &   5.5390818E-05\\
P 31     &  1.7451779e-05  &  1.71499494e-05   & 1.771E-05  &     2.28230E-05     &   1.9017965E-05\\
S 33     &  7.9125730e-06  &  7.74800508e-06   & 8.331E-06  &     1.04160E-05     &   7.6937777E-06\\
S 34     &  4.5194069e-05  &  4.42301243e-05   & 4.601E-05  &     5.91400E-05     &   4.3391171E-04\\
FE 54    &  1.8268714E-04  &  1.78490960e-04   & 1.874E-04  &     2.49280E-04     &   1.6390771E-04\\
FE 56    &  3.0096567E-03  &  2.94198314e-03   & 3.100E-03  &     4.08090E-03     &   2.7071363E-03\\
FE 57    &  8.1660429e-05  &  8.00810684e-05   & 8.781E-05  &     1.02140E-04     &   7.2351380E-05\\
FE 58    &  2.9018441e-05  &  2.90010573e-05   & 3.211E-05  &     1.76610E-05     &   1.1919641E-05\\
CO 59    &  1.3762730e-05  &  1.36096972e-05   & 1.441E-05  &     1.33220E-05     &   8.5931824E-06\\
NI 58    &  1.2853806E-04  &  1.25593342e-04   & 1.317E-04  &     1.71330E-04     &   1.1363259E-04\\
NI 60    &  5.5129775e-05  &  5.40953096e-05   & 5.698E-05  &     6.93570E-05     &   4.5602490E-05\\
NI 61    &  3.9916327e-06  &  3.97199615e-06   & 4.122E-06  &     3.46150E-06     &   8.8770785E-06\\
NI 62    &  1.0936648e-05  &  1.08258788e-05   & 1.056E-05  &     1.04870E-05     &   5.0042019E-08\\
NI 64    &  4.0636092e-06  &  4.02730786e-06   & 3.165E-06  &     3.32430E-06     &    -\\
SR 88    &  3.3440015e-07  &  3.80870156e-07   & 1.978E-07  &     1.63060E-06     &    -\\
Y 89     &  9.2306586e-08  &  1.06181661e-07   & 5.072E-08  &     3.17410E-07     &    -\\
ZR 90    &  9.6840845e-08  &  1.11106954e-07   & 5.726E-08  &     3.24600E-07     &    -\\
BA136    &  2.1123224e-08  &  2.61090422e-08   & 9.317E-09  &     2.79920E-08     &    -\\
BA138    &  1.9605483e-07  &  2.43881700e-07   & 7.581E-08  &     1.68590E-07     &    -\\
LA139    &  2.2851196e-08  &  2.82865328e-08   & 9.138E-09  &     2.00170E-08     &    -\\
PB208    &  3.2787968e-08  &  3.75103148e-08   & 2.243E-08  &     4.82470E-08     &    -\\ 
\noalign{\smallskip}
\hline
\end{tabular}
\label{tab:yields}
\end{center}
\end{table}



\end{document}